\RequirePackage{filecontents}

\RequirePackage{snapshot}
\documentclass[reprint,aps,pra,floatfix,longbibliography,secnumarabic]{preprint}

\renewcommand\documentclass[2][]{}
\let\origparagraph=\paragraph
\AtBeginDocument{%
  \let\paragraph=\origparagraph
}
\makeatother
\documentclass[reprint,aps,pra,floatfix,longbibliography]{revtex4-1}

\makeatletter
\def\@bibdataout@aps{%
 \immediate\write\@bibdataout{%
  @CONTROL{%
   apsrev41Control%
   \longbibliography@sw{%
    ,author="08",editor="1",pages="0",title="0",year="0"%
   }{%
    ,author="08",editor="1",pages="0",title="",year="1"%
   }%
  }%
 }%
 \if@filesw
  \immediate\write\@auxout{\string\citation{apsrev41Control}}%
 \fi
}%
\makeatother

\usepackage[T1]{fontenc}
\usepackage{amsmath}
\usepackage{textcomp}
\usepackage{txfonts}
\usepackage{tgtermes}
\usepackage[altg]{qtxmath}
\renewcommand*{\vec}[1]{\boldsymbol{#1}}
\usepackage{microtype}
\usepackage{graphicx}
\usepackage{braket}
\usepackage[usenames,dvipsnames]{color}
\graphicspath{{figures/}}
\usepackage[breaklinks=true,
pdfauthor={Meng Wen, Christoph H. Keitel, and Heiko Bauke},
pdftitle={Spin-one-half particles in strong electromagnetic fields: Spin effects and radiation reaction}]{hyperref}

\newcommand*{\D}{\mathrm{d}}
\newcommand*{\e}{\mathrm{e}}
\newcommand*{\I}{\mathrm{i}}
\newcommand*{\trans}[1]{#1^{\mathsf{T}}}

\newcommand*{\op}[1]{\hat{#1}}

\newcommand*{\opp}{\op{\vec p}}
\newcommand*{\oppo}{\op{p}_0}

\begin{document}

\title{Spin-one-half particles in strong electromagnetic fields:
  Spin effects and radiation reaction}

\author{Meng Wen}
\affiliation{Max-Planck-Institut f{\"u}r Kernphysik, Saupfercheckweg~1, 69117~Heidelberg, Germany}
\author{Christoph H. Keitel}
\affiliation{Max-Planck-Institut f{\"u}r Kernphysik, Saupfercheckweg~1, 69117~Heidelberg, Germany}
\author{Heiko Bauke}
\email{heiko.bauke@mpi-hd.mpg.de}
\affiliation{Max-Planck-Institut f{\"u}r Kernphysik, Saupfercheckweg~1, 69117~Heidelberg, Germany}

\begin{abstract}
  Various classical models of electrons including their spin degrees of
  freedom are commonly applied to describe the electron dynamics in
  strong electromagnetic fields.  We demonstrate that different models
  can lead to different or even contradicting predictions regarding how
  the spin degree of freedom modifies the electron's orbital motion when
  the electron moves in strong electromagnetic fields.  This discrepancy
  is rooted in the model-specific energy dependency of the spin-induced
  Stern-Gerlach force acting on the electron.  The Frenkel model and the
  classical Foldy-Wouthuysen model are compared exemplarily in the
  nonrelativistic and the relativistic limits in order to identify
  parameter regimes where these classical models make different
  predictions.  This allows for experimental tests of these models.  In
  ultra strong laser setups in parameter regimes where effects of the
  Stern-Gerlach force become relevant, radiation-reaction effects are
  also expected to set in.  We incorporate the radiation reaction
  classically via the Landau-Lifshitz equation and demonstrate that
  although radiation-reaction effects can have a significant effect on
  the electron trajectory, the Frenkel model and the classical
  Foldy-Wouthuysen model remain distinguishable also if
  radiation-reaction effects are taken into account.  Our calculations
  are also suitable to verify the Landau-Lifshitz equation for the
  radiation reaction of electrons and other spin-$1/2$ particles.
\end{abstract}


\maketitle

\allowdisplaybreaks

\section{Introduction}

The concept of spin was introduced by Uhlenbeck and Goudsmit
\cite{Uhlenbeck:Goudsmit:1925:Ersetzung_Hypothese,Uhlenbeck:Nature:1926}
as an internal angular momentum degree of freedom of elementary
particles in order to explain some experimental findings such as the
emission spectra of alkali metals.  A charged particle with spin can
interact with external electromagnetic fields via a coupling to its
charge as well as via its spin degree of freedom.  In addition to the
Lorentz force, the particle experiences a spin-dependent force induced
by the gradients of the electromagnetic fields.  Thus, a theoretical
description of such particles, e.\,g., electrons, must model how the
electromagnetic fields affect the dynamics of the spin (spin precession)
as well as the electron's orbital motion, which in general also depends
on the spin.  Such a spin-dependent motion is realized in the seminal
Stern-Gerlach experiment
\cite{Gerlach:Stern:1922:experimentelle_Nachweis} and variants thereof
\cite{Heinemann:1996:Stern-Gerlach,Garraway:Stenholm:1999:Observing_spin,Batelaan:Gay:etal:1997:Stern-Gerlach_Effect,Gallup:Batelaan:etal:2001:Quantum-Mechanical_Analysis,McGregor:Bach:etal:2011:Transverse_quantum}.
Effects of a spin-dependent force can be found, e.\,g., in astrophysical
systems \cite{Mahajan:Asenjo:etal:2014:Comparison_of} and in quantum
plasmas
\cite{Marklund:Brodin:2007:Dynamics_of,Brodin:Marklund:2007:Spin_magnetohydrodynamics,Brodin:Marklund:etal:2011:Spin_and}.
Stimulated by the advent of high-intensity laser facilities, the
interplay between spin precession and electron motion has also been
studied for electrons in strong electromagnetic fields
\cite{Rathe:Keitel:etal:1997:Intense_laser,Hu:Keitel:1999:Spin_Signatures,Walser:Keitel:2000:Spin-induced_force,Walser:Urbach:etal:2002:Spin_and,Roman:Roso:etal:2003:complete_description}. The
role of the spin may become significant in similar regimes where also
the radiation reaction sets in.  For strongly laser-driven electrons,
the radiation reaction has been investigated intensely by use of the
Landau-Lifshitz equation
\cite{Zhidkov:Koga:etal:2002:Radiation_Damping,Di-Piazza:Hatsagortsyan:etal:2009:Strong_Signatures,Tamburini:Pegoraro:etal:2010:Radiation_reaction,Ji:Pukhov:etal:2014:Radiation-Reaction_Trapping}
but also by quantum mechanical methods
\cite{Di-Piazza:Hatsagortsyan:etal:2010:Quantum_Radiation,Ilderton:Torgrimsson:2013:Radiation_reaction,Neitz:Di-Piazza:2013:Stochasticity_Effects,Blackburn:Ridgers:etal:2014:Quantum_Radiation,Vranic:Martins:etal:2014:All-Optical_Radiation,Li:Hatsagortsyan:etal:2014:Robust_Signatures,Gonoskov:Bashinov:etal:2014:Anomalous_Radiative,Vranic:Grismayer:etal:2016:Quantum_radiation}.

The electron's spin degree of freedom appears naturally in the framework
of relativistic quantum mechanics governed by the Dirac equation
\cite{Dirac:1928:Quantum_Theory}.  A classical description of the
electron spin may be found phenomenologically or via a correspondence
principle, which is applicable when the typical length scale of the
electromagnetic fields is larger than the position uncertainty of the
particle.  In this way, various classical models have been devised.
From a mathematical point of view, classical models of charged point
particles with spin are appealing because they are usually simpler and
easier to interpret than relativistic quantum theory.  Furthermore, a
classical description of spin may be incorporated into classical
many-particle theories and classical many-particle simulations, e.\,g.,
particle-in-cell codes \cite{Pfund:AIPCP:1998,BRODIN:JPP:2013}.

A fully relativistic classical description of the spin precession in the
presence of static homogeneous electromagnetic fields was provided by
Thomas \cite{Thomas:1926:Motion_of,Thomas:1927:I_kinematics}, Bargmann,
Michel, and Telegdi \cite{Bargmann:Michel:etal:1959:Precession_of}; see
also Ref.~\cite{Jackson:1999:Classical_Electrodynamics}.  This is today
a commonly accepted classical model, which has been applied in many
studies.  For the question how the spin modifies the electron's
trajectory in electromagnetic fields the situation is not as clear.  Two
fundamentally different approaches to the incorporation of
spin-dependent forces into classical theories can be found in the
literature.  On the one hand, one may start from a classical theory and
include possible Stern-Gerlach forces by accounting for quantum effects
or by classical considerations.  On the other hand, it is also possible
to derive a classical model from quantum theory by examining the
classical limit.  The first classical theory including a covariant
spin-induced Stern-Gerlach force was proposed by Frenkel
\cite{Frenkel:1926:Spinning_Electrons,Frenkel:1926:Elektrodynamik_des}.
The Frenkel model and similar classical models
\cite{Good:1962:Classical_Equations,Barducci:Casalbuoni:etal:1976:Supersymmetries_and,Barut:1980:Electrodynamics,Ravndal:1980:Supersymmetric_Dirac,Holten:1991:On_electrodynamics,Yee:Bander:1993:Equations_of,Chaichian:Gonzalez-Felipe:etal:1997:Spinning_relativistic,Pomeranskii:Khriplovich:1998:Equations_of,Pomeranskii:Senkov:etal:2000:Spinning_relativistic,Mahajan:Asenjo:etal:2014:Comparison_of,Gralla:Harte:etal:2009:Rigorous_derivation,Karabali:Nair:2014:Relativistic_Particle}
are mainly based on classical considerations.  For example, the Frenkel
model has been derived from different fundamental laws as the principle
of least action \cite{Barut:1980:Electrodynamics}, the conservation of
energy
\cite{Bhabha:Corben:1941:General_Classical,Nagy:1957:Relativistic_equation,Corben:1961:Spin_in,Nyborg:1962:On_classical,Nyborg:1962:Thomas_precession,Ternov:Bordovitsyn:1980:Modern_interpretation},
and the on-shell condition
\cite{Good:1962:Classical_Equations,Nyborg:1964:Approximate_relativistic,Bagrov:Bordovitsyn:1980:Classical_spin,Ternov:Bordovitsyn:1980:Modern_interpretation,Teitelboim:Villarroel:etal:1980:Classical_electrodynamics}.
Bearing in mind that the spin was introduced as an intrinsic quantum
feature of the electron
\cite{Jammer:1966:Conceptual_Development,Giulini:2008:spin}, it may
appear more appropriate to start from the Dirac equation to find
classical models of charged particles with spin
\cite{Bagrov:Belov:etal:1998}.  Such a classical model with a
spin-dependent force can be derived from relativistic quantum theory by
applying the correspondence principle to the von~Neumann equation in the
Foldy-Wouthuysen representation of the Dirac equation
\cite{Silenko:2008:Foldy-Wouthyusen_transformation,Chen:Chiou:2010:Foldy-Wouthuysen_transformation,Chen:Chiou:2014:Correspondence_between,Chen:Chiou:2014:High-order_Foldy-Wouthuysen}.
We call this the classical Foldy-Wouthuysen model below.

Spin-induced Stern-Gerlach forces in the classical Foldy-Wouthuysen
model and the Frenkel model, which are representatives of the two
families of classical models with spin-dependent forces as indicated
above, have been benchmarked against the Dirac theory in our recent
publication \cite{Wen:2016:Identifying}.  Currently, radiation-reaction
effects are investigated in many publications, where the electron's spin
is usually neglected assuming that this is appropriate for unpolarized
electron beams \cite{Bogdanov:Kazinski:2015:Properties_of} as well as
for the radiation-pressure-dominated regime of extreme optical laser
intensities $>10^{23}\,\mathrm{W/cm^2}$ and highly relativistic
electrons \cite{Tamburini:Pegoraro:etal:2010:Radiation_reaction}.  In
general, however, both spin-induced Stern-Gerlach forces and
radiation-reaction effects must be anticipated in light-matter
interaction at relativistic intensities.

In the present paper, we extend the study of classical models by
identifying spin effects in trajectories of electrons with different
spin states taking into account also classical radiation reactions via
the Landau-Lifshitz force \cite{Landau:Lifshitz:II}.  This article is
organized as follows: In Sec.~\ref{sec:models}, we introduce all
required notations and specify the classical Foldy-Wouthuysen and the
Frenkel models as well as the Landau-Lifshitz force.  These two models
are applied to various setups with strong electromagnetic fields in
Sec.~\ref{sec:dynamics}.  Examining homogeneous static magnetic fields
first, we continue with setups of increasing complexity: inhomogeneous
static magnetic fields, time-dependent electromagnetic plane waves and
focused pulses.  Regimes are identified where the classical models yield
different trajectories.  Furthermore, the models are benchmarked to
relativistic quantum theory by comparing classical trajectories to the
center-of-mass motion as predicted by the Dirac theory for regimes where
a solution of the Dirac equation is feasible.  Our main results are
summarized and discussed in Sec.~\ref{sec:conclusions}.

\section{Quantum and classical models of spin one-half particles}
\label{sec:models}

In the following, the theoretical foundations of our study are
established and all required notations are introduced.  We summarize
various semiclassical and quantum models for electron motion that take
into account also the electron's spin degree of freedom as well as
radiation reaction effects.
\enlargethispage{4ex}

\subsection{Basic notations}

The state of an electron at time $t$ is described by its position
$\vec r(t)$, its velocity $\vec v(t)$, and its spin orientation
$\vec S(t)$, where $\vec S(t)$ is a vector of length $\hbar/2$, with
$\hbar$ denoting the reduced Planck constant.  Sometimes it is
convenient to specify the electron's kinematic momentum
$\vec p_\mathrm{kin}(t)=m\gamma\vec v(t)$ instead of the velocity
$\vec v(t)$, where $m$ denotes the electron's rest mass and $\gamma$ is
the relativistic Lorentz factor.  Introducing the speed of light $c$, it
may be expressed as a function of the electron's velocity or,
equivalently, by its momentum:
\begin{equation}
  \label{eq:gamma}
  \gamma =
  \frac{1}{\sqrt{1-\vec v(t)^2/c^2}} =
  \sqrt{1 + \frac{\vec p_\mathrm{kin}(t)^2}{m^2c^2}} \,.
\end{equation}
The electron couples via its charge $q$ to the electromagnetic fields,
which are denoted $\vec E(\vec r, t)$ and $\vec B(\vec r, t)$.  These
may be expressed in terms of the electromagnetic potentials
$\varphi(\vec r, t)$ and $\vec A(\vec r, t)$ as
\begin{subequations}
  \begin{align}
    \vec E(\vec r, t) & =
    -\vec\nabla\varphi(\vec r, t) - \frac{\partial \vec A(\vec r, t)}{\partial t} \,, \\
    \vec B(\vec r, t) & =
    \vec\nabla\times\vec A(\vec r, t)\,.
  \end{align}
\end{subequations}

\subsection{Dirac equation}

A fully relativistic quantum mechanical description of the evolution of
an electron of mass $m$ and charge $q$ in the potentials $\vec A(\vec r,
t)$ and $\varphi(\vec r, t)$ is provided by the Dirac equation for the
electron's four-component wave function $\Psi(\vec r, t)$:\pagebreak[1]
\allowdisplaybreaks[0]%
\begin{multline}
  \label{eq:Dirac_equation}
  \I \hbar\frac{\partial \Psi(\vec r, t)}{\partial t}  = \\
  \left(
    c \vec \alpha\cdot\big(
    \opp - q \vec A(\vec r, t)
    \big) + m c^2 \beta + q\varphi(\vec r, t)
  \right) \Psi(\vec r, t) \,.
\end{multline}
\allowdisplaybreaks%
Here, $\opp=-\I\hbar\vec\nabla$ denotes the canonical momentum operator
and $\vec\alpha=\trans{(\alpha_x,\alpha_y,\alpha_z)}$ and $\beta$
indicate the Dirac matrices
\cite{Gross:2004:Relativistic_Quantum,Thaller:2005:Advanced_Visual}. In
the standard representation, these $4\times 4$ matrices are given by
\begin{equation}
  \label{eq:dirac_matices}
  \begin{aligned}
    \alpha_x & =
    \begin{pmatrix}
      0 & \sigma_x \\ \sigma_x & 0
    \end{pmatrix}\,, & \alpha_y & =
    \begin{pmatrix}
      0 & \sigma_y \\ \sigma_y & 0
    \end{pmatrix}\,, \\[0.5ex]
    \alpha_z & =
    \begin{pmatrix}
      0 & \sigma_z \\ \sigma_z & 0
    \end{pmatrix}\,, & \beta & =
    \begin{pmatrix}
      1 & 0 \\ 0 & -1
    \end{pmatrix}\,,
  \end{aligned}
\end{equation}
with $\sigma_x$, $\sigma_y$, and $\sigma_z$ denoting the three $2\times
2$ Pauli matrices in their standard representation.  The electron's
dynamics is completely determined by the wave function $\Psi(\vec r,
t)$.  The time-dependent expectation values of the electron's position,
its kinematic momentum, and its spin vector are given by
\begin{align}
  \label{eq:std_pos}
  \braket{\op{\vec r}}(t) & =
  \braket{\Psi(\vec r, t)|\vec r|\Psi(\vec r, t)} \,,\\
  \braket{\op{\vec p}_\mathrm{kin}}(t) & =
  \braket{\Psi(\vec r, t)|\opp - q \vec A(\vec r, t)|\Psi(\vec r, t)} \,,\\
  \label{eq:std_spin}
  \braket{\op{\vec S}}(t) & =
  \braket{\Psi(\vec r, t)|\hbar\vec\Sigma/2|\Psi(\vec r, t)} \,,
\end{align}
with $\vec\Sigma=\trans{(\Sigma_x,\Sigma_y,\Sigma_z)}$ and
\begin{align}
  \label{eq:Sigma_matices}
  \Sigma_x & =
  \begin{pmatrix}
    \sigma_x & 0 \\ 0 & \sigma_x
  \end{pmatrix}\,, & \Sigma_y & =
  \begin{pmatrix}
    \sigma_y & 0 \\ 0 & \sigma_y
  \end{pmatrix}\,, & \Sigma_z & =
  \begin{pmatrix}
    \sigma_z & 0 \\ 0 & \sigma_z
  \end{pmatrix}\,.
\end{align}

The standard operators for the position and the spin as employed in
\eqref{eq:std_pos} and \eqref{eq:std_spin} are the straight forward
generalizations of the position and spin operators of nonrelativistic
quantum theory to the relativistic domain.  They exhibit some defects,
however
\cite{Bauke:Ahrens:etal:2014:Relativistic_spin,Bauke:etal:2014:What_is}. For
example, the operator $\vec r$ leads to a velocity operator that is not
conserved under free motion, and similarly, the spin operator
$\hbar\vec\Sigma/2$ is also not conserved under free motion.  To remedy
these issues, Foldy and Wouthuysen
\cite{Foldy:Wouthuysen:1950:Non-Relativistic_Limit} introduced new
position and spin operators, by which the expectation values of the
position and the spin vector become
  \begin{align}
    \label{eq:FW_pos}
    \braket{\op{\vec r}}(t) & = \nonumber\\
    & \hspace{-6ex}\Braket{\Psi(\vec r, t)|\vec r +
      \I\hbar\left( \frac{\I\vec\Sigma\times\opp}{2\oppo(\oppo+mc)} -
        \frac{\beta(\vec\alpha\cdot\opp)\opp}{2\oppo^2(\oppo+mc)} +
        \frac{\beta\vec\alpha}{2\oppo}
      \right)|\Psi(\vec r, t)} \,,\\
    \label{eq:FW_spin}
    \braket{\op{\vec S}}(t) & = \Braket{\Psi(\vec r,
      t)|\frac{\hbar}{2}\vec\Sigma +
      \frac{\I\hbar\beta}{2\oppo}\op{\vec{p}}\times\vec\alpha -
      \frac{\hbar\op{\vec{p}}\times(\vec\Sigma\times\op{\vec{p}})}{2\oppo(\oppo+mc)}|\Psi(\vec r, t)} \,,
  \end{align}
where $\oppo = \sqrt{\smash[b]{m^2c^2 + \smash{\opp}^2}}$.
Expressions \eqref{eq:FW_pos} and \eqref{eq:FW_spin} are not gauge
invariant and therefore only meaningful for a vanishing vector
potential.  The Foldy-Wouthuysen position and spin operators would
become gauge independent by the substitution
$\opp\to\opp-q\vec A(\vec r, t)$.

In general, the expectation values for the position \eqref{eq:std_pos}
and \eqref{eq:FW_pos} yield different numerical values as well as
expectation values for the spin \eqref{eq:std_spin} and
\eqref{eq:FW_spin}.  For the parameters applied here, however, the
differences are insignificant.  Thus, we mainly employ
\eqref{eq:std_pos} and \eqref{eq:std_spin} when calculating position and
spin expectation values as a function of time.  As the Foldy-Wouthuysen
spin operator commutes with the free Dirac Hamiltonian, it is possible
to superimpose positive-energy free-particle states to a wave packet
with well-defined spin.  Thus, for the construction of initial (free)
wave packets the Foldy-Wouthuysen spin operator is more advantageous.

\subsection{Thomas-Bargmann-Michel-Telegdi equation}
\label{sec:TBMT_equation}

For a covariant classical description of the electron's spin, the spin
is characterized in the laboratory frame by the four-vector
$\Xi^\alpha = \trans{(\Xi^0, \vec{\Xi})}$
\cite{Jackson:1999:Classical_Electrodynamics}.  It is related to the
spin four-vector in the electron's rest frame
$\xi{^\alpha} = \trans{(\xi^0, \vec \xi)}=\trans{(0, \vec \xi)}$ via the
Lorentz transformation
\begin{subequations}
  \begin{align}
    \Xi^0 & = \frac{\gamma}{c} \vec{v}\cdot \vec{\xi} \,, \\
    \vec{\Xi} & =
    \vec{\xi} + \frac{\gamma^2}{(\gamma+1)c^2} \left(
       \vec{v}\cdot\vec{\xi}
    \right)\vec{v}\,.
  \end{align}
\end{subequations}
The classical description of an electron's spin vector in the rest frame
$\vec{\xi}=\hbar\vec{\pi}/2$ has the same definition as the standard
quantum spin~\eqref{eq:std_spin}
\cite{Mane:Shatunov:etal:2005:Spin-polarized_charged}; both of them are
proportional to the polarization $\vec{\pi}$. In nonrelativistic quantum
mechanics, it is calculated by
$\vec{\pi} = \braket{\Phi|\vec{\sigma}|\Phi}$, with $\Phi$ denoting the
normalized two-component spinor.  A classical description of the spin
precession of relativistic electrons in homogeneous time-independent
electromagnetic fields was first given by Thomas
\cite{Thomas:1926:Motion_of,Thomas:1927:I_kinematics}; this was later
also derived by Bargmann, Michel, and Telegdi
\cite{Bargmann:Michel:etal:1959:Precession_of} taking into account also
the anomalous magnetic moment
\cite{Schwinger:1948:On_Quantum-Electrodynamics}.  Using the
antisymmetric electromagnetic-field tensor
\begin{equation}
  F^{\alpha\beta} =
  \begin{pmatrix}
    0 & -E_x/c & -E_y/c & -E_z/c \\
    E_x/c & 0 & -B_z & B_y \\
    E_y/c & B_z & 0 & -B_x \\
    E_z/c & -B_y & B_x & 0
  \end{pmatrix}\,,
\end{equation}
Bargmann, Michel, and Telegdi derived, by classical arguments, the equation
\begin{equation}
  \label{eq:TBMT0}
  \frac{\D \Xi^\alpha}{\D \tau} =
  \frac{gq}{2m} \left(
    F^{\alpha\beta}\Xi_\beta + \frac{1}{c^2}u^\alpha \Xi_\lambda F^{\lambda\mu}u_\mu
  \right) -
  \frac{1}{c^2} u^\alpha \Xi_\lambda \frac{\D u^\lambda}{\D \tau}
  \,,
\end{equation}
which is known today as the Thomas-Bargmann-Michel-Telegdi (TBMT)
equation. Here $\tau$ denotes the proper time of the electron with
$\D \tau = \D t / \gamma$, four-velocity $u^\alpha = \gamma(c,\vec{v})$,
and $g$ identifying the gyromagnetic factor.

Assuming homogeneous electromagnetic fields or neglecting forces due to
the electron's spin, which are functions of the field gradients, the
motion of the electron is described by the Lorentz force
\begin{equation}
  \frac{\D u^\alpha}{\D \tau} =
  \frac{q}{m} F^{\alpha\beta} u_\beta
\end{equation}
only. Then the TBMT equation for $\Xi^\alpha$ simplifies to
\begin{equation}
  \label{eq::TBMT1}
  \frac{\D \Xi^\alpha}{\D \tau} =
  \frac{q}{m} \left(
    F^{\alpha\beta}\Xi_\beta +
    \frac{1}{c^2}\left(\frac g 2 - 1 \right) u^\alpha
    \left(\Xi_\lambda F^{\lambda\mu}u_\mu\right)
  \right)\,.
\end{equation}
Thus, for $g\ne 2$ the spin dynamics depends on the electron's velocity.
An analytical solution of \eqref{eq::TBMT1} can be obtained for the
motion in uniform fields \cite{Lobanov:Pavlova:1999:Solutions_of}.  In
agreement with the Dirac theory, we set $g=2$ in the following.  Then
the corresponding equation for the rest-frame spin vector
$\vec{\xi}(t)$, which follows from \eqref{eq::TBMT1}, is given by
\begin{equation}
  \label{eq:TBMT}
  \frac{\D \vec{\xi}(t)}{\D t} =
  \frac{q}{m} \vec{\xi}(t) \times \left(
    \frac{1}{\gamma}\vec{B} -
    \frac{1}{\gamma+1}\frac{\vec{v}\times\vec{E}}{c^2}
  \right) \,,
\end{equation}
where $\vec E$ and $\vec B$ denote here the constant homogeneous
electromagnetic fields in the laboratory frame and $\vec v$ is the
electron's velocity.  Introducing generalized classical Poisson brackets
as in Ref.~\cite{Yang:Hirschfelder:1980:Generalizations_of},
Eq.~\eqref{eq:TBMT} may be written as
\begin{equation}
  \frac{\D \vec{\xi}(t)}{\D t} = \left\{\vec{\xi}(t), H_\mathrm{TBMT} \right\}\,,
\end{equation}
with the TBMT Hamilton function
\cite{Hegstrom:Lhuillier:1977:Reinterpretation_of,Mane:Shatunov:etal:2005:Spin-polarized_charged}
\begin{equation}
  \label{eq:Htbmt}
  H_\mathrm{TBMT} =
  -\frac{q}{m} \vec{\xi}(t) \cdot \left(
    \frac{1}{\gamma}\vec{B} -
    \frac{1}{\gamma+1}\frac{\vec{v}\times\vec{E}}{c^2}
  \right) \,.
\end{equation}

\subsection{Frenkel's equation of motion}

Soon after Thomas derived his equation of motion for a spin-$1/2$
particle \cite{Thomas:1926:Motion_of}, Frenkel published his
investigation on the same problem but also including forces due to the
spin \cite{Frenkel:1926:Elektrodynamik_des}.  He utilized the
antisymmetric tensor $\Xi_{\alpha\beta}$ as a relativistic generation of
the spin vector.  It is defined in terms of the electron's electric
dipole moment $\vec{d}$ and its magnetic moment $\vec{m}$ in the
laboratory frame
\begin{subequations}
  \begin{align}
    \vec{d} & =
    \frac{\gamma}{c}\vec{v}\times \vec{\xi} \,, \\
    \vec{m} & =
    \gamma \vec{\xi} -
    \frac{\gamma^2}{\left(\gamma+1\right)c^2}\left(\vec{v}\cdot \vec{\xi}\right)\vec{v}
  \end{align}
\end{subequations}
as
\begin{equation}
  \Xi_{\alpha\beta} =
  \begin{pmatrix}
    0 & d_x & d_y & d_z \\
    -d_x & 0 & m_z & -m_y \\
    -d_y & -m_z & 0 & m_x \\
    -d_z & m_y & -m_x & 0
  \end{pmatrix}\,.
\end{equation}
Frenkel's equations of motion may be written as (neglecting possible
terms proportional to $\hbar^2$ and higher order corrections)
\cite{Frenkel:1926:Elektrodynamik_des,Ternov:Bordovitsyn:1980:Modern_interpretation}
\begin{subequations}
  \label{eq:Frenkel_covariant}
  \begin{align}
  \label{eq:Frenkel_covariant:a}
    M \frac{\D u^\alpha}{\D \tau} & =
    q F^{\alpha\beta}u_\beta + \frac{q}{2m}\Xi_{\mu\nu} \hat{D}^\alpha F^{\mu\nu} \,, \\
    m \frac{\D \Xi_{\alpha\beta}}{\D \tau} & =
    q\Xi_\alpha^\mu F_{\beta\mu} - q\Xi_\beta^\mu F_{\alpha \mu} \,,
  \end{align}
\end{subequations}
where
\begin{equation}
\label{eq:effm}
  M=m-\frac{q}{2mc^2}\Xi_{\alpha\beta}F^{\alpha\beta}
\end{equation}
represents the effective mass of the electron in electromagnetic fields
and
\begin{equation}
  \label{eq:Dalpha}
  \hat{D}^\alpha =
  -\frac{\partial}{\partial r_\alpha} +
  \frac{u^\alpha u^\beta}{c^2} \frac{\partial}{\partial r^\beta}
\end{equation}
(with the four-vector $r_\alpha=\trans{(ct,-\vec r)}$) denotes the
covariant generalization of the nabla operator, which may be found by
enforcing conformance with the on-shell condition
$u_\alpha u^\alpha = c^2$
\cite{Ternov:Bordovitsyn:1980:Modern_interpretation} or by applying the
method of geometric perturbation theory
\cite{Walse:Keitel:2001:Geometric_and}.  Because for the Lorentz force
$u_\alpha F^{\alpha\beta}u_\beta=0$ holds and furthermore
$u_\alpha\hat{D}^\alpha=0$, it follows from the orbital equation of
motion \eqref{eq:Frenkel_covariant:a} that this model satisfies the
on-shell condition $u_\alpha u^\alpha=c^2$.  For actual calculations it
is convenient to express Eq.~\eqref{eq:Frenkel_covariant} in terms of
the position, momentum, and spin vectors
\cite{Good:1962:Classical_Equations}.  With the magnetic field in the
rest frame of the electron,
\begin{multline}
  \vec{B}'(\vec r(t), t) =
  \gamma\vec{B}(\vec r(t), t) -
  \frac{\vec{p}_\mathrm{kin}(t)\times \vec{E}(\vec r(t), t)}{mc^2} \\ -
  \frac{(\vec{p}_\mathrm{kin}(t)\cdot \vec{B}(\vec r(t), t))\vec{p}_\mathrm{kin}(t)}{(\gamma+1)m^2c^2}
\end{multline}
and
\begin{equation}
\label{addm}
  \frac{q}{2mc^2}\Xi_{\alpha\beta}F^{\alpha\beta}=\frac{q}{mc^2}\vec
  \xi(t)\cdot\vec B'(\vec r(t), t) \,,
\end{equation}
the Frenkel's equations of motion can be written as
\begin{subequations}
  \label{eq:class_F}
  \begin{align}
    \label{eq:class_F_r}
    \frac{\D \vec r(t)}{\D t}
    & = \frac{\vec p_\mathrm{kin}(t)}{m\gamma} \,, \\
    \label{eq:class_F_p}
    \frac{\D \vec p_\mathrm{kin}(t)}{\D t}
    & = \frac{qm}{M} \left(
      \vec E(\vec r(t), t) +
      \frac{\vec p_\mathrm{kin}(t)\times\vec B(\vec r(t), t)}{m\gamma}
      \right) \nonumber \\
    & \qquad\qquad\quad +\frac{q}{\gamma M} \hat{\vec{D}}
      (\vec{\xi}(t)\cdot\vec{B}'(\vec r(t), t))\,, \\
    \label{eq:class_F_s}
    \frac{\D \vec{\xi}(t)}{\D t}
    & =\frac{q}{m} \vec{\xi}(t) \times \vec B_\mathrm{eff}(\vec r(t), t) \,,
  \end{align}
\end{subequations}
where the operator $\hat{\vec{D}}= \vec{\nabla} +
{\gamma^2}\vec{v}(t)(\partial_t + \vec{v}(t) \cdot \vec{\nabla})/{c^2}$
acts on the electromagnetic fields only and
\begin{equation}
  \vec B_\mathrm{eff}(\vec r, t) =
  \frac{1}{\gamma}\vec{B}(\vec r, t) -
  \frac{1}{\gamma(\gamma+1)}
  \frac{\vec{p}_\mathrm{kin}(t)\times\vec{E}(\vec r, t)}{mc^2}
  \,.
\end{equation}

\subsection{Classical model via Foldy-Wouthuysen transformation}

The Foldy-Wouthuysen transformation
\cite{Foldy:Wouthuysen:1950:Non-Relativistic_Limit} is a unitary
transformation of the Dirac equation into a block diagonal form yielding
some transformed Hamilton operator $\op{H}_\mathrm{FW}$.  In this
representation, operators for the position, the momentum, and the spin
have the same form as in the nonrelativistic quantum theory.  Except for
the case of free particles and some other special cases, the
Foldy-Wouthuysen transformation can not be carried out exactly.  If the
electromagnetic fields do not vanish, one can, however, construct a series
of unitary transforms, where each transform implements the
Foldy-Wouthuysen transformation up to some order in $1/c$.  In this way,
one may study low-order relativistic corrections of the Dirac equation
to the nonrelativistic Pauli equation or weakly relativistic effects in
quantum plasmas
\cite{Asenjo:Zamanian:etal:2012:Semi-relativistic_effects}.

A block diagonal Hamilton operator $\op{H}_\mathrm{FW}$, which is exact
in all orders of $1/c$ but accounts only for effects that are linear in
the electromagnetic fields, can be derived from a relativistic
generalization of the standard Foldy-Wouthuysen transformation
\cite{Eriksen:Kolsrud:1960:Canonical_transformations,Dixon:1965:Classical_theory,Suttorp:De-Groot:1970:Covariant_equations,Silenko:2008:Foldy-Wouthyusen_transformation}.
The resulting representation of the Dirac equation is applicable to
highly relativistic particles.  Silenko
\cite{Silenko:2008:Foldy-Wouthyusen_transformation} derived equations of
motion for the quantum mechanical operators for the position, the
kinematic momentum, and the spin by employing the transformed Dirac
Hamiltonian and the Heisenberg picture.  More precisely, the
time evolution of some observable, which is represented by some possibly
explicitly time-dependent operator $\op{O}(t)$, is given by
\begin{equation}
  \label{eq:Heisenberg}
  \frac{\D \op{O}(t)}{\D t } =
  \frac{1}{\I\hbar}\left[\op{O}(t), \op{H}_\mathrm{FW}\right] +
  \frac{\partial \op{O}(t)}{\partial t} \,.
\end{equation}
Substituting for $\op{O}$ in Eq.~\eqref{eq:Heisenberg} the operators
$\vec r$, $-\I\hbar\vec\nabla - q\vec A(\vec r, t)$, and
$\hbar\vec\Sigma/2$, respectively, which are the position
\eqref{eq:FW_pos}, the kinematic momentum, and the spin in the
Foldy-Wouthuysen representation~\eqref{eq:FW_spin}, gives the quantum
mechanical equations of motion for these observables.  The classical
equations of motion then follow from these equations by applying the
correspondence principle.  This means considering the limit $\hbar\to0$
and replacing operators with commuting numbers in the quantum mechanical
equation of motion \eqref{eq:Heisenberg}.  In this way, the evolution
equations of the classical Foldy-Wouthuysen model \footnote{Note that
  the model considered does not have a commonly used standard name.
  Although it was not derived by Foldy and Wouthuysen, we call it the
  ``classical Foldy-Wouthuysen model'' here because it originates from a
  Dirac Hamiltonian in the Foldy-Wouthuysen representation.
  Alternatively, one might just speak of the semiclassical equations of
  motion for the Dirac theory.} for the position, the kinematic
momentum, and the spin result in (in our notation)
\begin{subequations}
  \label{eq:class_FW}
  \begin{align}
    \label{eq:class_FW_r}
    \frac{\D \vec r(t)}{\D t} & =
    \frac{\vec p_\mathrm{kin}(t)}{m\gamma} \,, \\
    \label{eq:class_FW_p}
    \frac{\D \vec p_\mathrm{kin}(t)}{\D t} & =
    q\left(
      \vec E(\vec r(t), t) +
      \frac{\vec p_\mathrm{kin}(t)\times\vec B(\vec r(t), t)}{m\gamma}
    \right)  \nonumber \\
    & \qquad\qquad\quad +\frac{q}{m}\vec\nabla\left(
      \vec S(t)\cdot\vec B_\mathrm{eff}(\vec r(t), t)
    \right)\,, \\
    \label{eq:class_FW_S}
    \frac{\D \vec S(t)}{\D t} & =
    \frac{q}{m} \vec{S}(t) \times \vec B_\mathrm{eff}(\vec r(t), t) \,;
  \end{align}
\end{subequations}
see Ref.~\cite{Silenko:2008:Foldy-Wouthyusen_transformation} for
details.  These equations have also been derived by other means in
Ref.~\cite{Bagrov:Belov:etal:1998}, see Eq.~(5.16) there.  For
homogeneous electromagnetic fields the Stern-Gerlach force
\mbox{$\sim\vec\nabla\left( \vec S(t)\cdot\vec B_\mathrm{eff}(\vec r(t),
    t) \right)$} vanishes, and consequently
Eqs.~\eqref{eq:class_FW_r}--\eqref{eq:class_FW_S} reduce to the Lorentz
equation plus the TBMT equation \eqref{eq:TBMT} for the spin
\cite{Chen:Chiou:2010:Foldy-Wouthuysen_transformation}.

\subsection{Lorentz and Stern-Gerlach forces in the Frenkel and the
  classical Foldy-Wouthuysen models}
\label{sec:spin_forces}

Comparing the classical equations of motion \eqref{eq:class_F} and
\eqref{eq:class_FW}, we see that the spin follows the TBMT
equation~\eqref{eq:TBMT} in both cases, in the Frenkel model as well as
in the classical Foldy-Wouthuysen model.  Note that the Frenkel
equations of motion \eqref{eq:class_F} are formulated in terms of the
electron's spin $\vec \xi(t)$ in its rest frame, whereas the classical
Foldy-Wouthuysen model \eqref{eq:class_FW} describes the spin
$\vec S(t)$ in the laboratory frame.  Both vectors correspond to the
polarization of the particle as claimed in the respective models.

The forces that determine the electron's orbital motion are different in
the considered classical models.  These forces can be split into a
spin-independent and a spin-dependent part.
Equation~\eqref{eq:class_FW_p} may be written as the sum
${\D {\vec p}_\mathrm{kin}}/{\D t}=\vec F_\mathrm{1,\,FW}+\vec
F_\mathrm{2,\,FW}$ with
\allowdisplaybreaks%
\begin{subequations}
  \label{eq:F1F2_FW}
  \begin{align}
    \label{eq:F_1_FW}
    \vec F_\mathrm{1,\,FW} & = q\left( \vec E(\vec r(t), t) + \frac{\vec
        p_\mathrm{kin}(t)\times\vec B(\vec r(t), t)}{m\gamma} \right)
    \intertext{and}
    \vec F_\mathrm{2,\,FW} & =
    \frac{q}{m}\vec\nabla\left( \vec S(t)\cdot\vec B_\mathrm{eff}(\vec
      r(t), t) \right)\,,
  \end{align}
\end{subequations}
where the spin-independent force $\vec F_\mathrm{1,\,FW}$ equals the
standard Lorentz force and the spin-dependent force
$ \vec F_\mathrm{2,\,FW}$ represents the relativistic Stern-Gerlach
force within the classical Foldy-Wouthuysen model.  In the case of the
Frenkel model, Eq.~\eqref{eq:class_F_p} may be written as the sum
${\D {\vec p}_\mathrm{kin}}/{\D t}=\vec F_\mathrm{1,\,F}+\vec
F_\mathrm{2,\,F}$ of the modified Lorentz force
\begin{subequations}
  \begin{align}
    \label{eq:F_1_F}
    \vec F_\mathrm{1,\,F} & =
    \frac{qm}{M} \left( \vec E(\vec r(t), t) +
      \frac{\vec p_\mathrm{kin}(t)\times\vec B(\vec r(t), t)}{m\gamma}
    \right)
    \intertext{and the Stern-Gerlach force}
    \vec F_\mathrm{2,\,F} & =
    \frac{q}{\gamma M}
    \hat{\vec{D}} (\vec{\xi}(t)\cdot\vec{B}'(\vec r(t), t))\,.
  \end{align}
\end{subequations}

The forces $\vec F_\mathrm{1,\,FW}$ and $\vec F_\mathrm{1,\,F}$ are
equal up to the factor $m/M$.  Note that it renders the modified Lorentz
force of the Frenkel model spin dependent via the additional prefactor
$m/M$ in Eq.~\eqref{eq:F_1_F}
\cite{Ternov:Bordovitsyn:1980:Modern_interpretation,Takabayasi:1981:Relativistic_Dynamics,Kassandrov:Markova:etal:2009:On_model}.
The effective mass~\eqref{eq:effm} has been explained by a
magnetic-energy contribution to the electron's total relativistic energy
\cite{Lebedev:2016:Spin_radiative}.  The Lorentz force~\eqref{eq:F_1_F}
is a consequence of Eq.~\eqref{eq:Frenkel_covariant:a}.  Here, we have
on the right-hand side the standard Lorentz four-force
$q F^{\alpha\beta}u_\beta$ (plus the spin-dependent term) but on the
left-hand side the modified four-momentum $M\,{\D u^\alpha}/{\D \tau}$
instead of the standard four-momentum $m\,{\D u^\alpha}/{\D \tau}$.  In
the weak-field limit, i.\,e.,
$|q|\hbar(\gamma|\vec B| + |\vec p_\mathrm{kin}\times \vec
E|/(mc^2))/(m^2c^2)\to 0$, the modified Lorentz force of the Frenkel
model goes to the standard Lorentz force.  Further possible
discrepancies between the two models originate from the forces
$\vec F_\mathrm{2,\,FW}$ and $\vec F_\mathrm{2,\,F}$, which represent
the model-dependent Stern-Gerlach forces.  It may be instructive to
compare the four-vector forms of the Stern-Gerlach forces
$\vec F_\mathrm{2,\,FW}$ and $\vec F_\mathrm{2,\,F}$.  These are
\begin{subequations}
  \begin{equation}
    \label{eq:F2_FW_cov}
    F^\alpha_\mathrm{2,\,FW} =
    - \frac{q}{m} \frac{\partial}{\partial r_\alpha} U_\mathrm{FW}
  \end{equation}
  and
  \begin{equation}
   F^\alpha_\mathrm{2,\,F} =
    \frac{q}{M} \hat{D}^\alpha U_\mathrm{F} =
    -\frac{q}{M}\frac{\partial}{\partial r_\alpha}U_\mathrm{F} +
    \frac{q}{M}\frac{u^\alpha u^\beta}{c^2} \frac{\partial}{\partial r^\beta} U_\mathrm{F}\,,
  \end{equation}
\end{subequations}
respectively, with the scalars
\mbox{$U_\mathrm{FW}=\vec S(t)\cdot\vec B_\mathrm{eff}(\vec r(t), t)$}
and \mbox{$U_\mathrm{F}=\vec \xi(t)\cdot\vec B'(\vec r(t), t)/\gamma$}.
Here $\hat{D}^\alpha$ is defined in Eq.~\eqref{eq:Dalpha}, which may be
found by enforcing conformance with the on-shell condition
$u_\alpha u^\alpha = c^2$ via $u_\alpha F^\alpha_\mathrm{2,\,F}=0$.
Note that Eq.~\eqref{eq:F2_FW_cov} does not hold the on-shell condition
because $u_\alpha F^\alpha_\mathrm{2,\,FW}$ does not vanish identically.
Furthermore, in order to identify the origins of possible discrepancies
in the trajectories of the classical Foldy-Wouthuysen model and the
Frenkel model, it is instructive to write the Stern-Gerlach forces more
explicitly.  For the classical Foldy-Wouthuysen model the force is
\begin{multline}
  \label{eq:F_2_FW}
  \vec F_\mathrm{2,\,FW} =
  \frac{q}{\gamma m}\vec\nabla\left(
    \vec S(t)\cdot\vec{B}(\vec r(t), t)
  \right) \\
  -\frac{q}{\gamma(\gamma+1) m^2c^2}\vec\nabla\left(
    \vec S(t)\cdot\left(\vec{p}_\mathrm{kin}(t)\times\vec{E}(\vec r(t), t)\right)
  \right)
\end{multline}
and for the Frenkel model we obtain
\begin{widetext}\vspace{-3ex}
\begin{multline}
  \label{eq:F_2_F}
  \vec F_\mathrm{2,\,F} =
  \frac{q}{M}\vec\nabla\left(
    \vec \xi(t)\cdot\vec{B}(\vec r(t), t)
  \right) - \frac{q}{\gamma Mmc^2} \vec\nabla\left(
    \vec \xi(t)\cdot(\vec{p}_\mathrm{kin}(t)\times\vec{E}(\vec r(t), t))
  \right) -
  \frac{q}{\gamma (\gamma+1)Mm^2c^2}\vec\nabla\left(
    \vec \xi(t)\cdot(\vec{p}_\mathrm{kin}(t)\cdot \vec{B}(\vec r(t), t))\vec{p}_\mathrm{kin}(t)
  \right)\\
  + \frac{q\vec{p}_\mathrm{kin}(t)}{Mmc}\cdot
  \left(
    \gamma \frac{\partial}{\partial t} +
    \frac{\vec{p}_\mathrm{kin}(t)}{m} \cdot \vec{\nabla}
  \right)\left(
    \vec \xi(t)\cdot\vec{B}(\vec r(t), t)
  \right)
  - \frac{q\vec{p}_\mathrm{kin}(t)}{\gamma Mm^2c^3}\cdot
  \left(
    \gamma \frac{\partial}{\partial t} +
    \frac{\vec{p}_\mathrm{kin}(t)}{m} \cdot \vec{\nabla}
  \right)
  \left(
    \vec \xi(t)\cdot(\vec{p}_\mathrm{kin}(t)\times\vec{E}(\vec r(t), t))
  \right) \\
  -\frac{q\vec{p}_\mathrm{kin}(t)}{\gamma (\gamma+1)Mm^3c^3}\cdot
  \left(
    \gamma \frac{\partial}{\partial t} + \frac{\vec{p}_\mathrm{kin}(t)}{m} \cdot \vec{\nabla}
  \right)\left(
    \vec \xi(t)\cdot(\vec{p}_\mathrm{kin}(t)\cdot \vec{B}(\vec r(t), t))\vec{p}_\mathrm{kin}(t)
  \right) \,.
\end{multline}
\end{widetext}

Comparing \eqref{eq:F_2_FW} and \eqref{eq:F_2_F} term by term, we find
that the first summands of both forces are almost equivalent but differ
by a factor $\gamma m/M$.  In particular, the approximation $M\approx m$
holds if $|\vec B'|$ is small compared to $2m^2c^2/|q\hbar|$, which is
$\approx10^{10}\mathrm{\,T}$ for electrons (corresponding to an
electromagnetic wave with intensity of about
$10^{30}\,\mathrm{W/cm^2}$).  Consequently, the first term of the spin
force is decreased by the factor $1/\gamma$ in the classical
Foldy-Wouthuysen model compared to the Frenkel model in this
regime. Thus, the discrepancy between these models becomes large if the
electron's velocity approaches the speed of light.  Similarly, the
second terms differ by a factor \mbox{$M/m/(\gamma+1)$}.  The further
terms in the Stern-Gerlach force~\eqref{eq:F_2_F} of the Frenkel model
do not have a counterpart in the classical Foldy-Wouthuysen model.  Also
these terms become particularly large if the electron's velocity
approaches the speed of light, leading to completely different
dependencies of $\vec F_\mathrm{2,\,FW}$ and $\vec F_\mathrm{2,\,F}$ on
$\gamma$.  Consequently, the classical Foldy-Wouthuysen model and the
Frenkel model differ mainly in the high-velocity limit.  Thus, we study
in the following the interaction of relativistic electrons in strong
electromagnetic fields.

\subsection{Radiation reaction via the Landau-Lifshitz equation}

For a highly relativistic electron motion driven by an ultraintense
laser not only are spin effects expected to set in, but also radiation
reaction may become important.  The electron's emission of radiation can
be incorporated into the classical modeling via an additional force.
The radiation reaction force as it has been derived for spinless
particles is given by the Landau-Lifshitz force \cite{Landau:Lifshitz:II}
\allowdisplaybreaks[0]%
\begin{widetext}\vspace{-3ex}
  \begin{align}
    \label{eq:RR}
    \vec F_\mathrm{RR} & = \frac{q^3}{6\pi\varepsilon_0 mc^3} \gamma \left( \left(
    \frac{\partial}{\partial t} + \frac{\vec{p}_\mathrm{kin}(t)}{ \gamma m} \cdot \vec{\nabla}
  \right)\vec{E}(\vec r(t), t)  + \frac{\vec{p}_\mathrm{kin}(t)}{\gamma m} \times \left(
   \frac{\partial}{\partial t} + \frac{\vec{p}_\mathrm{kin}(t)}{ \gamma m} \cdot \vec{\nabla}
  \right)\vec{B}(\vec r(t), t) \right) \nonumber \\
    & \qquad+ \frac{q^4}{6\pi\varepsilon_0 m^2c^3} \left(\left(  \frac{\vec{p}_\mathrm{kin}(t)}{\gamma m} \cdot \frac{\vec{E}(\vec r(t), t)}{c} \right) \frac{\vec{E}(\vec r(t), t)}{c}  + \left( \vec{E}(\vec r(t), t) + \frac{\vec{p}_\mathrm{kin}(t)}{\gamma m} \times \vec{B}(\vec r(t), t) \right) \times \vec{B}(\vec r(t), t) \right) \nonumber \\
    & \qquad+ \frac{q^4}{6\pi\varepsilon_0 m^2c^5} \gamma^2 \left(  \left(  \frac{\vec{p}_\mathrm{kin}(t)}{\gamma m} \cdot \frac{\vec{E}(\vec r(t), t)}{c} \right)^2 - \left( \vec{E}(\vec r(t), t) + \frac{\vec{p}_\mathrm{kin}(t)}{\gamma m} \times \vec{B}(\vec r(t), t) \right)^2 \right) \frac{\vec{p}_\mathrm{kin}(t)}{\gamma m} \,,
  \end{align}
\end{widetext}
where $\varepsilon_0$ denotes the vacuum permittivity.  Heuristically
this radiation-reaction force can be incorporated into
Eq.~\eqref{eq:class_F_p} and Eq.~\eqref{eq:class_FW_p} as an additional
force term, yielding the equation of motion
\begin{equation}
  \label{eq:mo}
  \frac{\D {\vec p}_\mathrm{kin}}{\D t} =
  \vec F_\mathrm{1,\,F/FW} + \vec F_\mathrm{2,\,F/FW} + \vec
  F_\mathrm{3,\,F/FW} \,,
\end{equation}
with $\vec F_\mathrm{3,\,FW}=\vec F_\mathrm{RR}$ for the classical
Foldy-Wouthuysen model and
$\vec F_\mathrm{3,\,F}=\vec F_\mathrm{RR}\, m/M$ for the Frenkel model.
In the latter case, the prefactor $m/M$ in the radiation reaction has
been introduced in analogy to the modified Lorentz force of the Frenkel
model.  Here, we have assumed that the electron's effective mass $M$
affects the force on an electron irrespective of the nature of the
force.

One may compare the orders of magnitude of the different Lorentz,
Stern-Gerlach, and radiation-reaction forces in the equation of
motion~\eqref{eq:mo} by considering an electron in an electromagnetic
field with wavelength $\lambda$ and comparing the constant of
proportionality between the forces and the applied fields.  For example,
the constant of proportionality of the (modified) Lorentz force
(Eqs.~\eqref{eq:F_1_FW} and~\eqref{eq:F_1_F}) to the electric field is
in leading order $q$.  The constants of proportionality of the
Stern-Gerlach forces \eqref{eq:F_2_FW} and \eqref{eq:F_2_F} and the
Landau-Lifshitz force~\eqref{eq:RR} can be written as
$\tilde{q}_\mathrm{S}=q\lambda_\e/(2\lambda)$ and
$\tilde{q}_\mathrm{R}=2q\alpha\lambda_\e/(3\lambda)$, respectively,
where $\lambda_\e \approx 2.4\times 10^{-12}\,\mathrm{m}$ denotes the
Compton wavelength of the electron and $\alpha$ is the fine-structure
constant.  When the wavelength of the field is long, i.\,e.,
$\lambda \gg \lambda_\e$, coefficients are
$|q| \gg |\tilde{q}_\mathrm{S}| \approx 10^2 |\tilde{q}_\mathrm{R}|$.
In this regime the Lorentz force is the dominating force, while the
Stern-Gerlach force and Landau-Lifshitz forces are subordinate forces.
Furthermore, one may also incorporate additional force terms into the
equation of motion \eqref{eq:mo} with mixed contributions by the spin
and radiation reactions as in
Ref.~\cite{Teitelboim:Villarroel:etal:1980:Classical_electrodynamics},
whose coefficient of proportionality, which is of the order of
$|\tilde{q}_\mathrm{S}\tilde{q}_\mathrm{R}/q|$, is even much smaller
than those of the Stern-Gerlach and the Landau-Lifshitz forces.
Therefore, possible terms of spin radiation-reaction are neglected in
this paper.

\section{Dynamics of spin-\boldmath$1/2$ particles}
\label{sec:dynamics}

In Sec.~\ref{sec:spin_forces} it is demonstrated that the (generalized)
Lorentz forces, the spin-induced Stern-Gerlach forces, and the
Landau-Lifshitz forces differ in the Frenkel and the classical
Foldy-Wouthuysen models in leading order by factors of $m/M$,
$\gamma m/M$, and $m/M$, respectively.  Setups with static magnetic
fields as well as electromagnetic waves are examined in the following,
where the classical models lead to different predictions in these
setups.  A comparison of the classical models with the Dirac theory is
made for parameters where a numerical solution of the time-dependent
Dirac equation is computationally feasible.  For an adequate rating,
radiation-reaction effects are excluded in these comparisons from the
classical models because radiation-reaction effects are also beyond the
Dirac theory.  For experimentally relevant systems with strong laser
pulses, however, both spin forces and radiation reactions are expected
to have nonnegligible effects, and both are investigated in detail in
the following.

\subsection{Model-dependent trajectories of electrons in a homogeneous
  magnetic field: Effective mass effects}
\label{sec:magnetic_field}

\begin{figure}
  \centering
  \includegraphics[scale=0.45]{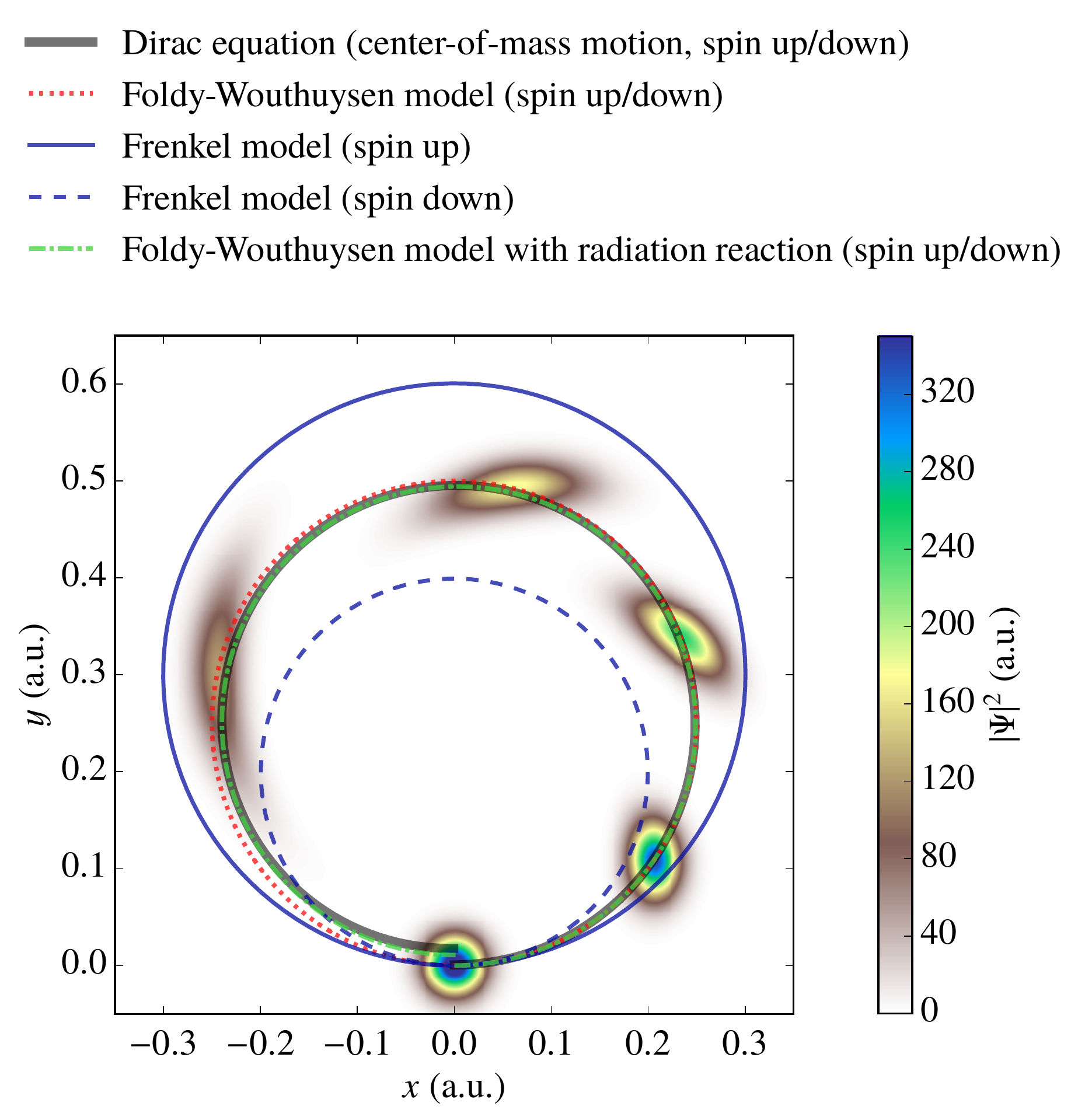}
  \caption{Trajectories of an electron with initial momentum $3.65mc$ in
    an ultrastrong homogeneous magnetic field (perpendicular to the
    plane of projection) of strength
    $2000\,\mathrm{a.u.}=4.7\times10^8\,\mathrm{T}$ as predicted by the
    Dirac equation (solid gray line), the classical Foldy-Wouthuysen
    model (dotted red line), and the Frenkel model (solid blue and dashed
    blue lines).  The electron's initial spin orientation is parallel or
    antiparallel to the magnetic field.  Trajectories of electrons with
    parallel spin (solid blue line) differ from trajectories of
    electrons with antiparallel spin (dashed blue line) within the
    Frenkel model, whereas the Dirac theory and the classical
    Foldy-Wouthuysen model do not lead to spin-dependent trajectories
    for homogeneous fields.  In order to illustrate radiation-reaction
    effects, the dot-dashed green line shows the trajectory for the
    classical Foldy-Wouthuysen model with radiation damping included via
    the Landau-Lifshitz equation.  As an illustration the density of the
    electron's quantum mechanical wave packet is also indicated at five
    points in time.}
  \label{fig:in_magnetic_field}
\end{figure}

In the case of homogeneous electromagnetic fields the gradient forces of
the classical Foldy-Wouthuysen model and the Frenkel model vanish and
the electron's motion depends only on the Lorentz force, which is
modified by a factor $m/M$ within the Frenkel model as outlined in
Sec.~\ref{sec:spin_forces}.  If the electron moves in the plane
perpendicular to the homogeneous magnetic field and if the electric
field vanishes, the effective mass is given by
\begin{equation}
  M = m - \frac{q\gamma}{mc^2}\vec \xi(t)\cdot\vec B \,.
\end{equation}
Consequently, the Lorentz force is modified by a factor
\mbox{$1/(1-{q\gamma}/({m^2c^2})\,\vec \xi(t)\cdot\vec B)$} and
therefore it depends on the electron's spin orientation relative to the
magnetic field.  Thus, the Frenkel theory predicts that trajectories of
electrons with parallel spin differ from trajectories of electrons with
antiparallel spin as illustrated in Fig.~\ref{fig:in_magnetic_field}.
Note that the factor
\mbox{$1/(1-{q\gamma}/({m^2c^2})\,\vec \xi(t)\cdot\vec B)$} may become
negative for ${q\gamma}/({m^2c^2})\,\vec \xi(t)\cdot\vec B>1$ or even
diverge for ${q\gamma}/({m^2c^2})\,\vec \xi(t)\cdot\vec B=1$, which
requires magnetic field strengths of the order of $2m^2c^2/|q\hbar|$.

In the following, we consider setups where the electron's spin is
initially parallel or anti parallel to the direction of the homogeneous
magnetic field, which maximizes possible spin-dependent effective-mass
effects.  As a consequence of Eqs.~\eqref{eq:class_F_s}
and~\eqref{eq:class_FW_S} the spin component in the magnetic-field
direction is conserved.  In the Foldy-Wouthuysen representation
\cite{Silenko:2008:Foldy-Wouthyusen_transformation}, it can be easily
shown that the Dirac Hamiltonian with
$\vec A(\vec r)=\frac{1}{2}\vec B\times \vec r$ commutes with the
$z$~component of the spin operator if the magnetic field $\vec B$ is
directed in the $z$~direction.  The spin is also conserved in the
quantum mechanical description of this setup.  Although the effect of
the spin-dependent effective mass
\cite{Ternov:Bordovitsyn:1980:Modern_interpretation,Takabayasi:1981:Relativistic_Dynamics,Lebedev:2016:Spin_radiative}
including the possibility of a negative mass
\cite{Kassandrov:Markova:etal:2009:On_model} has been considered in the
literature, this behavior is in contrast to the predictions of the
classical Foldy-Wouthuysen model as well as the quantum mechanical Dirac
theory.  In the classical Foldy-Wouthuysen model the force on the
electron does not depend on its spin orientation in the case of
homogeneous electromagnetic fields.  Consequently, homogeneous fields
cannot lead to spin-dependent trajectories.  In the classical
Foldy-Wouthuysen model electrons with different spin orientations,
i.\,e., parallel or antiparallel to the magnetic-field direction, follow
the same simple cyclotron gyration trajectory; they move in a circle
with the gyroradius $m|\vec p_\mathrm{kin}|/(|\gamma q\vec B|)$.
Similarly, an electron's center-of-mass trajectory, which follows from a
quantum mechanical description of the electron motion via the Dirac
theory, does not depend on the electron's spin state as also indicated
in Fig.~\ref{fig:in_magnetic_field}.  The two-dimensional electron wave
packet shown in Fig.~\ref{fig:in_magnetic_field} was initially composed
of a superposition of positive-energy eigenstates of the free Dirac
Hamiltonian and the $z$~component of the Foldy-Wouthuysen spin operator
\cite{Foldy:Wouthuysen:1950:Non-Relativistic_Limit,Bauke:Ahrens:etal:2014:Relativistic_spin}
such that the wave-packet's center of mass lies at the origin of the
coordinate system.  The numerically obtained
\cite{Bauke:Keitel:2011:Accelerating_Fourier} center-of-mass
trajectories for electron wave packets with parallel and antiparallel
initial spin are indistinguishable on the scale of
Fig.~\ref{fig:in_magnetic_field}.  Differences are of the order of the
estimated numerical errors.  This numerical result can also be
substantiated by the analytical solution of the Dirac equation for the
electron in a homogeneous magnetic field, which allows one to construct
coherent wave packets with a center-of-mass motion that follows the
classical Lorentz equation irrespective of the spin state
\cite{Bagrov:Bukhbinder:etal:1975:Coherent_states,Bagrov:Gitman:2014:Dirac_Equation}.
Note that the quantum mechanical wave packet does not have a sharp
kinematic momentum, thus the center-of-mass trajectory of the whole
quantum wave packet does not follow a circular motion with constant
radius \cite{Rakhimov:JOPCM:2011}.  Since the gyroradius is proportional
to the velocity, the part of the packet with a higher momentum moves
faster with a larger radius, while the slower part moves on a smaller
radius.  Thus the spatial distribution of the packet shrinks radially
and extends azimuthally as indicated in
Fig.~\ref{fig:in_magnetic_field}.

In order to allow for direct comparison of the Frenkel model and the
classical Foldy-Wouthuysen model, radiation-reaction effects have been
neglected so far.  This is also justified because the electron energy is
not in a strongly relativistic regime, and radiation-reaction effects
remain small in the chosen parameter regime as shown in
Fig.~\ref{fig:in_magnetic_field}.  It illustrates radiation-reaction
effects for the classical Foldy-Wouthuysen model with radiation damping
included via the Landau-Lifshitz force~\eqref{eq:RR} in addition to the
Stern-Gerlach force.  The dot-dashed green line shows how the electron
trajectory deviates slightly from circular motion due to the radiation
reaction, which decreases the electron's energy.  Note that deviations
from circular motion in the cases of the center-of-mass motion of the
quantum mechanical wave packet and of the Foldy-Wouthuysen model with
radiation damping are of completely different physical origins.
Radiation-reaction effects on the trajectory within the framework of the
Frenkel model are of the same small magnitude as for the classical
Foldy-Wouthuysen model and therefore are not shown in
Fig.~\ref{fig:in_magnetic_field}.

The fact that the electron center-of-mass trajectory does not depend on
the spin orientation suggests that the equivalence principle is
violated on the quantum level; i.\,e., the inertial mass, which enters
the Lorentz equation, does not equal the gravitational mass, which enters
the Einstein field equations via the energy-density component $T_{00}$
of the stress-energy tensor.  Note that the spin-dependent effective
mass~\eqref{eq:effm} of the Frenkel model results from postulating that
the on-shell condition
$m^2u_\alpha u^\alpha = (\mathcal{E}-q\varphi)^2/c^2 - \vec
p_\mathrm{kin}^2 = m^2c^2$ (with $\mathcal{E}$ denoting here the total
energy) for spinless charged particles also holds for electrons with
spin.  In the realm of relativistic quantum mechanics and in the
presence of magnetic fields, however, one cannot just replace physical
quantities that enter the classical on-shell condition by their
corresponding quantum mechanical operators to get the quantum mechanical
formulation of the on-shell condition.  In fact, the Dirac equation
yields the operator relation
$(\I\hbar\partial_t -q\varphi)^2/c^2 -
[\vec\alpha\cdot(-\I\hbar\vec\nabla-q\vec A)]^2=m^2c^2$, not
$(\I\hbar\partial_t -q\varphi)^2/c^2 - (-\I\hbar\vec\nabla-q\vec
A)^2=m^2c^2$.

In summary, the Frenkel model predicts a spin-dependent effective-mass
effect, which becomes observable in principle in ultrastrong magnetic
fields.  Such an effect is predicted neither by the quantum mechanical
Dirac theory nor by the classical Foldy-Wouthuysen model.  Therefore,
one may argue that the classical Foldy-Wouthuysen model is the superior
classical model for the electron in the investigated regime.  In
homogeneous magnetic fields, electron trajectories may be modified by
the radiation reaction but not by Stern-Gerlach forces.

\subsection{Spin-induced trajectory splitting in inhomogeneous magnetic
  fields}

Electrons moving in inhomogeneous magnetic fields experience a
spin-dependent Stern-Gerlach force in addition to the Lorentz force.
The effect of the Stern-Gerlach force is usually difficult to observe
due to the much larger Lorentz force and spreading of electron bunches.
In Refs.~\cite{Batelaan:Gay:etal:1997:Stern-Gerlach_Effect}
and~\cite{Gallup:Batelaan:etal:2001:Quantum-Mechanical_Analysis} it has
been demonstrated that in a longitudinal Stern-Gerlach setup the effect
of the Stern-Gerlach force can become observable.  Electrons traveling
though a current-carrying circular loop (with radius $\ell/\pi$) along
the symmetry axis experience a Stern-Gerlach force (anti-)parallel to
the direction of motion.  As a consequence, spin-forward electrons are
delayed relatively to spin-backward electrons.  The separation of
spin-forward and -backward electrons depends on the Stern-Gerlach force
and whether it has the form of~\eqref{eq:F_2_FW} or \eqref{eq:F_2_F} or
is even of another kind.

\begin{figure}
  \centering
  \includegraphics[scale=0.45]{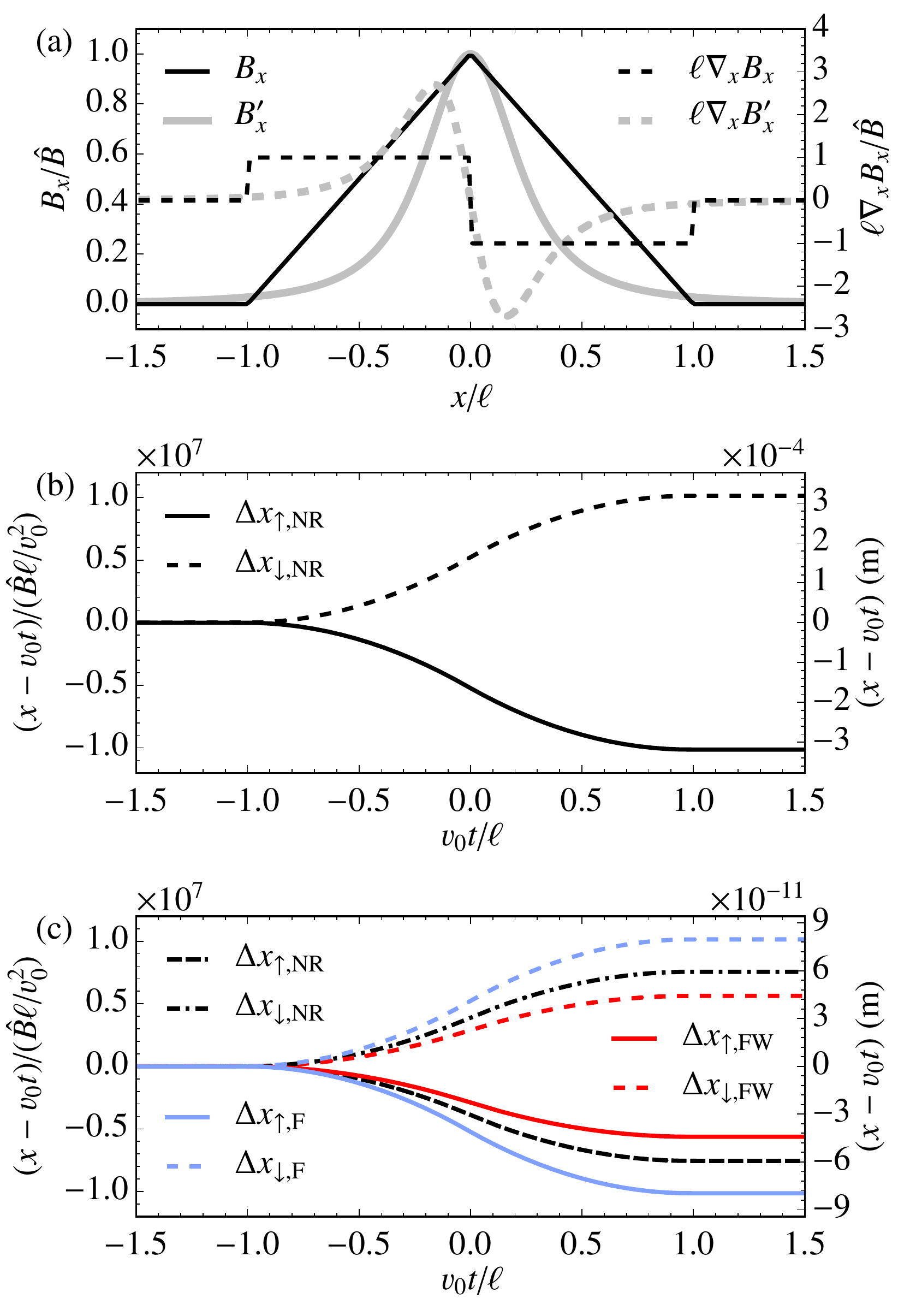}
  \caption{(a) Inhomogeneous magnetic field $B_x$ with gradient
    $\nabla_x B_x = \pm \hat B /\ell$ in the range $x \in [-\ell, \ell]$
    (solid black curve) and the field
    $B'_x=\hat B / \left(1+\pi^2x^2/\ell^2\right)^{3/2}$ (solid gray
    curve) generated by a single-loop solenoid with radius $\ell / \pi$.
    Their gradients are shown by dashed curves of the respective
    colors.  %
    (b) Trajectories of electrons with spin up and spin down predicted
    by the nonrelativistic spin model (indicated by the index NR).
    Electrons propagate in the magnetic field $B_x$ along the $x$ axis
    with an initial velocity $v_0 = 10^5\,\mathrm{m/s}$. The right
    $(x-v_0t)$ coordinate in SI units corresponds to a maximum magnetic
    field $\hat B = 10\,\mathrm{T}$ and the characteristic length
    $\ell/\pi = 1\,\mathrm{cm}$, similar to those presented in
    Ref.~\cite{Batelaan:Gay:etal:1997:Stern-Gerlach_Effect}. %
    (c) The same trajectories of electrons predicted by the
    nonrelativistic spin model, classical Foldy-Wouthuysen model
    (indicated by the index FW), and Frenkel model (indicated by the
    index F), respectively. The initial velocity of electrons is
    $v_0 = 2\times 10^8\,\mathrm{m/s}.$ }
  \label{fig:Bfield}
\end{figure}

Let us consider an electron moving in an inhomogeneous magnetic-field
configuration where the magnetic-field vectors point in a constant
direction.  The electron travels parallel to the magnetic-field
direction with the initial velocity $v_0$.  In this setup,
radiation-reaction effects are absent because the radiation reaction
force \eqref{eq:RR} vanishes.  Due to the Stern-Gerlach force this
electron moves in front or behind an electron moving freely at a
constant velocity $v_0$ by the distance
$\Delta x_{\uparrow/\downarrow}=x(t)-v_0t$, where the index in
$\Delta x_{\uparrow/\downarrow}$ indicates the electron's initial spin
orientation.  To estimate the dependence of this separation on the
magnetic-field gradient we examine the case of a constant field gradient
$\nabla_x B_x = \pm \hat B/\ell$ over the range $x\in[-\ell,\ell]$ as
indicated in Fig.~\ref{fig:Bfield}(a) by black curves, where $\hat B$
denotes the maximal field strength.  The asymptotic spin splitting
between electrons of different initial spin states can be estimated as
\begin{equation}
  \label{eq:dr}
  \Delta x =
  |\Delta x_{\downarrow}-\Delta x_{\uparrow}|
  \approx
  \frac{|q|\hbar}{\gamma m^2 v_0^2}\hat B \ell f_\mathrm{NR/FW/F} \,.
\end{equation}
In this equation, we introduced the model- and $\gamma$-dependent
coefficient $f$.  It is $f_\mathrm{NR}=1$ in the nonrelativistic (NR)
case (with Stern-Gerlach force
$F_\mathrm{2,\,NR} = (q/m)\nabla (\vec S \cdot \vec B)$),
$f_\mathrm{FW} = 1/\gamma$ in the classical Foldy-Wouthuysen model, and
$f_\mathrm{F}=\gamma$ in the Frenkel model, respectively.  For
nonrelativistic velocities $v_0$ both $f_\mathrm{FW}$ and $f_\mathrm{F}$
reduce to the nonrelativistic limit $f_\mathrm{NR}$.  In the
relativistic limit, however, the classical Foldy-Wouthuysen model and
the Frenkel model differ in their predictions for the considered setup.
As a consequence of \eqref{eq:dr}, the trajectory splittings in the
various models have different dependencies on the electron's Lorentz
factor, i.\,e., $\Delta x_\mathrm{NR} v_0^2 \sim 1/\gamma$,
$\Delta x_\mathrm{FW} v_0^2 \sim 1/\gamma^2$, and
$\Delta x_\mathrm{F} v_0^2$ is independent of $\gamma$, respectively.

Figures~\ref{fig:Bfield}(b) and \ref{fig:Bfield}(c) show the distance
$\Delta x_{\uparrow/\downarrow}$ for the nonrelativistic and
relativistic electron velocities for the various classical models and
the inhomogeneous magnetic field of a circular current.  The chosen
nonrelativistic parameters yield a spin splitting
$\Delta x \approx 634\, \mbox{\textmu m}$ as shown in
Fig.~\ref{fig:Bfield}(b), which coincides with that discussed in
Ref.~\cite{Batelaan:Gay:etal:1997:Stern-Gerlach_Effect}.  In the
relativistic regime, this scheme shows small but valuable numerical
differences
$\Delta x_{\uparrow/\downarrow}\lessapprox 10^{-10}\,\mathrm{m}$ among
different models; see Fig.~\ref{fig:Bfield}(c).

It should be noted that measuring the distances
$\Delta x_{\uparrow/\downarrow}$ or $\Delta x$ becomes very challenging
in the relativistic limit due to the $1/v_0^2$ dependency.  However, the
differences that we have found between the Stern-Gerlach forces of the
different classical models will help us to understand the effects of the
different forms of the Stern-Gerlach force in more complicated field
configurations, e.\,g., strong electromagnetic waves. One may not
attempt to amplify the spin splitting by enhancing the Stern-Gerlach
force via decreasing $\ell$.  A possible interpretation of
Eq.~\eqref{eq:dr} is that while the Stern-Gerlach force dominated by the
field gradient $\vec\nabla\cdot \vec B \sim 1/\ell$ decreases with the
characteristic length scale of the field variation $\ell$, the time to
cover the distance $\ell$ is $\Delta t \approx \ell / v_0$, where $v_0$
is the velocity in the direction of the field gradient.  Thus,
$\Delta x \sim (\vec\nabla\cdot \vec B) (\Delta t)^2$ is proportional to
$\ell$.

Compared to gradients in static magnetic fields, much stronger field
gradients can be produced by the fast-developing \mbox{x-ray} and
infrared high-intensity laser facilities.  Due to the short wavelength
the separation $\Delta x$ that can be created during a single cycle
tends to be small.  It can, however, become large (depending on the
model) for relativistic electrons as we show in the following section.

\subsection{Electron dynamics in strong laser pulses}
\label{sec:focused}

Ultra strong laser fields are promising candidates for systems in which
effects of Stern-Gerlach forces or of radiation reaction may be observed
\cite{Ruiz:Ellison:etal:2015:Relativistic_ponderomotive}.  Thus, in the
following we study the motion of free electrons in ultrastrong laser
pulses within the frameworks of the Dirac equation, the classical
Foldy-Wouthuysen model, and the Frenkel model.  Spin effects and
radiation reaction effects are enhanced at short wavelengths in the
\mbox{x-ray} regime, which is applied in the following.  Nevertheless,
spin effects and radiation reaction effects may become detectable also
in the optical or even in the near-infrared regime as will be shown.

The electric-field component of a linearly polarized
\mbox{$\sin^2$-shaped} plane-wave laser pulse with linear polarization
along the $y$~direction and propagating along the $x$~axis is given by
\begin{align}
  \label{eq:E_pulse}
  \vec E(\vec r, t) =
  \hat{E}\,\mathcal{H}\left(
    \frac{-\psi}{n}
  \right)\mathcal{H}\left(
    \frac{\psi}{n }+\pi
  \right)\,\sin^2\left(\frac{\psi}{n}\right)
 \sin\left(\psi\right) \vec{e}_y\,,
\end{align}
with $\hat E$ denoting the peak amplitude, the Heaviside step function
$\mathcal{H}(\psi)$, the phase $\psi=2 \pi(x-ct)/\lambda$, the
wavelength $\lambda$, and the pulse width $n$ measured in laser cycles.
The magnetic-field component follows via
$\vec B(\vec r, t)=\mbox{$\vec{e}_x\times \vec E(\vec r, t)/c$}$.  At
time $t=0$, the front of the laser pulse reaches the electron at the
origin of the coordinate system.  The electron's initial spin
orientation is parallel or antiparallel to the $z$~axis, i.\,e., along
the direction of the magnetic field.  Note that as a consequence of
Eqs.~\eqref{eq:class_F_s} and~\eqref{eq:class_FW_S} the spin remains in
its initial state for all times for the considered setup.

\begin{figure}
  \centering
  \includegraphics[scale=0.45]{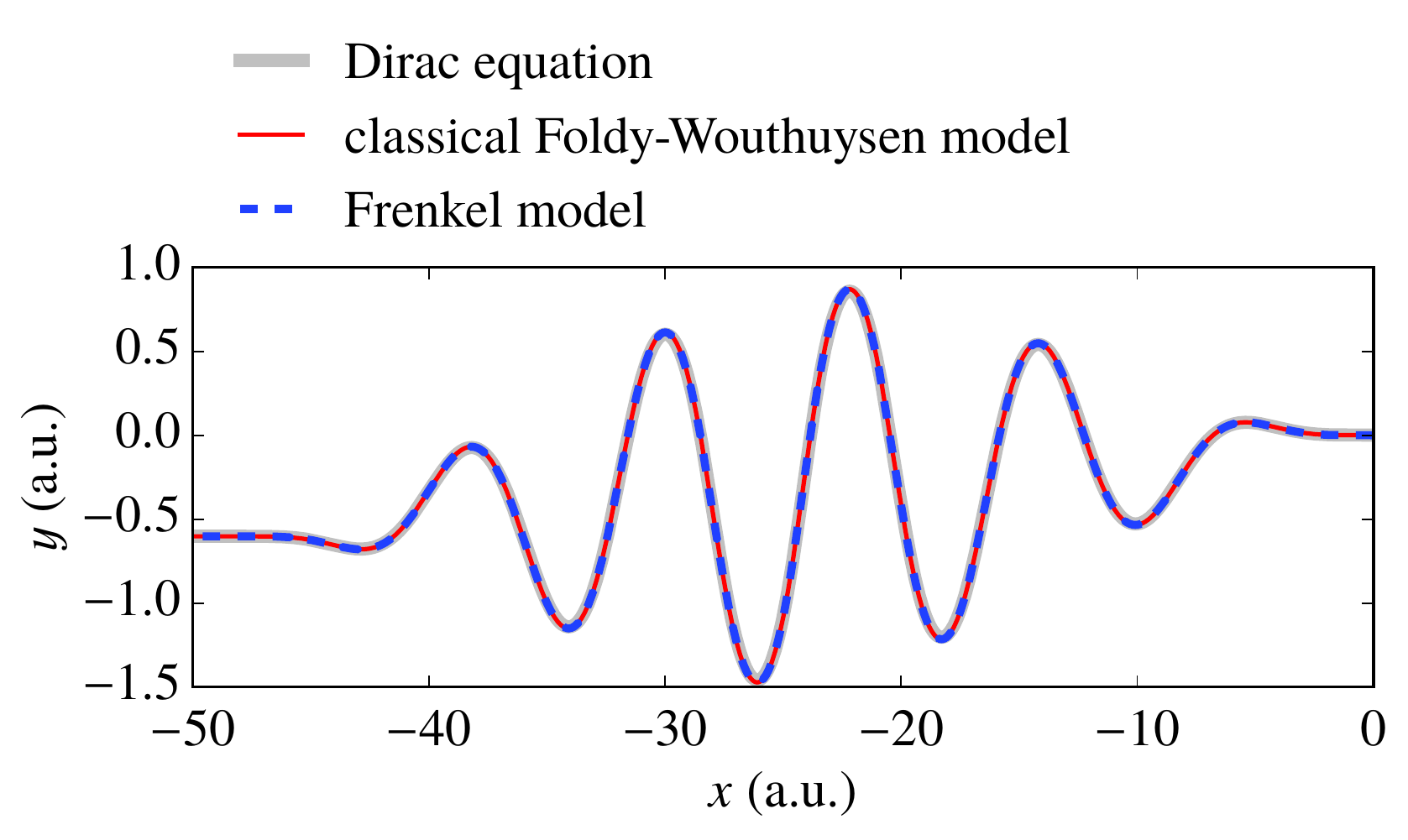}
  \caption{Trajectories of electrons in a plane-wave laser pulse
    \eqref{eq:E_pulse} with the initial spin orientation parallel to the
    $z$ direction and initial momentum
    $\vec p_\mathrm{kin}(0)=\trans{(-mc,0,0)}$ as predicted by the
    various considered models.  Laser parameters are
    $\hat
    E=5000\,\mathrm{a.u.}=\mbox{$2.57\times10^{15}\,\mathrm{V/m}$}$,
    $\lambda=20\,\mathrm{a.u.}=1.06\,\mathrm{nm}$, and $n=6$.  In the
    case of the Dirac equation the wave packet had a width of
    $0.5\,\mathrm{a.u.}=0.026\,\mathrm{nm}$ initially.}
  \label{fig:free_partice_laser}
\end{figure}

We solved the time-dependent Dirac equation as well as the equations of
motion of the classical Foldy-Wouthuysen model and the Frenkel model for
the setup described above.  As a consequence of the Stern-Gerlach
forces, the electron's trajectory depends on the electron's spin
orientation.  Although the influence on the shape of the trajectory can
be resolved within the numerical accuracy, it is very small as shown in
Fig.~\ref{fig:free_partice_laser}.  The trajectories as obtained by the
three models are indistinguishable from each other at the scale of
Fig.~\ref{fig:free_partice_laser}.  Note that the three trajectories in
Fig.~\ref{fig:free_partice_laser} are also indistinguishable from the
trajectory predicted by the pure Lorentz force (not shown in
Fig.~\ref{fig:free_partice_laser}), which is independent of the
electron's spin.  For a substantial effect of the Stern-Gerlach forces
on the electron trajectory, the Stern-Gerlach forces must reach the same
order of magnitude as the Lorentz force, as it is the case when the
wavelength of the laser field is of the order of the Compton wavelength.
At this scale, however, also other quantum effects due to the nonzero
width of the electron wave packet are expected to set in.

Although the spin effect on the trajectories of spin-$1/2$ particles is
not discernible at the scale of Fig.~\ref{fig:free_partice_laser}, it is
possible to determine and to compare the spin-induced Stern-Gerlach
forces predicted by the Dirac theory, the Frenkel model, and the
classical Foldy-Wouthuysen model in this parameter regime where
radiation reaction is negligibly weak as we have shown in
Ref.~\cite{Wen:2016:Identifying}.  It was demonstrated that the two
considered classical models feature different dynamics during the
interaction with a plane-wave x-ray laser field.  The net effect is,
however, too small to distinguish between the two models.  In contrast,
the interaction of electrons with focused infrared laser pulses of
finite transverse size leads to a distinguishable net dynamics of the
classical Foldy-Wouthuysen and the Frenkel models, as demonstrated in
this section.  Thus, it is potentially possible for the classical models
to be tested experimentally by employing infrared laser pulses of
upcoming high-power laser facilities.  While the key results of this
regime have been presented in Ref.~\cite{Wen:2016:Identifying}, in this
section we extend this investigation by including the radiation reaction
via the Landau-Lifshitz force~\eqref{eq:RR}.  Numerical solutions of the
Dirac equation are not feasible in this regime due to the long time
scale of infrared laser pulses~\footnote{The longer pulse length in
  combination with the wave-packet spreading leads to a $\lambda^3$ or
  $\lambda^4$ scaling of the computational demand to solve the Dirac
  equation in two or three dimensions, respectively.} and are not
suitable, as radiation reaction effects are now substantial.

In order to reach a high intensity, the laser pulse is often tightly
focused to a spot of several laser wavelengths.  The pulse profile is
modeled as a Gaussian beam with the transversal focus radius $w_0$.
Introducing the Rayleigh length $x_r=\pi w_0^2/\lambda$ and the
diffraction angle $\epsilon=w_0/x_r=\lambda/(\pi w_0)$, the components
of the electromagnetic fields are expressed to the fifth order of the
small diffraction angle $\epsilon$ as (see Eqs.~(1)--(9) in
Ref.~\cite{Salamin:PhysRevLett.88.095005:2002})
\begin{widetext}
  \allowdisplaybreaks%
  \begin{subequations}
    \label{field:yp}
    \begin{align}
      E_x & = E \nu \left(
        \epsilon C_1  + \epsilon^3\left(
          - \frac{C_2}{2} + \rho^2 C_3 -\frac{\rho^4 C_4}{4}
        \right) +
        \epsilon^5\left(
          -\frac{3 C_3}{8}-\frac{3\rho^2 C_4}{8} + \frac{17\rho^4 C_5}{16} -
          \frac{3 \rho^6 C_6}{8} + \frac{\rho^8 C_7}{32}
        \right)
      \right) \,,\\
      E_y & = E \left(
        S_0 + \epsilon^2\left(\nu^2 S_2 - \frac{\rho^4 S_3}{4}\right) +
        \epsilon^4\left(
          \frac{S_2}{8}-\frac{\rho^2 S_3}{4} -
          \frac{\rho^2(\rho^2-16\nu^2)S_4}{16} -
          \frac{\rho^4(\rho^2+2\nu^2)S_5}{8} +
          \frac{\rho^8S_6}{32}
        \right)
      \right)\,, \\
      E_z & = E\nu\zeta \left(
        \epsilon^2 S_2 +\epsilon^4\left(\rho^2 S_4 - \frac{\rho^4 S_5}{4}\right)
      \right)\,, \\
      B_x & = \frac{E \zeta}{c} \left(
        \epsilon C_1  + \epsilon^3\left(
          \frac{C_2}{2} + \frac{\rho^2 C_3}{2} - \frac{\rho^4 C_4}{4}
        \right)  +
        \epsilon^5\left(
          \frac{3 C_3}{8}+\frac{3\rho^2 C_4}{8} + \frac{3\rho^4 C_5}{16} -
          \frac{ \rho^6 C_6}{4} + \frac{\rho^8 C_7}{32}
        \right)
      \right) \,, \\
      B_y & = 0 \,, \\
      B_z & = \frac{E}{c} \left(
        S_0 + \epsilon^2\left(
          \frac{\rho^2 S_2}{2} - \frac{\rho^4 S_3}{4}
        \right) +  \epsilon^4\left(
          -\frac{S_2}{8}+\frac{\rho^2S_3}{4} +
          \frac{5\rho^4 S_4}{16} - \frac{\rho^6 S_5}{4} +
          \frac{\rho^8 S_6}{32}
        \right)
      \right) \,.
    \end{align}
  \end{subequations}
\end{widetext}
Here, we have introduced the functions
\begin{subequations}
  \begin{align}
    E & =
    \hat Ew_\parallel\left(-\frac{\psi}{2n}\right)
    \frac{w_0}{w_\perp(x)}\exp\left(-\frac{r^2}{w_\perp(x)^2}\right) \,, \\
    S_j & = \left(\frac{w_0}{w_\perp(x)}\right)^2
    \sin\left(\psi+j \arctan\left(\frac{x}{x_r}\right)\right) \,,\\
    C_j & = \left(\frac{w_0}{w_\perp(x)}\right)^2
    \cos\left(\psi+j \arctan\left(\frac{x}{x_r}\right)\right)  \,,
  \end{align}
\end{subequations}
for integer $j$, the \mbox{$\sin^2$-shaped} longitudinal profile
$w_\parallel(\eta) =
\mathcal{H}\left(-\eta\right)\mathcal{H}\left(\eta+\pi\right)\,\sin^2\left(\eta\right)$,
the pulse width $n$ measured in laser cycles, and the radius along the
propagation axis $w_\perp(x)=w_0\sqrt{1+(x/x_r)^2}$.  The phase
$\psi$ is defined as
\begin{equation}
  \psi =
  2\pi\left(\frac{ct-x}{\lambda}\right) -
  \frac{\pi r^2/\lambda}{x+x_r^2/x} +
  \arctan\left(\frac{x}{x_r}\right)
\end{equation}
and the equations above depend on the parameters $\nu=y/w_0$,
$\zeta=z/w_0$, $\rho=r/w_0$, and $r=\sqrt{y^2+z^2}$.

\begin{figure}[b]
  \centering
  \includegraphics[scale=0.45]{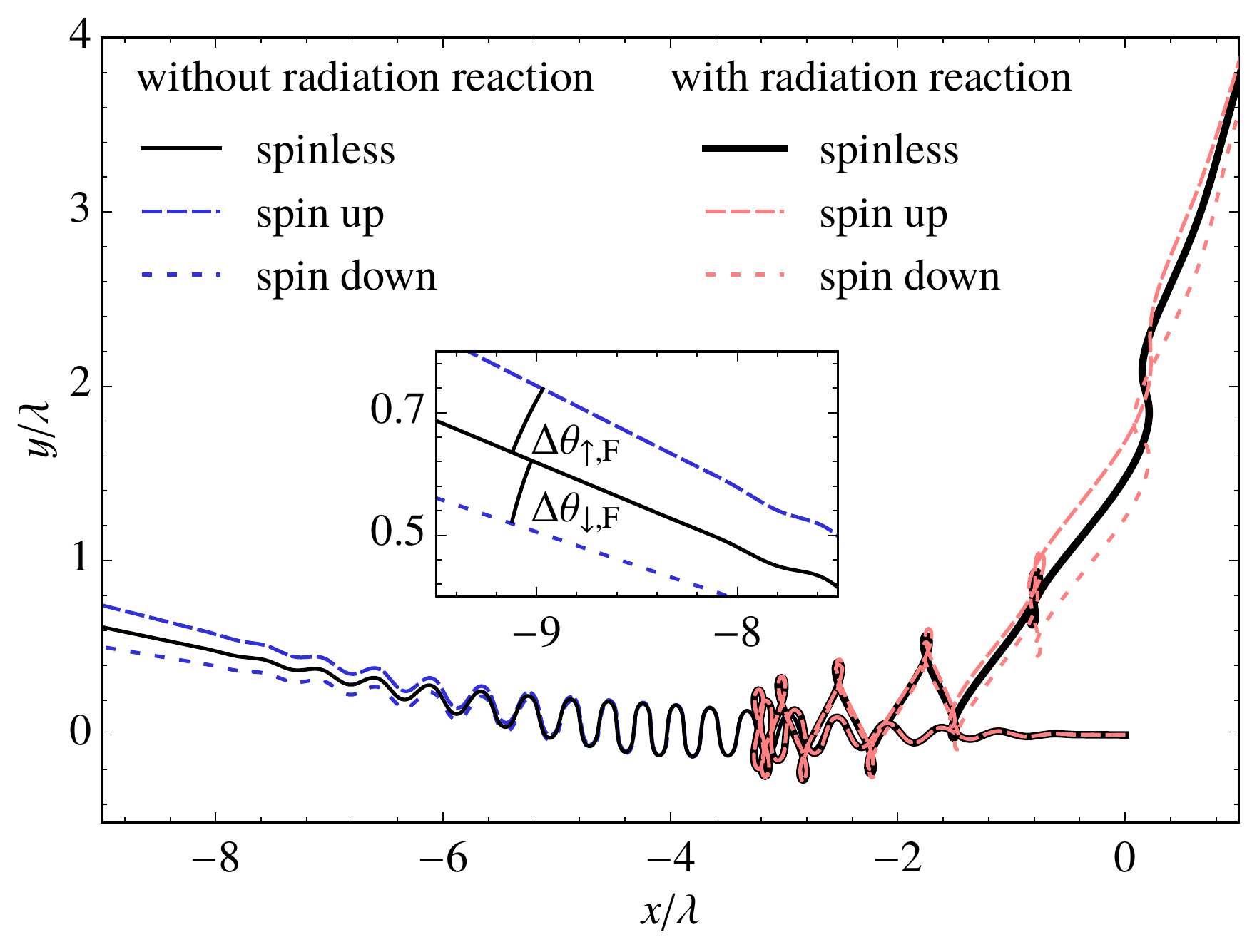}
  \caption{Trajectory of a highly energetic electron with the initial
    $\gamma=100$ in a focused laser pulse with wavelength
    $\lambda=1.51\times10^4\,\mathrm{a.u.}=800\,\mathrm{nm}$, amplitude
    of strength
    $\hat E = 1056\,\mathrm{a.u.} = 8.03\times10^{14}\,\mathrm{V/m}$,
    duration (number of cycles) $n=20$, and focus radius $w_0=2\lambda$.
    Black curves denote the trajectories of a spinless particle and
    dashed dark blue and dashed light red curves correspond to those
    described by the Frenkel model \eqref{eq:F_2_F} for electrons with
    the spin parallel (spin up) and antiparallel (spin down) to the
    $z$~axis.  Trajectories without radiation reaction force (thin black
    and dashed dark blue curves) show smaller reflection angle than
    those with radiation reaction (thick black and dashed light red
    curves).  Inset: Definition of the extra deflection angles
    $\Delta\theta_{\uparrow,\mathrm{F}}$ and
    $\Delta\theta_{\downarrow,\mathrm{F}}$ for spin-up and spin-down
    electrons.  All trajectories shown start from position $(0, 0)$,
    where the electron hits the front of the laser pulse.  Note that the
    field gradient is three orders of magnitude lower than for the setup
    in Fig.~\ref{fig:free_partice_laser}.}
  \label{fig:focussed_pulse}
\end{figure}
\begin{figure}
  \centering
  \includegraphics[scale=0.45]{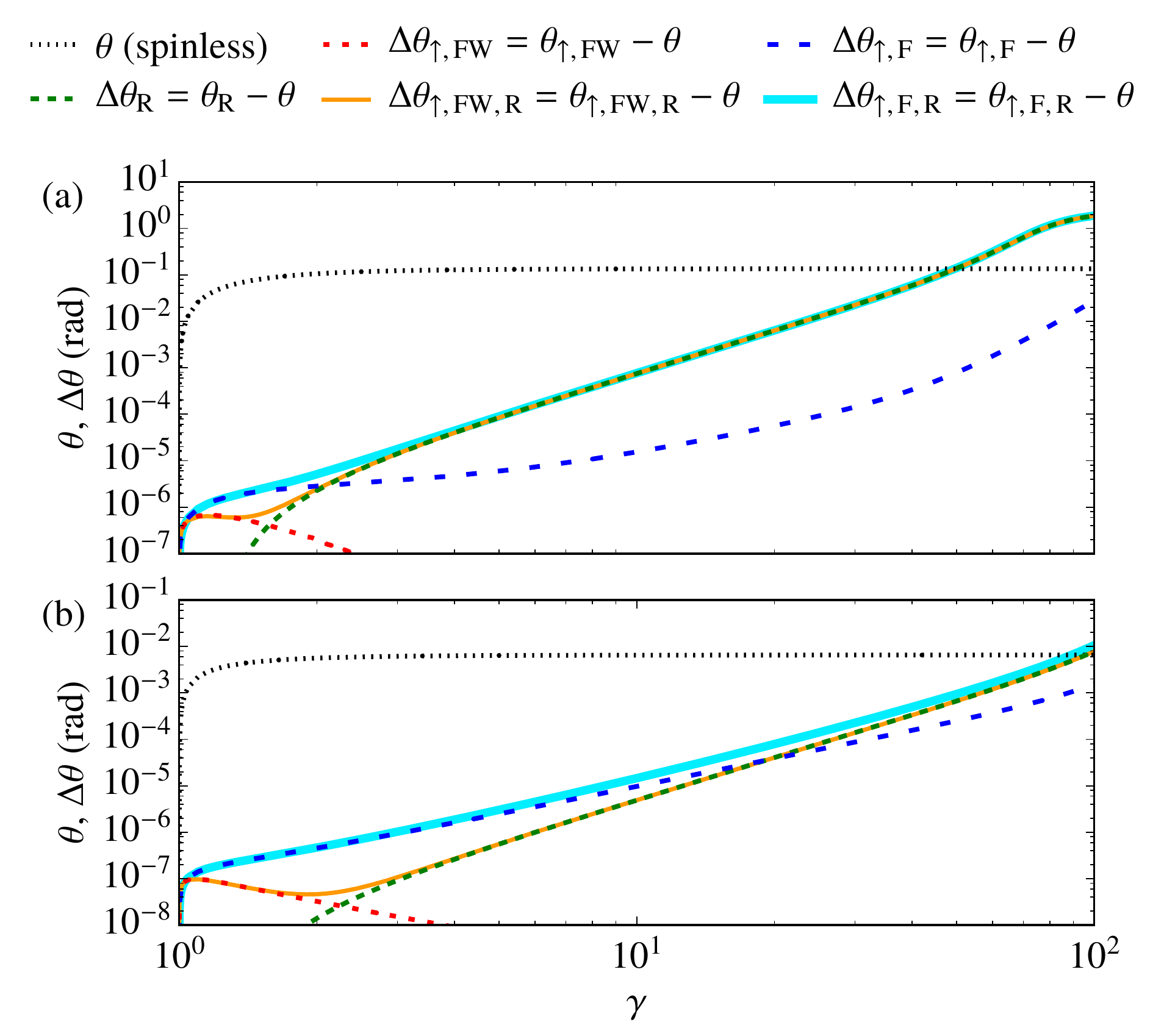}
  \caption{Ponderomotive deflection angle $\theta$ and extra angles
    $\Delta \theta$ induced by the radiation reaction force and spin
    forces of the Frenkel and the classical Foldy-Wouthuysen models as
    functions of the particle's initial energy, represented by the
    Lorentz factor $\gamma$, with the electric-field-strength scaling
    $\hat E=4\pi mc^2\sqrt{\gamma^2-1}/(|q|\lambda)$ in (a) and
    $\hat E=2\pi mc^2\sqrt{\gamma^2-1}/(|q|\lambda)$ in (b).  Dotted
    black curves correspond to the deflection of spinless particles
    without accounting for the radiation reaction.  Dashed green lines
    correspond to the extra deflection $\Delta \theta$ induced by the
    radiation reaction only, while dashed red and blue curves correspond
    to the extra deflection $\Delta \theta$ induced by the spin force
    within the classical Foldy-Wouthuysen model and the Frenkel model,
    respectively.  Solid orange and cyan curves correspond to extra
    deflections from the classical Foldy-Wouthuysen and Frenkel models
    including the radiation reaction.  In all cases, electrons in the
    spin-up state (parallel to the magnetic field) are considered.}
  \label{fig:focussed_pulse_theta0}
\end{figure}

An electron that is initially directed towards the focus of the
counterpropagating high-intensity laser pulse is displaced transversely
due to the (modified) Lorentz force
$\vec F_\mathrm{1,\,FW}\approx\vec F_\mathrm{1,\,F}$, which is induced
by the transverse electric-field component $E_y$.  When the oscillating
field changes sign, the force drives the electron back to its initial
transverse position.  However, this force is smaller than the expelling
force due to the longitudinal focusing inhomogeneity.  As illustrated by
the thin black line in Fig.~\ref{fig:focussed_pulse}, the oscillation
center of a spinless charged particle drifts radially from the spot
center, which is called ponderomotive scattering
\cite{Batelaan:Contemporary_Physics:2000,hartemann2001high}.  The
deflection of a spinless charged particle in the ponderomotive potential
of the focused laser pulse is defined by the angle between the final
transverse momentum and the longitudinal momentum
$\theta=-\arctan(p_{\mathrm{kin},y}/p_{\mathrm{kin},x})$ after the
particle is separated from the laser fields.  As shown in
Fig.~\ref{fig:focussed_pulse_theta0}(a), the ponderomotive deflection
$\theta$ increases with increasing initial energy of the electron and
increasing field strength, which was chosen to scale as
$\hat E=4\pi mc^2\sqrt{\gamma^2-1}/(|q|\lambda)$.  This particular
scaling causes a strong acceleration of the electron opposite to its
initial velocity but without reflecting it, i.\,e., $|\theta| < \pi/2$.
However, this nonreflecting condition can be broken due to damping when
the relativistic motion leads to strong radiation
\cite{Di-Piazza:Hatsagortsyan:etal:2009:Strong_Signatures}.  The
classical radiation reaction force~\eqref{eq:RR} leads, in the high
relativistic limit, to a reflection angle $|\theta_\mathrm{R}| > \pi/2$,
as can be seen by comparing the thin and the thick black lines in
Fig.~\ref{fig:focussed_pulse}.  The total reflection angle
$\theta_\mathrm{R}$ sensitively depends on the electric-field strength.
The electrons cannot be reflected even when the radiation reaction is
considered when the electric field is reduced by a factor of~2,
i.\,e., with the field-strength scaling
$\hat E=2\pi mc^2\sqrt{\gamma^2-1}/(|q|\lambda)$.  In this case, the
deflection angles in all models are less than $\pi/2$ in the applied
energy region, as shown in Fig.~\ref{fig:focussed_pulse_theta0}(b).

\begin{figure}
  \centering
  \includegraphics[scale=0.45]{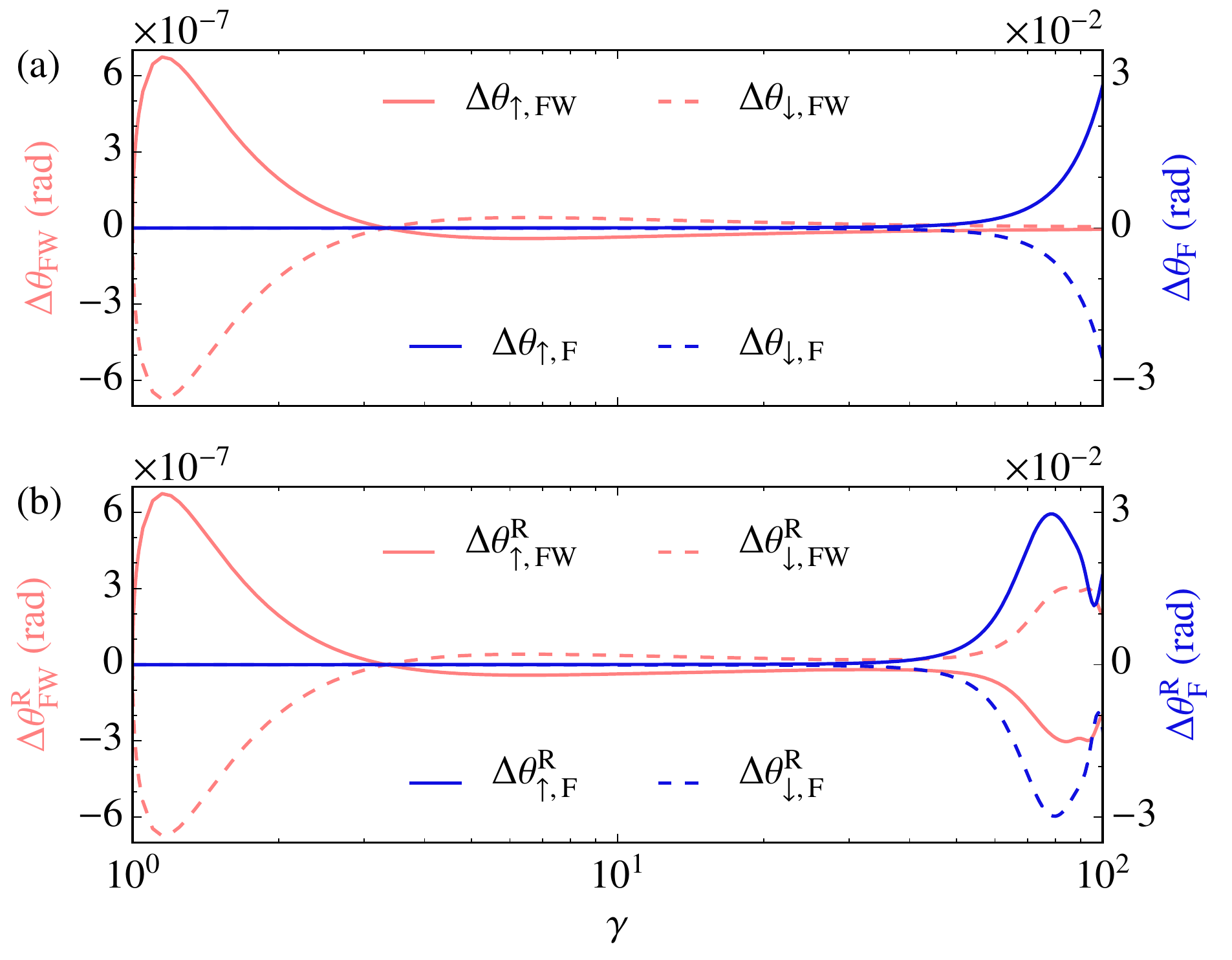}
  \caption{Spin contribution to the deflection angle as a function
    of the initial energy of the particle represented by $\gamma$ for
    spin-up and spin-down electrons.
    (a) The difference in the deflection angles of a spinless particle
    and a spin-$1/2$ electron for the Frenkel vs the classical
    Foldy-Wouthuysen models, i.\,e., $\Delta\theta_{\uparrow/\downarrow,\,\mathrm{FW/F}}=\theta_{\uparrow/\downarrow,\,\mathrm{FW/F}}-\theta$.
    (b) The difference in the deflection angles of a spinless particle
    and a spin-$1/2$ electron for the Frenkel and the classical
    Foldy-Wouthuysen models taking into account the radiation
    reaction force \eqref{eq:RR} for the spinless particles as well as
    for the electron, i.\,e., $\Delta\theta_{\uparrow/\downarrow,\,\mathrm{FW/F}}^\mathrm{R}=\theta_{\uparrow/\downarrow,\,\mathrm{FW/F},\,\mathrm{R}}-\theta_\mathrm{R}$.
    In both panels, the laser parameters are the same as in
    Fig.~\ref{fig:focussed_pulse} but the electric-field strength scales
    with the initial $\gamma$ as
    $\hat E=4\pi mc^2\sqrt{\gamma^2-1}/(|q|\lambda)$.  Note the
    different scales on the $y$~axes for the Frenkel and the classical
    Foldy-Wouthuysen models.}
  \label{fig:focussed_pulse_theta1}
\end{figure}

In addition to the deflection of a charged particle in the ponderomotive
potential of the laser fields, the spin may induce a further deflection
via the Stern-Gerlach forces \eqref{eq:F_2_FW} and \eqref{eq:F_2_F} if
the electron is polarized along the direction of the magnetic-field
component.  For the laser pulse \eqref{field:yp} this means that the
electron is polarized along the positive or negative $z$~axis,
representing spin-up (indicated by~$\uparrow$) and spin-down (indicated
by~$\downarrow$) states. Note that, the electron's spin is initially
parallel to the $z$ axis in the above considered setup and it travels
towards the center of the focused laser pulse; i.\,e., it moves along
$z=\zeta=0$, where $E_z/c=B_x=B_y=0$.  Thus
$\vec{B}_{\mathrm{eff},\,\perp z}=0$ in the \mbox{$x$-$y$} plane, and
consequently, the electron remains in the $x$-$y$ plane and the spin of
the electron is frozen in its initial state.  Moreover, because of the
smallness of the field gradients around the laser pulse's center the
motion of an electron bunch is still limited to the \mbox{$x$-$y$}
plane, when the transversal initial distance between each electron and
the focus center is much smaller than the size of the focus spot
$w_0$. The trajectories (given by the Frenkel model) of electrons with
spin up and down are indicated by the dashed dark blue curves in
Fig.~\ref{fig:focussed_pulse}.  The Stern-Gerlach force \eqref{eq:F_2_F}
enhances or decreases the deflection by the extra angle
$\Delta \theta_{\uparrow/\downarrow,\mathrm{F}} =
\theta_{\uparrow/\downarrow,\mathrm{F}} - \theta$, where
$\theta_{\uparrow/\downarrow,\mathrm{F}}$ denotes the deflection angle
within the Frenkel theory for spin-up electrons and spin-down electrons,
respectively.  The magnitude of
$\Delta \theta_{\uparrow/\downarrow,\mathrm{F}}$ increases with the
electron's energy and reaches the magnitude of $10^{-2}\,\mathrm{rad}$
for relativistic electrons in high-intensity laser fields of the applied
parameters as shown in Figs.~\ref{fig:focussed_pulse_theta0}(a) and
\ref{fig:focussed_pulse_theta1}(a).

The classical Foldy-Wouthuysen model behaves qualitatively similarly to
the Frenkel model and is therefore not considered in
Fig.~\ref{fig:focussed_pulse}.  The two spin models lead, however, to
quantitatively different extra deflection angles compared to the
spinless case.  This means that we find for the classical
Foldy-Wouthuysen model extra deflection angles that are much smaller
than those for the Frenkel model shown in Fig.~\ref{fig:focussed_pulse}.
The red and blue curves in Figs.~\ref{fig:focussed_pulse_theta0} and
\ref{fig:focussed_pulse_theta1}(a) correspond to energy-dependent
deflections from the classical Foldy-Wouthuysen model and the Frenkel
model, respectively.  In contrast to the Frenkel model, the deflection
as predicted by the classical Foldy-Wouthuysen model does not vary with
the electron's initial energy $\gamma$ monotonically. It may even change
sign, as indicated in Fig.~\ref{fig:focussed_pulse_theta1}(a).
Furthermore, the absolute value of the spin-induced additional
deflection angle $\Delta\theta_\mathrm{\uparrow/\downarrow,\mathrm{FW}}$
from the classical Foldy-Wouthuysen model remains under the magnitude of
$10^{-6}\,\mathrm{rad}$ and decreases with the electron's initial energy
$\gamma$ in the relativistic parameter regime as shown in
Fig.~\ref{fig:focussed_pulse_theta1}(a).  In an experiment, the
reference angle $\theta$ can not be measured, because it corresponds to
a hypothetical spinless electron.  But the total deflection angles
$\theta_{\uparrow,\,\mathrm{FW/F}}$ and
$\theta_{\downarrow,\,\mathrm{FW/F}}$ of oppositely polarized electrons
can be determined.  The difference %
$|\theta_{\uparrow,\,\mathrm{FW/F}}-\theta_{\downarrow,\,\mathrm{FW/F}}|$
is about twice the value of $|\Delta\theta_{\uparrow,\,\mathrm{FW/F}}|$
and $|\Delta\theta_{\downarrow,\,\mathrm{FW/F}}|$ in magnitude.

For completeness, we also considered the case where radiation-reaction
effects are included in the Frenkel and the classical Foldy-Wouthuysen
models as shown in Figs.~\ref{fig:focussed_pulse_theta0} and
\ref{fig:focussed_pulse_theta1}(b).  Comparing the curves for spin
one-half particles with and without radiation reaction in
Fig.~\ref{fig:focussed_pulse_theta0}, one sees that the total effect on
the deflection angle of a radiating spin-$1/2$ electron is dominated by
the Stern-Gerlach forces in the low-energy region and it is dominated by
the radiation reaction in the high-energy region.  The size of the
spin-force-dominated energy region depends on the applied scaling
between the field strength and the electron's energy.  It becomes larger
for weaker fields as shown Figs.~\ref{fig:focussed_pulse_theta0}(a) and
\ref{fig:focussed_pulse_theta0}(b).  Although the total deflection angle
can be rather sensitive to radiation-reaction effects in the high-energy
region, the spin contributions to the deflection angle are not sensitive
to the radiation reaction.  Including also radiation reaction effects in
the Frenkel and the classical Foldy-Wouthuysen models yields
spin-induced additional deflection angles of a similar magnitude to
those without the radiation reaction, which can be identified by
comparing Figs.~\ref{fig:focussed_pulse_theta1}(a)
and~\ref{fig:focussed_pulse_theta1}(b).  As discussed above, the
radiation reaction as well as the spin effects may alter the trajectory
of electrons in strong focused laser pulses compared to the classical
dynamics caused by the Lorentz force.  Spin effects and the radiation
reaction become relevant in similar parameter regimes.  Pure spin
effects can be isolated by comparing the dynamics of electrons with
opposite spin states.  The spin-induced contribution to the deflection
predicted by the classical Foldy-Wouthuysen model,
$<10^{-6}\,\mathrm{rad}$, for the applied parameters, is too small to be
demonstrated experimentally.  However, the Frenkel model leads to a
total deflection via the ponderomotive potential of the order of
$10^{-1}\,\mathrm{rad}$ and an additional spin-induced deflection of the
order of $10^{-2}\,\mathrm{rad}$, if an electron beam with an energy of
tens of MeV and an infrared laser of intensity
$\sim 10^{22}\,\mathrm{W/cm^2}$ as discussed above can be applied.
Considering that electron bunches with an emittance as low as
$10^{-3}\,\mathrm{rad}$ have been prepared
\cite{PhysRevLett.109.064802}, the deflection via the ponderomotive
potential of the order of $10^{-1}\mathrm{rad}$ and also the
spin-induced deflection of the order of $10^{-2}\,\mathrm{rad}$ are
potentially measurable in experiments.  In such an experiment, polarized
electron bunches of low emittance would be employed to measure the
deflection angle as a function of the polarization direction.  In
current head-on experiments, however, with focused fields of high
inhomogeneities and energetic electrons, e.\,g.,
\cite{Burke:Field:etal:1997:Positron_Production} and
\cite{Ta-Phuoc:Corde:etal:2012:All-optical_Compton}, no significant spin
effect in orbital motion was observed.  The lack of experimental
evidence of a nonnegligible spin-induced deflection may be seen again as
the superiority of the classical Foldy-Wouthuysen model regarding the
issue of spin-modified dynamics.

In common experimental setups, unpolarized electron beams are often
employed rather than polarized beams.  The (theoretical) deflection
angle for a beam of unpolarized electrons can be determined by averaging
over trajectories (with or without radiation reaction) of electrons with
initial spin states evenly distributed in all directions.  The spin
contribution to the deflection angle is almost symmetric under spin
inversion as shown in Fig.~\ref{fig:focussed_pulse_theta1}.  Thus, for
our parameters spin effects average out for unpolarized electron beams
and the beam trajectory is mainly determined by the Lorentz force and
radiation-reaction forces.  This means that experimental tests for
radiation-reaction effects via the Landau-Lifshitz force without spin
terms may be realized by employing unpolarized electron beams
\cite{Bogdanov:Kazinski:2015:Properties_of}.  In addition, considering,
e.\,g., Fig.~\ref{fig:focussed_pulse_theta1}, the radiation-reaction
terms in the Landau-Lifshitz equation may also be tested via our
spin-dependent descriptions by utilizing spin-polarized electrons.

\section{Discussion and conclusions}
\label{sec:conclusions}

Classical theories of charged particles with spin and radiation reaction
are valuable for describing light-matter interaction at high intensities
when quantum effects are not important.  They are appealing because
classical models can be employed to describe many-particle systems where
fully quantum mechanical models become intractable.  We have
investigated the dynamics of spin-$1/2$ particles in various setups by
applying two classical spin models, the classical Foldy-Wouthuysen model
and the Frenkel model, which are both supplemented by a classical
radiation-reaction model, the Landau-Lifshitz equation.  The Frenkel
model and the classical Foldy-Wouthuysen model have been introduced to
describe the fully relativistic dynamics of electrons in strong
electromagnetic fields and both are commonly applied in various branches
of physics.  The predictions of these classical models have been
compared to each other and to predictions by the Dirac equation, when a
solution of the Dirac equation was feasible.  Discrepancies in the
predictions of the two classical models may become experimentally
detectable in light-matter interactions in strong, highly focused
electron beams, where the radiation reaction is also expected to set in.

According to the Frenkel model, the potential energy of a spin in a
magnetic field induces a spin-dependent effective mass~\eqref{eq:effm},
which leads to a Stern-Gerlach-like splitting of trajectories of
electrons with different spin states even in homogeneous magnetic
fields.  This effect, however, is not predicted by the more fundamental
Dirac equation.  Thus, one may argue that the spin-dependent effective
mass effect is not physical.

In the setup of the longitudinal Stern-Gerlach effect, where
radiation-reaction effects are absent, the Frenkel model and classical
FreFoldy-Wouthuysen model lead, in the relativistic limit, to qualitatively
different spin effects on the electron trajectory.  Furthermore, we have
demonstrated that in tightly focused beams in the near-infrared the
effect of the Stern-Gerlach force of the Frenkel model becomes
sufficiently large to be potentially detectable in an experiment.
Depending on the electron's energy and the electromagnetic-field
configuration the radiation-reaction effect on the electron dynamics may
be stronger than possible spin effects.  Nevertheless, the spin-induced
contributions can be identified by employing electrons of opposite spin
state.  Modifications of the electron motion due to the spin are almost
symmetric under spin inversion, at least for the considered setups (see
Fig.~\ref{fig:focussed_pulse_theta1}).  Thus, for unpolarized electron
beams spin effects average out on the level of the trajectory of the
whole beam.  This means that radiation-reaction effects are identifiable
by comparing the spin-averaged electron trajectory to the (hypothetical)
trajectory of a charged spinless particle for the considered parameters
as given by the Lorentz force only.

Among the classical spin models, the Frenkel model is certainly
prominent for its long history and its wide application.  Our results,
however, suggest that the classical Foldy-Wouthuysen model is superior,
as it is qualitatively in better agreement with the quantum mechanical
Dirac equation.  An experimental evaluation of the classical spin
theories with currently available electron energies of tens of mega
electronvolts and a laser intensity $\sim 10^{22}\,\mathrm{W/cm^2}$ as
suggested in this paper would provide valuable support or disagreement
with the classical spin theories complementary to our comparison to the
Dirac equation.

\acknowledgments

We would like to thank Antonino~Di~Piazza for valuable discussions.

\bibliography{Frenkel}

\begin{thebibliography}{99}%
\makeatletter
\providecommand \@ifxundefined [1]{%
 \@ifx{#1\undefined}
}%
\providecommand \@ifnum [1]{%
 \ifnum #1\expandafter \@firstoftwo
 \else \expandafter \@secondoftwo
 \fi
}%
\providecommand \@ifx [1]{%
 \ifx #1\expandafter \@firstoftwo
 \else \expandafter \@secondoftwo
 \fi
}%
\providecommand \natexlab [1]{#1}%
\providecommand \enquote  [1]{``#1''}%
\providecommand \bibnamefont  [1]{#1}%
\providecommand \bibfnamefont [1]{#1}%
\providecommand \citenamefont [1]{#1}%
\providecommand \href@noop [0]{\@secondoftwo}%
\providecommand \href [0]{\begingroup \@sanitize@url \@href}%
\providecommand \@href[1]{\@@startlink{#1}\@@href}%
\providecommand \@@href[1]{\endgroup#1\@@endlink}%
\providecommand \@sanitize@url [0]{\catcode `\\12\catcode `\$12\catcode
  `\&12\catcode `\#12\catcode `\^12\catcode `\_12\catcode `\%12\relax}%
\providecommand \@@startlink[1]{}%
\providecommand \@@endlink[0]{}%
\providecommand \url  [0]{\begingroup\@sanitize@url \@url }%
\providecommand \@url [1]{\endgroup\@href {#1}{\urlprefix }}%
\providecommand \urlprefix  [0]{URL }%
\providecommand \Eprint [0]{\href }%
\providecommand \doibase [0]{http://dx.doi.org/}%
\providecommand \selectlanguage [0]{\@gobble}%
\providecommand \bibinfo  [0]{\@secondoftwo}%
\providecommand \bibfield  [0]{\@secondoftwo}%
\providecommand \translation [1]{[#1]}%
\providecommand \BibitemOpen [0]{}%
\providecommand \bibitemStop [0]{}%
\providecommand \bibitemNoStop [0]{.\EOS\space}%
\providecommand \EOS [0]{\spacefactor3000\relax}%
\providecommand \BibitemShut  [1]{\csname bibitem#1\endcsname}%
\let\auto@bib@innerbib\@empty
\bibitem [{\citenamefont {Uhlenbeck}\ and\ \citenamefont
  {Goudsmit}(1925)}]{Uhlenbeck:Goudsmit:1925:Ersetzung_Hypothese}%
  \BibitemOpen
  \bibfield  {author} {\bibinfo {author} {\bibfnamefont {G.~E.}\ \bibnamefont
  {Uhlenbeck}}\ and\ \bibinfo {author} {\bibfnamefont {S.}~\bibnamefont
  {Goudsmit}},\ }\bibfield  {title} {\enquote {\bibinfo {title} {{Ersetzung der
  Hypothese vom unmechanischen Zwang durch eine Forderung bez\"uglich des
  inneren Verhaltens jedes einzelnen Elektrons}},}\ }\href {\doibase
  10.1007/bf01558878} {\bibfield  {journal} {\bibinfo  {journal} {Die
  Naturwissenschaften}\ }\textbf {\bibinfo {volume} {13}},\ \bibinfo {pages}
  {953} (\bibinfo {year} {1925})}\BibitemShut {NoStop}%
\bibitem [{\citenamefont {Uhlenbeck}\ and\ \citenamefont
  {Goudsmit}(1926)}]{Uhlenbeck:Nature:1926}%
  \BibitemOpen
  \bibfield  {author} {\bibinfo {author} {\bibfnamefont {G.~E.}\ \bibnamefont
  {Uhlenbeck}}\ and\ \bibinfo {author} {\bibfnamefont {S.}~\bibnamefont
  {Goudsmit}},\ }\bibfield  {title} {\enquote {\bibinfo {title} {Spinning
  electrons and the structure of spectra},}\ }\href {\doibase 10.1038/117264a0}
  {\bibfield  {journal} {\bibinfo  {journal} {Nature (London)}\ }\textbf
  {\bibinfo {volume} {117}},\ \bibinfo {pages} {264} (\bibinfo {year}
  {1926})}\BibitemShut {NoStop}%
\bibitem [{\citenamefont {Gerlach}\ and\ \citenamefont
  {Stern}(1922)}]{Gerlach:Stern:1922:experimentelle_Nachweis}%
  \BibitemOpen
  \bibfield  {author} {\bibinfo {author} {\bibfnamefont {W.}~\bibnamefont
  {Gerlach}}\ and\ \bibinfo {author} {\bibfnamefont {O.}~\bibnamefont
  {Stern}},\ }\bibfield  {title} {\enquote {\bibinfo {title} {{Der
  experimentelle Nachweis der Richtungsquantelung im Magnetfeld}},}\ }\href
  {\doibase 10.1007/bf01326983} {\bibfield  {journal} {\bibinfo  {journal} {Z.
  Phys.}\ }\textbf {\bibinfo {volume} {9}},\ \bibinfo {pages} {349} (\bibinfo
  {year} {1922})}\BibitemShut {NoStop}%
\bibitem [{\citenamefont {Heinemann}(1996)}]{Heinemann:1996:Stern-Gerlach}%
  \BibitemOpen
  \bibfield  {author} {\bibinfo {author} {\bibfnamefont {K.}~\bibnamefont
  {Heinemann}},\ }\href@noop {} {\enquote {\bibinfo {title} {On {Stern-Gerlach}
  forces allowed by special relativity and the special case of the classical
  spinning particle of {Derbenev-Kondratenko}},}\ } (\bibinfo {year} {1996}),\
  \Eprint {http://arxiv.org/abs/physics/9611001} {arXiv:physics/9611001}
  \BibitemShut {NoStop}%
\bibitem [{\citenamefont {Garraway}\ and\ \citenamefont
  {Stenholm}(1999)}]{Garraway:Stenholm:1999:Observing_spin}%
  \BibitemOpen
  \bibfield  {author} {\bibinfo {author} {\bibfnamefont {B.~M.}\ \bibnamefont
  {Garraway}}\ and\ \bibinfo {author} {\bibfnamefont {S.}~\bibnamefont
  {Stenholm}},\ }\bibfield  {title} {\enquote {\bibinfo {title} {Observing the
  spin of a free electron},}\ }\href {\doibase 10.1103/physreva.60.63}
  {\bibfield  {journal} {\bibinfo  {journal} {Phys. Rev. A}\ }\textbf {\bibinfo
  {volume} {60}},\ \bibinfo {pages} {63} (\bibinfo {year} {1999})}\BibitemShut
  {NoStop}%
\bibitem [{\citenamefont {Batelaan}\ \emph {et~al.}(1997)\citenamefont
  {Batelaan}, \citenamefont {Gay},\ and\ \citenamefont
  {Schwendiman}}]{Batelaan:Gay:etal:1997:Stern-Gerlach_Effect}%
  \BibitemOpen
  \bibfield  {author} {\bibinfo {author} {\bibfnamefont {H.}~\bibnamefont
  {Batelaan}}, \bibinfo {author} {\bibfnamefont {T.~J.}\ \bibnamefont {Gay}}, \
  and\ \bibinfo {author} {\bibfnamefont {J.~J.}\ \bibnamefont {Schwendiman}},\
  }\bibfield  {title} {\enquote {\bibinfo {title} {{Stern-Gerlach} effect for
  electron beams},}\ }\href {\doibase 10.1103/physrevlett.79.4517} {\bibfield
  {journal} {\bibinfo  {journal} {Phys. Rev. Lett.}\ }\textbf {\bibinfo
  {volume} {79}},\ \bibinfo {pages} {4517} (\bibinfo {year}
  {1997})}\BibitemShut {NoStop}%
\bibitem [{\citenamefont {Gallup}\ \emph {et~al.}(2001)\citenamefont {Gallup},
  \citenamefont {Batelaan},\ and\ \citenamefont
  {Gay}}]{Gallup:Batelaan:etal:2001:Quantum-Mechanical_Analysis}%
  \BibitemOpen
  \bibfield  {author} {\bibinfo {author} {\bibfnamefont {G.~A.}\ \bibnamefont
  {Gallup}}, \bibinfo {author} {\bibfnamefont {H.}~\bibnamefont {Batelaan}}, \
  and\ \bibinfo {author} {\bibfnamefont {T.~J.}\ \bibnamefont {Gay}},\
  }\bibfield  {title} {\enquote {\bibinfo {title} {Quantum-mechanical analysis
  of a longitudinal {Stern-Gerlach} effect},}\ }\href {\doibase
  10.1103/physrevlett.86.4508} {\bibfield  {journal} {\bibinfo  {journal}
  {Phys. Rev. Lett.}\ }\textbf {\bibinfo {volume} {86}},\ \bibinfo {pages}
  {4508} (\bibinfo {year} {2001})}\BibitemShut {NoStop}%
\bibitem [{\citenamefont {McGregor}\ \emph {et~al.}(2011)\citenamefont
  {McGregor}, \citenamefont {Bach},\ and\ \citenamefont
  {Batelaan}}]{McGregor:Bach:etal:2011:Transverse_quantum}%
  \BibitemOpen
  \bibfield  {author} {\bibinfo {author} {\bibfnamefont {S.}~\bibnamefont
  {McGregor}}, \bibinfo {author} {\bibfnamefont {R.}~\bibnamefont {Bach}}, \
  and\ \bibinfo {author} {\bibfnamefont {H.}~\bibnamefont {Batelaan}},\
  }\bibfield  {title} {\enquote {\bibinfo {title} {Transverse quantum
  {Stern-Gerlach} magnets for electrons},}\ }\href {\doibase
  10.1088/1367-2630/13/6/065018} {\bibfield  {journal} {\bibinfo  {journal}
  {New J. Phys.}\ }\textbf {\bibinfo {volume} {13}},\ \bibinfo {pages} {065018}
  (\bibinfo {year} {2011})}\BibitemShut {NoStop}%
\bibitem [{\citenamefont {Mahajan}\ \emph {et~al.}(2014)\citenamefont
  {Mahajan}, \citenamefont {Asenjo},\ and\ \citenamefont
  {Hazeltine}}]{Mahajan:Asenjo:etal:2014:Comparison_of}%
  \BibitemOpen
  \bibfield  {author} {\bibinfo {author} {\bibfnamefont {S.~M.}\ \bibnamefont
  {Mahajan}}, \bibinfo {author} {\bibfnamefont {F.~A.}\ \bibnamefont {Asenjo}},
  \ and\ \bibinfo {author} {\bibfnamefont {R.~D.}\ \bibnamefont {Hazeltine}},\
  }\bibfield  {title} {\enquote {\bibinfo {title} {Comparison of the
  electron-spin force and radiation reaction force},}\ }\href {\doibase
  10.1093/mnras/stu2381} {\bibfield  {journal} {\bibinfo  {journal} {Mon. Not.
  R. Astron. Soc.}\ }\textbf {\bibinfo {volume} {446}},\ \bibinfo {pages}
  {4112} (\bibinfo {year} {2014})}\BibitemShut {NoStop}%
\bibitem [{\citenamefont {Marklund}\ and\ \citenamefont
  {Brodin}(2007)}]{Marklund:Brodin:2007:Dynamics_of}%
  \BibitemOpen
  \bibfield  {author} {\bibinfo {author} {\bibfnamefont {M.}~\bibnamefont
  {Marklund}}\ and\ \bibinfo {author} {\bibfnamefont {G.}~\bibnamefont
  {Brodin}},\ }\bibfield  {title} {\enquote {\bibinfo {title} {Dynamics of
  spin-$\frac{1}{2}$ quantum plasmas},}\ }\href {\doibase
  10.1103/physrevlett.98.025001} {\bibfield  {journal} {\bibinfo  {journal}
  {Phys. Rev. Lett.}\ }\textbf {\bibinfo {volume} {98}},\ \bibinfo {pages}
  {025001} (\bibinfo {year} {2007})}\BibitemShut {NoStop}%
\bibitem [{\citenamefont {Brodin}\ and\ \citenamefont
  {Marklund}(2007)}]{Brodin:Marklund:2007:Spin_magnetohydrodynamics}%
  \BibitemOpen
  \bibfield  {author} {\bibinfo {author} {\bibfnamefont {G.}~\bibnamefont
  {Brodin}}\ and\ \bibinfo {author} {\bibfnamefont {M.}~\bibnamefont
  {Marklund}},\ }\bibfield  {title} {\enquote {\bibinfo {title} {Spin
  magnetohydrodynamics},}\ }\href {\doibase 10.1088/1367-2630/9/8/277}
  {\bibfield  {journal} {\bibinfo  {journal} {New J. Phys.}\ }\textbf {\bibinfo
  {volume} {9}},\ \bibinfo {pages} {277} (\bibinfo {year} {2007})}\BibitemShut
  {NoStop}%
\bibitem [{\citenamefont {Brodin}\ \emph {et~al.}(2011)\citenamefont {Brodin},
  \citenamefont {Marklund}, \citenamefont {Zamanian},\ and\ \citenamefont
  {Stefan}}]{Brodin:Marklund:etal:2011:Spin_and}%
  \BibitemOpen
  \bibfield  {author} {\bibinfo {author} {\bibfnamefont {G.}~\bibnamefont
  {Brodin}}, \bibinfo {author} {\bibfnamefont {M.}~\bibnamefont {Marklund}},
  \bibinfo {author} {\bibfnamefont {J.}~\bibnamefont {Zamanian}}, \ and\
  \bibinfo {author} {\bibfnamefont {M.}~\bibnamefont {Stefan}},\ }\bibfield
  {title} {\enquote {\bibinfo {title} {Spin and magnetization effects in
  plasmas},}\ }\href {\doibase 10.1088/0741-3335/53/7/074013} {\bibfield
  {journal} {\bibinfo  {journal} {Plasma Phys. and Controlled Fusion}\ }\textbf
  {\bibinfo {volume} {53}},\ \bibinfo {pages} {074013} (\bibinfo {year}
  {2011})}\BibitemShut {NoStop}%
\bibitem [{\citenamefont {Rathe}\ \emph {et~al.}(1997)\citenamefont {Rathe},
  \citenamefont {Keitel}, \citenamefont {Protopapas},\ and\ \citenamefont
  {Knight}}]{Rathe:Keitel:etal:1997:Intense_laser}%
  \BibitemOpen
  \bibfield  {author} {\bibinfo {author} {\bibfnamefont {U.~W.}\ \bibnamefont
  {Rathe}}, \bibinfo {author} {\bibfnamefont {C.~H.}\ \bibnamefont {Keitel}},
  \bibinfo {author} {\bibfnamefont {M.}~\bibnamefont {Protopapas}}, \ and\
  \bibinfo {author} {\bibfnamefont {P.~L.}\ \bibnamefont {Knight}},\ }\bibfield
   {title} {\enquote {\bibinfo {title} {Intense laser-atom dynamics with the
  two-dimensional {Dirac} equation},}\ }\href {\doibase
  10.1088/0953-4075/30/15/004} {\bibfield  {journal} {\bibinfo  {journal} {J.
  Phys. B: At., Mol. Opt. Phys.}\ }\textbf {\bibinfo {volume} {30}},\ \bibinfo
  {pages} {L531} (\bibinfo {year} {1997})}\BibitemShut {NoStop}%
\bibitem [{\citenamefont {Hu}\ and\ \citenamefont
  {Keitel}(1999)}]{Hu:Keitel:1999:Spin_Signatures}%
  \BibitemOpen
  \bibfield  {author} {\bibinfo {author} {\bibfnamefont {S.~X.}\ \bibnamefont
  {Hu}}\ and\ \bibinfo {author} {\bibfnamefont {C.~H.}\ \bibnamefont
  {Keitel}},\ }\bibfield  {title} {\enquote {\bibinfo {title} {Spin signatures
  in intense laser-ion interaction},}\ }\href {\doibase
  10.1103/physrevlett.83.4709} {\bibfield  {journal} {\bibinfo  {journal}
  {Phys. Rev. Lett.}\ }\textbf {\bibinfo {volume} {83}},\ \bibinfo {pages}
  {4709} (\bibinfo {year} {1999})}\BibitemShut {NoStop}%
\bibitem [{\citenamefont {Walser}\ and\ \citenamefont
  {Keitel}(2000)}]{Walser:Keitel:2000:Spin-induced_force}%
  \BibitemOpen
  \bibfield  {author} {\bibinfo {author} {\bibfnamefont {M.~W.}\ \bibnamefont
  {Walser}}\ and\ \bibinfo {author} {\bibfnamefont {C.~H.}\ \bibnamefont
  {Keitel}},\ }\bibfield  {title} {\enquote {\bibinfo {title} {Spin-induced
  force in intense laser-electron interaction},}\ }\href {\doibase
  10.1088/0953-4075/33/6/105} {\bibfield  {journal} {\bibinfo  {journal} {J.
  Phys. B: At., Mol. Opt. Phys.}\ }\textbf {\bibinfo {volume} {33}},\ \bibinfo
  {pages} {L221} (\bibinfo {year} {2000})}\BibitemShut {NoStop}%
\bibitem [{\citenamefont {Walser}\ \emph {et~al.}(2002)\citenamefont {Walser},
  \citenamefont {Urbach}, \citenamefont {Hatsagortsyan}, \citenamefont {Hu},\
  and\ \citenamefont {Keitel}}]{Walser:Urbach:etal:2002:Spin_and}%
  \BibitemOpen
  \bibfield  {author} {\bibinfo {author} {\bibfnamefont {M.~W.}\ \bibnamefont
  {Walser}}, \bibinfo {author} {\bibfnamefont {D.~J.}\ \bibnamefont {Urbach}},
  \bibinfo {author} {\bibfnamefont {K.~Z.}\ \bibnamefont {Hatsagortsyan}},
  \bibinfo {author} {\bibfnamefont {S.~X.}\ \bibnamefont {Hu}}, \ and\ \bibinfo
  {author} {\bibfnamefont {C.~H.}\ \bibnamefont {Keitel}},\ }\bibfield  {title}
  {\enquote {\bibinfo {title} {Spin and radiation in intense laser fields},}\
  }\href {\doibase 10.1103/physreva.65.043410} {\bibfield  {journal} {\bibinfo
  {journal} {Phys. Rev. A}\ }\textbf {\bibinfo {volume} {65}},\ \bibinfo
  {pages} {043410} (\bibinfo {year} {2002})}\BibitemShut {NoStop}%
\bibitem [{\citenamefont {Roman}\ \emph {et~al.}(2003)\citenamefont {Roman},
  \citenamefont {Roso},\ and\ \citenamefont
  {Plaja}}]{Roman:Roso:etal:2003:complete_description}%
  \BibitemOpen
  \bibfield  {author} {\bibinfo {author} {\bibfnamefont {J.~S.}\ \bibnamefont
  {Roman}}, \bibinfo {author} {\bibfnamefont {L.}~\bibnamefont {Roso}}, \ and\
  \bibinfo {author} {\bibfnamefont {L.}~\bibnamefont {Plaja}},\ }\bibfield
  {title} {\enquote {\bibinfo {title} {A complete description of the spin
  force},}\ }\href {\doibase 10.1088/0953-4075/37/2/011} {\bibfield  {journal}
  {\bibinfo  {journal} {J. Phys. B: At., Mol. Opt. Phys.}\ }\textbf {\bibinfo
  {volume} {37}},\ \bibinfo {pages} {435} (\bibinfo {year} {2003})}\BibitemShut
  {NoStop}%
\bibitem [{\citenamefont {Zhidkov}\ \emph {et~al.}(2002)\citenamefont
  {Zhidkov}, \citenamefont {Koga}, \citenamefont {Sasaki},\ and\ \citenamefont
  {Uesaka}}]{Zhidkov:Koga:etal:2002:Radiation_Damping}%
  \BibitemOpen
  \bibfield  {author} {\bibinfo {author} {\bibfnamefont {A.}~\bibnamefont
  {Zhidkov}}, \bibinfo {author} {\bibfnamefont {J.}~\bibnamefont {Koga}},
  \bibinfo {author} {\bibfnamefont {A.}~\bibnamefont {Sasaki}}, \ and\ \bibinfo
  {author} {\bibfnamefont {M.}~\bibnamefont {Uesaka}},\ }\bibfield  {title}
  {\enquote {\bibinfo {title} {Radiation damping effects on the interaction of
  ultraintense laser pulses with an overdense plasma},}\ }\href {\doibase
  10.1103/physrevlett.88.185002} {\bibfield  {journal} {\bibinfo  {journal}
  {Phys. Rev. Lett.}\ }\textbf {\bibinfo {volume} {88}},\ \bibinfo {pages}
  {185002} (\bibinfo {year} {2002})}\BibitemShut {NoStop}%
\bibitem [{\citenamefont {Di~Piazza}\ \emph {et~al.}(2009)\citenamefont
  {Di~Piazza}, \citenamefont {Hatsagortsyan},\ and\ \citenamefont
  {Keitel}}]{Di-Piazza:Hatsagortsyan:etal:2009:Strong_Signatures}%
  \BibitemOpen
  \bibfield  {author} {\bibinfo {author} {\bibfnamefont {A.}~\bibnamefont
  {Di~Piazza}}, \bibinfo {author} {\bibfnamefont {K.~Z.}\ \bibnamefont
  {Hatsagortsyan}}, \ and\ \bibinfo {author} {\bibfnamefont {C.~H.}\
  \bibnamefont {Keitel}},\ }\bibfield  {title} {\enquote {\bibinfo {title}
  {Strong signatures of radiation reaction below the radiation-dominated
  regime},}\ }\href {\doibase 10.1103/physrevlett.102.254802} {\bibfield
  {journal} {\bibinfo  {journal} {Phys. Rev. Lett.}\ }\textbf {\bibinfo
  {volume} {102}},\ \bibinfo {pages} {254802} (\bibinfo {year}
  {2009})}\BibitemShut {NoStop}%
\bibitem [{\citenamefont {Tamburini}\ \emph {et~al.}(2010)\citenamefont
  {Tamburini}, \citenamefont {Pegoraro}, \citenamefont {Di~Piazza},
  \citenamefont {Keitel},\ and\ \citenamefont
  {Macchi}}]{Tamburini:Pegoraro:etal:2010:Radiation_reaction}%
  \BibitemOpen
  \bibfield  {author} {\bibinfo {author} {\bibfnamefont {M.}~\bibnamefont
  {Tamburini}}, \bibinfo {author} {\bibfnamefont {F.}~\bibnamefont {Pegoraro}},
  \bibinfo {author} {\bibfnamefont {A.}~\bibnamefont {Di~Piazza}}, \bibinfo
  {author} {\bibfnamefont {C.~H.}\ \bibnamefont {Keitel}}, \ and\ \bibinfo
  {author} {\bibfnamefont {A.}~\bibnamefont {Macchi}},\ }\bibfield  {title}
  {\enquote {\bibinfo {title} {Radiation reaction effects on radiation pressure
  acceleration},}\ }\href {\doibase 10.1088/1367-2630/12/12/123005} {\bibfield
  {journal} {\bibinfo  {journal} {New J. Phys.}\ }\textbf {\bibinfo {volume}
  {12}},\ \bibinfo {pages} {123005} (\bibinfo {year} {2010})}\BibitemShut
  {NoStop}%
\bibitem [{\citenamefont {Ji}\ \emph {et~al.}(2014)\citenamefont {Ji},
  \citenamefont {Pukhov}, \citenamefont {Kostyukov}, \citenamefont {Shen},\
  and\ \citenamefont {Akli}}]{Ji:Pukhov:etal:2014:Radiation-Reaction_Trapping}%
  \BibitemOpen
  \bibfield  {author} {\bibinfo {author} {\bibfnamefont {L.~L.}\ \bibnamefont
  {Ji}}, \bibinfo {author} {\bibfnamefont {A.}~\bibnamefont {Pukhov}}, \bibinfo
  {author} {\bibfnamefont {I.~Y.}\ \bibnamefont {Kostyukov}}, \bibinfo {author}
  {\bibfnamefont {B.~F.}\ \bibnamefont {Shen}}, \ and\ \bibinfo {author}
  {\bibfnamefont {K.}~\bibnamefont {Akli}},\ }\bibfield  {title} {\enquote
  {\bibinfo {title} {Radiation-reaction trapping of electrons in extreme laser
  fields},}\ }\href {\doibase 10.1103/physrevlett.112.145003} {\bibfield
  {journal} {\bibinfo  {journal} {Phys. Rev. Lett.}\ }\textbf {\bibinfo
  {volume} {112}},\ \bibinfo {pages} {145003} (\bibinfo {year}
  {2014})}\BibitemShut {NoStop}%
\bibitem [{\citenamefont {Di~Piazza}\ \emph {et~al.}(2010)\citenamefont
  {Di~Piazza}, \citenamefont {Hatsagortsyan},\ and\ \citenamefont
  {Keitel}}]{Di-Piazza:Hatsagortsyan:etal:2010:Quantum_Radiation}%
  \BibitemOpen
  \bibfield  {author} {\bibinfo {author} {\bibfnamefont {A.}~\bibnamefont
  {Di~Piazza}}, \bibinfo {author} {\bibfnamefont {K.~Z.}\ \bibnamefont
  {Hatsagortsyan}}, \ and\ \bibinfo {author} {\bibfnamefont {C.~H.}\
  \bibnamefont {Keitel}},\ }\bibfield  {title} {\enquote {\bibinfo {title}
  {Quantum radiation reaction effects in multiphoton compton scattering},}\
  }\href {\doibase 10.1103/physrevlett.105.220403} {\bibfield  {journal}
  {\bibinfo  {journal} {Phys. Rev. Lett.}\ }\textbf {\bibinfo {volume} {105}},\
  \bibinfo {pages} {220403} (\bibinfo {year} {2010})}\BibitemShut {NoStop}%
\bibitem [{\citenamefont {Ilderton}\ and\ \citenamefont
  {Torgrimsson}(2013)}]{Ilderton:Torgrimsson:2013:Radiation_reaction}%
  \BibitemOpen
  \bibfield  {author} {\bibinfo {author} {\bibfnamefont {A.}~\bibnamefont
  {Ilderton}}\ and\ \bibinfo {author} {\bibfnamefont {G.}~\bibnamefont
  {Torgrimsson}},\ }\bibfield  {title} {\enquote {\bibinfo {title} {Radiation
  reaction in strong field {QED}},}\ }\href {\doibase
  10.1016/j.physletb.2013.07.045} {\bibfield  {journal} {\bibinfo  {journal}
  {Phys. Lett. B}\ }\textbf {\bibinfo {volume} {725}},\ \bibinfo {pages} {481}
  (\bibinfo {year} {2013})}\BibitemShut {NoStop}%
\bibitem [{\citenamefont {Neitz}\ and\ \citenamefont
  {Di~Piazza}(2013)}]{Neitz:Di-Piazza:2013:Stochasticity_Effects}%
  \BibitemOpen
  \bibfield  {author} {\bibinfo {author} {\bibfnamefont {N.}~\bibnamefont
  {Neitz}}\ and\ \bibinfo {author} {\bibfnamefont {A.}~\bibnamefont
  {Di~Piazza}},\ }\bibfield  {title} {\enquote {\bibinfo {title} {Stochasticity
  effects in quantum radiation reaction},}\ }\href {\doibase
  10.1103/physrevlett.111.054802} {\bibfield  {journal} {\bibinfo  {journal}
  {Phys. Rev. Lett.}\ }\textbf {\bibinfo {volume} {111}},\ \bibinfo {pages}
  {054802} (\bibinfo {year} {2013})}\BibitemShut {NoStop}%
\bibitem [{\citenamefont {Blackburn}\ \emph {et~al.}(2014)\citenamefont
  {Blackburn}, \citenamefont {Ridgers}, \citenamefont {Kirk},\ and\
  \citenamefont {Bell}}]{Blackburn:Ridgers:etal:2014:Quantum_Radiation}%
  \BibitemOpen
  \bibfield  {author} {\bibinfo {author} {\bibfnamefont {T.~G.}\ \bibnamefont
  {Blackburn}}, \bibinfo {author} {\bibfnamefont {C.~P.}\ \bibnamefont
  {Ridgers}}, \bibinfo {author} {\bibfnamefont {J.~G.}\ \bibnamefont {Kirk}}, \
  and\ \bibinfo {author} {\bibfnamefont {A.~R.}\ \bibnamefont {Bell}},\
  }\bibfield  {title} {\enquote {\bibinfo {title} {Quantum radiation reaction
  in laser-electron-beam collisions},}\ }\href {\doibase
  10.1103/physrevlett.112.015001} {\bibfield  {journal} {\bibinfo  {journal}
  {Phys. Rev. Lett.}\ }\textbf {\bibinfo {volume} {112}},\ \bibinfo {pages}
  {015001} (\bibinfo {year} {2014})}\BibitemShut {NoStop}%
\bibitem [{\citenamefont {Vranic}\ \emph {et~al.}(2014)\citenamefont {Vranic},
  \citenamefont {Martins}, \citenamefont {Vieira}, \citenamefont {Fonseca},\
  and\ \citenamefont {Silva}}]{Vranic:Martins:etal:2014:All-Optical_Radiation}%
  \BibitemOpen
  \bibfield  {author} {\bibinfo {author} {\bibfnamefont {M.}~\bibnamefont
  {Vranic}}, \bibinfo {author} {\bibfnamefont {J.~L.}\ \bibnamefont {Martins}},
  \bibinfo {author} {\bibfnamefont {J.}~\bibnamefont {Vieira}}, \bibinfo
  {author} {\bibfnamefont {R.~A.}\ \bibnamefont {Fonseca}}, \ and\ \bibinfo
  {author} {\bibfnamefont {L.~O.}\ \bibnamefont {Silva}},\ }\bibfield  {title}
  {\enquote {\bibinfo {title} {All-optical radiation reaction at
  {$10^{21}\mathrm{\,W/cm^2}$}},}\ }\href {\doibase
  10.1103/physrevlett.113.134801} {\bibfield  {journal} {\bibinfo  {journal}
  {Phys. Rev. Lett.}\ }\textbf {\bibinfo {volume} {113}},\ \bibinfo {pages}
  {134801} (\bibinfo {year} {2014})}\BibitemShut {NoStop}%
\bibitem [{\citenamefont {Li}\ \emph {et~al.}(2014)\citenamefont {Li},
  \citenamefont {Hatsagortsyan},\ and\ \citenamefont
  {Keitel}}]{Li:Hatsagortsyan:etal:2014:Robust_Signatures}%
  \BibitemOpen
  \bibfield  {author} {\bibinfo {author} {\bibfnamefont {J.-X.}\ \bibnamefont
  {Li}}, \bibinfo {author} {\bibfnamefont {K.~Z.}\ \bibnamefont
  {Hatsagortsyan}}, \ and\ \bibinfo {author} {\bibfnamefont {C.~H.}\
  \bibnamefont {Keitel}},\ }\bibfield  {title} {\enquote {\bibinfo {title}
  {Robust signatures of quantum radiation reaction in focused ultrashort laser
  pulses},}\ }\href {\doibase 10.1103/physrevlett.113.044801} {\bibfield
  {journal} {\bibinfo  {journal} {Phys. Rev. Lett.}\ }\textbf {\bibinfo
  {volume} {113}},\ \bibinfo {pages} {044801} (\bibinfo {year}
  {2014})}\BibitemShut {NoStop}%
\bibitem [{\citenamefont {Gonoskov}\ \emph {et~al.}(2014)\citenamefont
  {Gonoskov}, \citenamefont {Bashinov}, \citenamefont {Gonoskov}, \citenamefont
  {Harvey}, \citenamefont {Ilderton}, \citenamefont {Kim}, \citenamefont
  {Marklund}, \citenamefont {Mourou},\ and\ \citenamefont
  {Sergeev}}]{Gonoskov:Bashinov:etal:2014:Anomalous_Radiative}%
  \BibitemOpen
  \bibfield  {author} {\bibinfo {author} {\bibfnamefont {A.}~\bibnamefont
  {Gonoskov}}, \bibinfo {author} {\bibfnamefont {A.}~\bibnamefont {Bashinov}},
  \bibinfo {author} {\bibfnamefont {I.}~\bibnamefont {Gonoskov}}, \bibinfo
  {author} {\bibfnamefont {C.}~\bibnamefont {Harvey}}, \bibinfo {author}
  {\bibfnamefont {A.}~\bibnamefont {Ilderton}}, \bibinfo {author}
  {\bibfnamefont {A.}~\bibnamefont {Kim}}, \bibinfo {author} {\bibfnamefont
  {M.}~\bibnamefont {Marklund}}, \bibinfo {author} {\bibfnamefont
  {G.}~\bibnamefont {Mourou}}, \ and\ \bibinfo {author} {\bibfnamefont
  {A.}~\bibnamefont {Sergeev}},\ }\bibfield  {title} {\enquote {\bibinfo
  {title} {Anomalous radiative trapping in laser fields of extreme
  intensity},}\ }\href {\doibase 10.1103/physrevlett.113.014801} {\bibfield
  {journal} {\bibinfo  {journal} {Phys. Rev. Lett.}\ }\textbf {\bibinfo
  {volume} {113}},\ \bibinfo {pages} {014801} (\bibinfo {year}
  {2014})}\BibitemShut {NoStop}%
\bibitem [{\citenamefont {Vranic}\ \emph {et~al.}(2016)\citenamefont {Vranic},
  \citenamefont {Grismayer}, \citenamefont {Fonseca},\ and\ \citenamefont
  {Silva}}]{Vranic:Grismayer:etal:2016:Quantum_radiation}%
  \BibitemOpen
  \bibfield  {author} {\bibinfo {author} {\bibfnamefont {M.}~\bibnamefont
  {Vranic}}, \bibinfo {author} {\bibfnamefont {T.}~\bibnamefont {Grismayer}},
  \bibinfo {author} {\bibfnamefont {R.~A.}\ \bibnamefont {Fonseca}}, \ and\
  \bibinfo {author} {\bibfnamefont {L.~O.}\ \bibnamefont {Silva}},\ }\bibfield
  {title} {\enquote {\bibinfo {title} {Quantum radiation reaction in head-on
  laser-electron beam interaction},}\ }\href {\doibase
  10.1088/1367-2630/18/7/073035} {\bibfield  {journal} {\bibinfo  {journal}
  {New J. Phys.}\ }\textbf {\bibinfo {volume} {18}},\ \bibinfo {pages} {073035}
  (\bibinfo {year} {2016})}\BibitemShut {NoStop}%
\bibitem [{\citenamefont {Dirac}(1928)}]{Dirac:1928:Quantum_Theory}%
  \BibitemOpen
  \bibfield  {author} {\bibinfo {author} {\bibfnamefont {P.~A.~M.}\
  \bibnamefont {Dirac}},\ }\bibfield  {title} {\enquote {\bibinfo {title} {The
  quantum theory of the electron},}\ }\href {\doibase 10.1098/rspa.1928.0023}
  {\bibfield  {journal} {\bibinfo  {journal} {Proc. R. Soc. London, Ser. A}\
  }\textbf {\bibinfo {volume} {117}},\ \bibinfo {pages} {610} (\bibinfo {year}
  {1928})}\BibitemShut {NoStop}%
\bibitem [{\citenamefont {Pfund}\ \emph {et~al.}(1998)\citenamefont {Pfund},
  \citenamefont {Lichters},\ and\ \citenamefont
  {{Meyer-ter-Vehn}}}]{Pfund:AIPCP:1998}%
  \BibitemOpen
  \bibfield  {author} {\bibinfo {author} {\bibfnamefont {R.~E.~W.}\
  \bibnamefont {Pfund}}, \bibinfo {author} {\bibfnamefont {R.}~\bibnamefont
  {Lichters}}, \ and\ \bibinfo {author} {\bibfnamefont {J.}~\bibnamefont
  {{Meyer-ter-Vehn}}},\ }\bibfield  {title} {\enquote {\bibinfo {title}
  {{LPIC++} a parallel one-dimensional relativistic electromagnetic
  particle-in-cell code for simulating laser-plasma-interaction},}\ }\href
  {\doibase 10.1063/1.55199} {\bibfield  {journal} {\bibinfo  {journal} {AIP
  Conf. Proc.}\ }\textbf {\bibinfo {volume} {426}},\ \bibinfo {pages} {141}
  (\bibinfo {year} {1998})}\BibitemShut {NoStop}%
\bibitem [{\citenamefont {Brodin}\ \emph {et~al.}(2013)\citenamefont {Brodin},
  \citenamefont {Holkundkar},\ and\ \citenamefont
  {Marklund}}]{BRODIN:JPP:2013}%
  \BibitemOpen
  \bibfield  {author} {\bibinfo {author} {\bibfnamefont {G.}~\bibnamefont
  {Brodin}}, \bibinfo {author} {\bibfnamefont {A.}~\bibnamefont {Holkundkar}},
  \ and\ \bibinfo {author} {\bibfnamefont {M.}~\bibnamefont {Marklund}},\
  }\bibfield  {title} {\enquote {\bibinfo {title} {Particle-in-cell simulations
  of electron spin effects in plasmas},}\ }\href {\doibase
  10.1017/s0022377813000093} {\bibfield  {journal} {\bibinfo  {journal} {J.
  Plasma Phys.}\ }\textbf {\bibinfo {volume} {79}},\ \bibinfo {pages} {377}
  (\bibinfo {year} {2013})}\BibitemShut {NoStop}%
\bibitem [{\citenamefont {Thomas}(1926)}]{Thomas:1926:Motion_of}%
  \BibitemOpen
  \bibfield  {author} {\bibinfo {author} {\bibfnamefont {L.~H.}\ \bibnamefont
  {Thomas}},\ }\bibfield  {title} {\enquote {\bibinfo {title} {The motion of
  the spinning electron},}\ }\href {\doibase 10.1038/117514a0} {\bibfield
  {journal} {\bibinfo  {journal} {Nature (London)}\ }\textbf {\bibinfo {volume}
  {117}},\ \bibinfo {pages} {514} (\bibinfo {year} {1926})}\BibitemShut
  {NoStop}%
\bibitem [{\citenamefont {Thomas}(1927)}]{Thomas:1927:I_kinematics}%
  \BibitemOpen
  \bibfield  {author} {\bibinfo {author} {\bibfnamefont {L.~H.}\ \bibnamefont
  {Thomas}},\ }\bibfield  {title} {\enquote {\bibinfo {title} {I. {The}
  kinematics of an electron with an axis},}\ }\href {\doibase
  10.1080/14786440108564170} {\bibfield  {journal} {\bibinfo  {journal} {The
  London, Edinburgh, and Dublin Philosophical Magazine and Journal of Science}\
  }\textbf {\bibinfo {volume} {3}},\ \bibinfo {pages} {1} (\bibinfo {year}
  {1927})}\BibitemShut {NoStop}%
\bibitem [{\citenamefont {Bargmann}\ \emph {et~al.}(1959)\citenamefont
  {Bargmann}, \citenamefont {Michel},\ and\ \citenamefont
  {Telegdi}}]{Bargmann:Michel:etal:1959:Precession_of}%
  \BibitemOpen
  \bibfield  {author} {\bibinfo {author} {\bibfnamefont {V.}~\bibnamefont
  {Bargmann}}, \bibinfo {author} {\bibfnamefont {L.}~\bibnamefont {Michel}}, \
  and\ \bibinfo {author} {\bibfnamefont {V.~L.}\ \bibnamefont {Telegdi}},\
  }\bibfield  {title} {\enquote {\bibinfo {title} {Precession of the
  polarization of particles moving in a homogeneous electromagnetic field},}\
  }\href {\doibase 10.1103/physrevlett.2.435} {\bibfield  {journal} {\bibinfo
  {journal} {Phys. Rev. Lett.}\ }\textbf {\bibinfo {volume} {2}},\ \bibinfo
  {pages} {435} (\bibinfo {year} {1959})}\BibitemShut {NoStop}%
\bibitem [{\citenamefont
  {Jackson}(1999)}]{Jackson:1999:Classical_Electrodynamics}%
  \BibitemOpen
  \bibfield  {author} {\bibinfo {author} {\bibfnamefont {J.~D.}\ \bibnamefont
  {Jackson}},\ }\href@noop {} {\emph {\bibinfo {title} {Classical
  Electrodynamics}}}\ (\bibinfo  {publisher} {John Wiley \& Sons},\ \bibinfo
  {address} {New York},\ \bibinfo {year} {1999})\BibitemShut {NoStop}%
\bibitem [{\citenamefont
  {Frenkel}(1926{\natexlab{a}})}]{Frenkel:1926:Spinning_Electrons}%
  \BibitemOpen
  \bibfield  {author} {\bibinfo {author} {\bibfnamefont {J.}~\bibnamefont
  {Frenkel}},\ }\bibfield  {title} {\enquote {\bibinfo {title} {Spinning
  electrons},}\ }\href {\doibase 10.1038/117653a0} {\bibfield  {journal}
  {\bibinfo  {journal} {Nature (London)}\ }\textbf {\bibinfo {volume} {117}},\
  \bibinfo {pages} {653} (\bibinfo {year} {1926}{\natexlab{a}})}\BibitemShut
  {NoStop}%
\bibitem [{\citenamefont
  {Frenkel}(1926{\natexlab{b}})}]{Frenkel:1926:Elektrodynamik_des}%
  \BibitemOpen
  \bibfield  {author} {\bibinfo {author} {\bibfnamefont {J.}~\bibnamefont
  {Frenkel}},\ }\bibfield  {title} {\enquote {\bibinfo {title} {{Die
  Elektrodynamik des rotierenden Elektrons}},}\ }\href {\doibase
  10.1007/bf01397099} {\bibfield  {journal} {\bibinfo  {journal} {Z. Phys.}\
  }\textbf {\bibinfo {volume} {37}},\ \bibinfo {pages} {243} (\bibinfo {year}
  {1926}{\natexlab{b}})}\BibitemShut {NoStop}%
\bibitem [{\citenamefont {Good}(1962)}]{Good:1962:Classical_Equations}%
  \BibitemOpen
  \bibfield  {author} {\bibinfo {author} {\bibfnamefont {R.~H.}\ \bibnamefont
  {Good}},\ }\bibfield  {title} {\enquote {\bibinfo {title} {Classical
  equations of motion for a polarized particle in an electromagnetic field},}\
  }\href {\doibase 10.1103/physrev.125.2112} {\bibfield  {journal} {\bibinfo
  {journal} {Phys. Rev.}\ }\textbf {\bibinfo {volume} {125}},\ \bibinfo {pages}
  {2112} (\bibinfo {year} {1962})}\BibitemShut {NoStop}%
\bibitem [{\citenamefont {Barducci}\ \emph {et~al.}(1976)\citenamefont
  {Barducci}, \citenamefont {Casalbuoni},\ and\ \citenamefont
  {Lusanna}}]{Barducci:Casalbuoni:etal:1976:Supersymmetries_and}%
  \BibitemOpen
  \bibfield  {author} {\bibinfo {author} {\bibfnamefont {A.}~\bibnamefont
  {Barducci}}, \bibinfo {author} {\bibfnamefont {R.}~\bibnamefont
  {Casalbuoni}}, \ and\ \bibinfo {author} {\bibfnamefont {L.}~\bibnamefont
  {Lusanna}},\ }\bibfield  {title} {\enquote {\bibinfo {title} {Supersymmetries
  and the pseudoclassical relativistic electron},}\ }\href {\doibase
  10.1007/bf02730291} {\bibfield  {journal} {\bibinfo  {journal} {Il Nuovo
  Cimento A}\ }\textbf {\bibinfo {volume} {35}},\ \bibinfo {pages} {377}
  (\bibinfo {year} {1976})}\BibitemShut {NoStop}%
\bibitem [{\citenamefont {Barut}(1980)}]{Barut:1980:Electrodynamics}%
  \BibitemOpen
  \bibfield  {author} {\bibinfo {author} {\bibfnamefont {A.~O.}\ \bibnamefont
  {Barut}},\ }\href@noop {} {\emph {\bibinfo {title} {Electrodynamics and
  Classical Theory of Fields and Particles}}}\ (\bibinfo  {publisher} {Dover},\
  \bibinfo {address} {Mineola, NY},\ \bibinfo {year} {1980})\BibitemShut
  {NoStop}%
\bibitem [{\citenamefont {Ravndal}(1980)}]{Ravndal:1980:Supersymmetric_Dirac}%
  \BibitemOpen
  \bibfield  {author} {\bibinfo {author} {\bibfnamefont {F.}~\bibnamefont
  {Ravndal}},\ }\bibfield  {title} {\enquote {\bibinfo {title} {Supersymmetric
  {Dirac} particles in external fields},}\ }\href {\doibase
  10.1103/physrevd.21.2823} {\bibfield  {journal} {\bibinfo  {journal} {Phys.
  Rev. D}\ }\textbf {\bibinfo {volume} {21}},\ \bibinfo {pages} {2823}
  (\bibinfo {year} {1980})}\BibitemShut {NoStop}%
\bibitem [{\citenamefont {van Holten}(1991)}]{Holten:1991:On_electrodynamics}%
  \BibitemOpen
  \bibfield  {author} {\bibinfo {author} {\bibfnamefont {J.}~\bibnamefont {van
  Holten}},\ }\bibfield  {title} {\enquote {\bibinfo {title} {On the
  electrodynamics of spinning particles},}\ }\href {\doibase
  10.1016/0550-3213(91)90139-o} {\bibfield  {journal} {\bibinfo  {journal}
  {Nucl. Phys. B}\ }\textbf {\bibinfo {volume} {356}},\ \bibinfo {pages} {3}
  (\bibinfo {year} {1991})}\BibitemShut {NoStop}%
\bibitem [{\citenamefont {Yee}\ and\ \citenamefont
  {Bander}(1993)}]{Yee:Bander:1993:Equations_of}%
  \BibitemOpen
  \bibfield  {author} {\bibinfo {author} {\bibfnamefont {K.}~\bibnamefont
  {Yee}}\ and\ \bibinfo {author} {\bibfnamefont {M.}~\bibnamefont {Bander}},\
  }\bibfield  {title} {\enquote {\bibinfo {title} {Equations of motion for
  spinning particles in external electromagnetic and gravitational fields},}\
  }\href {\doibase 10.1103/physrevd.48.2797} {\bibfield  {journal} {\bibinfo
  {journal} {Phys. Rev. D}\ }\textbf {\bibinfo {volume} {48}},\ \bibinfo
  {pages} {2797} (\bibinfo {year} {1993})}\BibitemShut {NoStop}%
\bibitem [{\citenamefont {Chaichian}\ \emph {et~al.}(1997)\citenamefont
  {Chaichian}, \citenamefont {Gonz\'alez~Felipe},\ and\ \citenamefont {{Louis
  Martinez}}}]{Chaichian:Gonzalez-Felipe:etal:1997:Spinning_relativistic}%
  \BibitemOpen
  \bibfield  {author} {\bibinfo {author} {\bibfnamefont {M.}~\bibnamefont
  {Chaichian}}, \bibinfo {author} {\bibfnamefont {R.}~\bibnamefont
  {Gonz\'alez~Felipe}}, \ and\ \bibinfo {author} {\bibfnamefont
  {D.}~\bibnamefont {{Louis Martinez}}},\ }\bibfield  {title} {\enquote
  {\bibinfo {title} {Spinning relativistic particle in an external
  electromagnetic field},}\ }\href {\doibase 10.1016/s0375-9601(97)00801-3}
  {\bibfield  {journal} {\bibinfo  {journal} {Phys. Lett. A}\ }\textbf
  {\bibinfo {volume} {236}},\ \bibinfo {pages} {188} (\bibinfo {year}
  {1997})}\BibitemShut {NoStop}%
\bibitem [{\citenamefont {Pomeranski\u{\i}}\ and\ \citenamefont
  {Khriplovich}(1998)}]{Pomeranskii:Khriplovich:1998:Equations_of}%
  \BibitemOpen
  \bibfield  {author} {\bibinfo {author} {\bibfnamefont {A.~A.}\ \bibnamefont
  {Pomeranski\u{\i}}}\ and\ \bibinfo {author} {\bibfnamefont {I.~B.}\
  \bibnamefont {Khriplovich}},\ }\bibfield  {title} {\enquote {\bibinfo {title}
  {Equations of motion of a spinning relativistic particle in external
  fields},}\ }\href {\doibase 10.1134/1.558554} {\bibfield  {journal} {\bibinfo
   {journal} {J. Exp. Theor. Phys.}\ }\textbf {\bibinfo {volume} {86}},\
  \bibinfo {pages} {839} (\bibinfo {year} {1998})}\BibitemShut {NoStop}%
\bibitem [{\citenamefont {Pomeranski\u{\i}}\ \emph {et~al.}(2000)\citenamefont
  {Pomeranski\u{\i}}, \citenamefont {Sen'kov},\ and\ \citenamefont
  {Khriplovich}}]{Pomeranskii:Senkov:etal:2000:Spinning_relativistic}%
  \BibitemOpen
  \bibfield  {author} {\bibinfo {author} {\bibfnamefont {A.~A.}\ \bibnamefont
  {Pomeranski\u{\i}}}, \bibinfo {author} {\bibfnamefont {R.~A.}\ \bibnamefont
  {Sen'kov}}, \ and\ \bibinfo {author} {\bibfnamefont {I.~B.}\ \bibnamefont
  {Khriplovich}},\ }\bibfield  {title} {\enquote {\bibinfo {title} {Spinning
  relativistic particles in external fields},}\ }\href {\doibase
  10.1070/pu2000v043n10abeh000674} {\bibfield  {journal} {\bibinfo  {journal}
  {Phys.-Usp.}\ }\textbf {\bibinfo {volume} {43}},\ \bibinfo {pages} {1055}
  (\bibinfo {year} {2000})}\BibitemShut {NoStop}%
\bibitem [{\citenamefont {Gralla}\ \emph {et~al.}(2009)\citenamefont {Gralla},
  \citenamefont {Harte},\ and\ \citenamefont
  {Wald}}]{Gralla:Harte:etal:2009:Rigorous_derivation}%
  \BibitemOpen
  \bibfield  {author} {\bibinfo {author} {\bibfnamefont {S.~E.}\ \bibnamefont
  {Gralla}}, \bibinfo {author} {\bibfnamefont {A.~I.}\ \bibnamefont {Harte}}, \
  and\ \bibinfo {author} {\bibfnamefont {R.~M.}\ \bibnamefont {Wald}},\
  }\bibfield  {title} {\enquote {\bibinfo {title} {Rigorous derivation of
  electromagnetic self-force},}\ }\href {\doibase 10.1103/physrevd.80.024031}
  {\bibfield  {journal} {\bibinfo  {journal} {Phys. Rev. D}\ }\textbf {\bibinfo
  {volume} {80}},\ \bibinfo {pages} {024031} (\bibinfo {year}
  {2009})}\BibitemShut {NoStop}%
\bibitem [{\citenamefont {Karabali}\ and\ \citenamefont
  {Nair}(2014)}]{Karabali:Nair:2014:Relativistic_Particle}%
  \BibitemOpen
  \bibfield  {author} {\bibinfo {author} {\bibfnamefont {D.}~\bibnamefont
  {Karabali}}\ and\ \bibinfo {author} {\bibfnamefont {V.~P.}\ \bibnamefont
  {Nair}},\ }\bibfield  {title} {\enquote {\bibinfo {title} {Relativistic
  particle and relativistic fluids: Magnetic moment and spin-orbit
  interactions},}\ }\href {\doibase 10.1103/physrevd.90.105018} {\bibfield
  {journal} {\bibinfo  {journal} {Phys. Rev. D}\ }\textbf {\bibinfo {volume}
  {90}},\ \bibinfo {pages} {105018} (\bibinfo {year} {2014})}\BibitemShut
  {NoStop}%
\bibitem [{\citenamefont {Bhabha}\ and\ \citenamefont
  {Corben}(1941)}]{Bhabha:Corben:1941:General_Classical}%
  \BibitemOpen
  \bibfield  {author} {\bibinfo {author} {\bibfnamefont {H.~J.}\ \bibnamefont
  {Bhabha}}\ and\ \bibinfo {author} {\bibfnamefont {H.~C.}\ \bibnamefont
  {Corben}},\ }\bibfield  {title} {\enquote {\bibinfo {title} {General
  classical theory of spinning particles in a {Maxwell} field},}\ }\href
  {\doibase 10.1098/rspa.1941.0056} {\bibfield  {journal} {\bibinfo  {journal}
  {Proc. R. Soc. A}\ }\textbf {\bibinfo {volume} {178}},\ \bibinfo {pages}
  {273} (\bibinfo {year} {1941})}\BibitemShut {NoStop}%
\bibitem [{\citenamefont {Nagy}(1957)}]{Nagy:1957:Relativistic_equation}%
  \BibitemOpen
  \bibfield  {author} {\bibinfo {author} {\bibfnamefont {K.}~\bibnamefont
  {Nagy}},\ }\bibfield  {title} {\enquote {\bibinfo {title} {Relativistic
  equation of motion for spinning particles},}\ }\href {\doibase
  10.1007/bf03156343} {\bibfield  {journal} {\bibinfo  {journal} {Acta Physica
  Academiae Scientiarum Hungaricae}\ }\textbf {\bibinfo {volume} {7}},\
  \bibinfo {pages} {325} (\bibinfo {year} {1957})}\BibitemShut {NoStop}%
\bibitem [{\citenamefont {Corben}(1961)}]{Corben:1961:Spin_in}%
  \BibitemOpen
  \bibfield  {author} {\bibinfo {author} {\bibfnamefont {H.~C.}\ \bibnamefont
  {Corben}},\ }\bibfield  {title} {\enquote {\bibinfo {title} {Spin in
  classical and quantum theory},}\ }\href {\doibase 10.1103/physrev.121.1833}
  {\bibfield  {journal} {\bibinfo  {journal} {Phys. Rev.}\ }\textbf {\bibinfo
  {volume} {121}},\ \bibinfo {pages} {1833} (\bibinfo {year}
  {1961})}\BibitemShut {NoStop}%
\bibitem [{\citenamefont
  {Nyborg}(1962{\natexlab{a}})}]{Nyborg:1962:On_classical}%
  \BibitemOpen
  \bibfield  {author} {\bibinfo {author} {\bibfnamefont {P.}~\bibnamefont
  {Nyborg}},\ }\bibfield  {title} {\enquote {\bibinfo {title} {On classical
  theories of spinning particles},}\ }\href {\doibase 10.1007/bf02733541}
  {\bibfield  {journal} {\bibinfo  {journal} {Il Nuovo Cimento}\ }\textbf
  {\bibinfo {volume} {23}},\ \bibinfo {pages} {47} (\bibinfo {year}
  {1962}{\natexlab{a}})}\BibitemShut {NoStop}%
\bibitem [{\citenamefont
  {Nyborg}(1962{\natexlab{b}})}]{Nyborg:1962:Thomas_precession}%
  \BibitemOpen
  \bibfield  {author} {\bibinfo {author} {\bibfnamefont {P.}~\bibnamefont
  {Nyborg}},\ }\bibfield  {title} {\enquote {\bibinfo {title} {Thomas
  precession and classical theories of spinning particles},}\ }\href {\doibase
  10.1007/bf02731260} {\bibfield  {journal} {\bibinfo  {journal} {Il Nuovo
  Cimento}\ }\textbf {\bibinfo {volume} {23}},\ \bibinfo {pages} {1057}
  (\bibinfo {year} {1962}{\natexlab{b}})}\BibitemShut {NoStop}%
\bibitem [{\citenamefont {Ternov}\ and\ \citenamefont
  {Bordovitsyn}(1980)}]{Ternov:Bordovitsyn:1980:Modern_interpretation}%
  \BibitemOpen
  \bibfield  {author} {\bibinfo {author} {\bibfnamefont {I.~M.}\ \bibnamefont
  {Ternov}}\ and\ \bibinfo {author} {\bibfnamefont {V.~A.}\ \bibnamefont
  {Bordovitsyn}},\ }\bibfield  {title} {\enquote {\bibinfo {title} {Modern
  interpretation of {J.~I. Frenkel's} classical spin theory},}\ }\href
  {\doibase 10.1070/pu1980v023n10abeh005040} {\bibfield  {journal} {\bibinfo
  {journal} {Soviet Phys.-Uspekhi}\ }\textbf {\bibinfo {volume} {23}},\
  \bibinfo {pages} {679} (\bibinfo {year} {1980})}\BibitemShut {NoStop}%
\bibitem [{\citenamefont
  {Nyborg}(1964)}]{Nyborg:1964:Approximate_relativistic}%
  \BibitemOpen
  \bibfield  {author} {\bibinfo {author} {\bibfnamefont {P.}~\bibnamefont
  {Nyborg}},\ }\bibfield  {title} {\enquote {\bibinfo {title} {Approximate
  relativistic equations of motion for an extended charged particle in an
  inhomogeneous external electromagnetic field},}\ }\href {\doibase
  10.1007/bf02733592} {\bibfield  {journal} {\bibinfo  {journal} {Il Nuovo
  Cimento}\ }\textbf {\bibinfo {volume} {31}},\ \bibinfo {pages} {1209}
  (\bibinfo {year} {1964})}\BibitemShut {NoStop}%
\bibitem [{\citenamefont {Bagrov}\ and\ \citenamefont
  {Bordovitsyn}(1980)}]{Bagrov:Bordovitsyn:1980:Classical_spin}%
  \BibitemOpen
  \bibfield  {author} {\bibinfo {author} {\bibfnamefont {V.~G.}\ \bibnamefont
  {Bagrov}}\ and\ \bibinfo {author} {\bibfnamefont {V.~A.}\ \bibnamefont
  {Bordovitsyn}},\ }\bibfield  {title} {\enquote {\bibinfo {title} {Classical
  spin theory},}\ }\href {\doibase 10.1007/bf00892416} {\bibfield  {journal}
  {\bibinfo  {journal} {Soviet Phys. J.}\ }\textbf {\bibinfo {volume} {23}},\
  \bibinfo {pages} {128} (\bibinfo {year} {1980})}\BibitemShut {NoStop}%
\bibitem [{\citenamefont {Teitelboim}\ \emph {et~al.}(1980)\citenamefont
  {Teitelboim}, \citenamefont {Villarroel},\ and\ \citenamefont {van
  Weert}}]{Teitelboim:Villarroel:etal:1980:Classical_electrodynamics}%
  \BibitemOpen
  \bibfield  {author} {\bibinfo {author} {\bibfnamefont {C.}~\bibnamefont
  {Teitelboim}}, \bibinfo {author} {\bibfnamefont {D.}~\bibnamefont
  {Villarroel}}, \ and\ \bibinfo {author} {\bibfnamefont {C.~G.}\ \bibnamefont
  {van Weert}},\ }\bibfield  {title} {\enquote {\bibinfo {title} {Classical
  electrodynamics of retarded fields and point particles},}\ }\href {\doibase
  10.1007/bf02895735} {\bibfield  {journal} {\bibinfo  {journal} {La Rivista
  del Nuovo Cimento}\ }\textbf {\bibinfo {volume} {3}},\ \bibinfo {pages} {1}
  (\bibinfo {year} {1980})}\BibitemShut {NoStop}%
\bibitem [{\citenamefont {Jammer}(1966)}]{Jammer:1966:Conceptual_Development}%
  \BibitemOpen
  \bibfield  {author} {\bibinfo {author} {\bibfnamefont {M.}~\bibnamefont
  {Jammer}},\ }\href@noop {} {\emph {\bibinfo {title} {The Conceptual
  Development of Quantum Mechanics}}},\ Pure \& Applied Physics\ (\bibinfo
  {publisher} {McGraw-Hill},\ \bibinfo {address} {New York},\ \bibinfo {year}
  {1966})\ \bibinfo {note} {pages 152--153}\BibitemShut {NoStop}%
\bibitem [{\citenamefont {Giulini}(2008)}]{Giulini:2008:spin}%
  \BibitemOpen
  \bibfield  {author} {\bibinfo {author} {\bibfnamefont {D.}~\bibnamefont
  {Giulini}},\ }\bibfield  {title} {\enquote {\bibinfo {title} {Electron spin
  or ``classically non-describable two-valuedness''},}\ }\href {\doibase
  10.1016/j.shpsb.2008.03.005} {\bibfield  {journal} {\bibinfo  {journal}
  {Studies In History and Philosophy of Science Part B: Studies In History and
  Philosophy of Modern Physics}\ }\textbf {\bibinfo {volume} {39}},\ \bibinfo
  {pages} {557} (\bibinfo {year} {2008})}\BibitemShut {NoStop}%
\bibitem [{\citenamefont {Bagrov}\ \emph {et~al.}(1998)\citenamefont {Bagrov},
  \citenamefont {Belov},\ and\ \citenamefont
  {Trifonov}}]{Bagrov:Belov:etal:1998}%
  \BibitemOpen
  \bibfield  {author} {\bibinfo {author} {\bibfnamefont {V.~G.}\ \bibnamefont
  {Bagrov}}, \bibinfo {author} {\bibfnamefont {V.~V.}\ \bibnamefont {Belov}}, \
  and\ \bibinfo {author} {\bibfnamefont {A.~Y.}\ \bibnamefont {Trifonov}},\
  }\href@noop {} {\enquote {\bibinfo {title} {{Semiclassical
  trajectory-coherent approximation in quantum mechanics: II. High order
  corrections to the Dirac operators in external electromagnetic field}},}\ }
  (\bibinfo {year} {1998}),\ \Eprint {http://arxiv.org/abs/quant-ph/9806017}
  {arXiv:quant-ph/9806017} \BibitemShut {NoStop}%
\bibitem [{\citenamefont
  {Silenko}(2008)}]{Silenko:2008:Foldy-Wouthyusen_transformation}%
  \BibitemOpen
  \bibfield  {author} {\bibinfo {author} {\bibfnamefont {A.~J.}\ \bibnamefont
  {Silenko}},\ }\bibfield  {title} {\enquote {\bibinfo {title}
  {{Foldy-Wouthyusen} transformation and semiclassical limit for relativistic
  particles in strong external fields},}\ }\href {\doibase
  10.1103/physreva.77.012116} {\bibfield  {journal} {\bibinfo  {journal} {Phys.
  Rev. A}\ }\textbf {\bibinfo {volume} {77}},\ \bibinfo {pages} {012116}
  (\bibinfo {year} {2008})}\BibitemShut {NoStop}%
\bibitem [{\citenamefont {Chen}\ and\ \citenamefont
  {Chiou}(2010)}]{Chen:Chiou:2010:Foldy-Wouthuysen_transformation}%
  \BibitemOpen
  \bibfield  {author} {\bibinfo {author} {\bibfnamefont {T.-W.}\ \bibnamefont
  {Chen}}\ and\ \bibinfo {author} {\bibfnamefont {D.-W.}\ \bibnamefont
  {Chiou}},\ }\bibfield  {title} {\enquote {\bibinfo {title}
  {{Foldy-Wouthuysen} transformation for a {Dirac-Pauli} dyon and the
  {Thomas-Bargmann-Michel-Telegdi} equation},}\ }\href {\doibase
  10.1103/physreva.82.012115} {\bibfield  {journal} {\bibinfo  {journal} {Phys.
  Rev. A}\ }\textbf {\bibinfo {volume} {82}},\ \bibinfo {pages} {012115}
  (\bibinfo {year} {2010})}\BibitemShut {NoStop}%
\bibitem [{\citenamefont {Chen}\ and\ \citenamefont
  {Chiou}(2014{\natexlab{a}})}]{Chen:Chiou:2014:Correspondence_between}%
  \BibitemOpen
  \bibfield  {author} {\bibinfo {author} {\bibfnamefont {T.-W.}\ \bibnamefont
  {Chen}}\ and\ \bibinfo {author} {\bibfnamefont {D.-W.}\ \bibnamefont
  {Chiou}},\ }\bibfield  {title} {\enquote {\bibinfo {title} {Correspondence
  between classical and {Dirac-Pauli} spinors in view of the {Foldy-Wouthuysen}
  transformation},}\ }\href {\doibase 10.1103/physreva.89.032111} {\bibfield
  {journal} {\bibinfo  {journal} {Phys. Rev. A}\ }\textbf {\bibinfo {volume}
  {89}},\ \bibinfo {pages} {032111} (\bibinfo {year}
  {2014}{\natexlab{a}})}\BibitemShut {NoStop}%
\bibitem [{\citenamefont {Chen}\ and\ \citenamefont
  {Chiou}(2014{\natexlab{b}})}]{Chen:Chiou:2014:High-order_Foldy-Wouthuysen}%
  \BibitemOpen
  \bibfield  {author} {\bibinfo {author} {\bibfnamefont {T.-W.}\ \bibnamefont
  {Chen}}\ and\ \bibinfo {author} {\bibfnamefont {D.-W.}\ \bibnamefont
  {Chiou}},\ }\bibfield  {title} {\enquote {\bibinfo {title} {High-order
  {Foldy-Wouthuysen} transformations of the {Dirac} and {Dirac-Pauli}
  {Hamiltonians} in the weak-field limit},}\ }\href {\doibase
  10.1103/physreva.90.012112} {\bibfield  {journal} {\bibinfo  {journal} {Phys.
  Rev. A}\ }\textbf {\bibinfo {volume} {90}},\ \bibinfo {pages} {012112}
  (\bibinfo {year} {2014}{\natexlab{b}})}\BibitemShut {NoStop}%
\bibitem [{\citenamefont {Wen}\ \emph {et~al.}(2016)\citenamefont {Wen},
  \citenamefont {Bauke},\ and\ \citenamefont {Keitel}}]{Wen:2016:Identifying}%
  \BibitemOpen
  \bibfield  {author} {\bibinfo {author} {\bibfnamefont {M.}~\bibnamefont
  {Wen}}, \bibinfo {author} {\bibfnamefont {H.}~\bibnamefont {Bauke}}, \ and\
  \bibinfo {author} {\bibfnamefont {C.~H.}\ \bibnamefont {Keitel}},\ }\bibfield
   {title} {\enquote {\bibinfo {title} {Identifying the {Stern-Gerlach} force
  of classical electron dynamics},}\ }\href {\doibase 10.1038/srep31624}
  {\bibfield  {journal} {\bibinfo  {journal} {Sci. Rep.}\ }\textbf {\bibinfo
  {volume} {6}},\ \bibinfo {pages} {31624} (\bibinfo {year}
  {2016})}\BibitemShut {NoStop}%
\bibitem [{\citenamefont {Bogdanov}\ and\ \citenamefont
  {Kazinski}(2015)}]{Bogdanov:Kazinski:2015:Properties_of}%
  \BibitemOpen
  \bibfield  {author} {\bibinfo {author} {\bibfnamefont {O.~V.}\ \bibnamefont
  {Bogdanov}}\ and\ \bibinfo {author} {\bibfnamefont {P.~O.}\ \bibnamefont
  {Kazinski}},\ }\bibfield  {title} {\enquote {\bibinfo {title} {Properties of
  electrons scattered by a strong plane electromagnetic wave with a linear
  polarization: Semiclassical treatment},}\ }\href {\doibase
  10.1134/s0021364015030030} {\bibfield  {journal} {\bibinfo  {journal} {JETP
  Lett.}\ }\textbf {\bibinfo {volume} {101}},\ \bibinfo {pages} {206} (\bibinfo
  {year} {2015})}\BibitemShut {NoStop}%
\bibitem [{\citenamefont {Landau}\ and\ \citenamefont
  {Lifshitz}(1980)}]{Landau:Lifshitz:II}%
  \BibitemOpen
  \bibfield  {author} {\bibinfo {author} {\bibfnamefont {L.~D.}\ \bibnamefont
  {Landau}}\ and\ \bibinfo {author} {\bibfnamefont {E.~M.}\ \bibnamefont
  {Lifshitz}},\ }\href@noop {} {\emph {\bibinfo {title} {The Classical Theory
  of Fields}}}\ (\bibinfo  {publisher} {Butterworth-Heinemann},\ \bibinfo
  {address} {Oxford, UK},\ \bibinfo {year} {1980})\BibitemShut {NoStop}%
\bibitem [{\citenamefont {Gross}(2004)}]{Gross:2004:Relativistic_Quantum}%
  \BibitemOpen
  \bibfield  {author} {\bibinfo {author} {\bibfnamefont {F.}~\bibnamefont
  {Gross}},\ }\href@noop {} {\emph {\bibinfo {title} {Relativistic Quantum
  Mechanics and Field Theory}}}\ (\bibinfo  {publisher} {Wiley-VCH},\ \bibinfo
  {address} {Weinheim},\ \bibinfo {year} {2004})\BibitemShut {NoStop}%
\bibitem [{\citenamefont {Thaller}(2005)}]{Thaller:2005:Advanced_Visual}%
  \BibitemOpen
  \bibfield  {author} {\bibinfo {author} {\bibfnamefont {B.}~\bibnamefont
  {Thaller}},\ }\href@noop {} {\emph {\bibinfo {title} {Advanced Visual Quantum
  Mechanics}}}\ (\bibinfo  {publisher} {Springer},\ \bibinfo {address}
  {Heidelberg},\ \bibinfo {year} {2005})\BibitemShut {NoStop}%
\bibitem [{\citenamefont {Bauke}\ \emph
  {et~al.}(2014{\natexlab{a}})\citenamefont {Bauke}, \citenamefont {Ahrens},
  \citenamefont {Keitel},\ and\ \citenamefont
  {Grobe}}]{Bauke:Ahrens:etal:2014:Relativistic_spin}%
  \BibitemOpen
  \bibfield  {author} {\bibinfo {author} {\bibfnamefont {H.}~\bibnamefont
  {Bauke}}, \bibinfo {author} {\bibfnamefont {S.}~\bibnamefont {Ahrens}},
  \bibinfo {author} {\bibfnamefont {C.~H.}\ \bibnamefont {Keitel}}, \ and\
  \bibinfo {author} {\bibfnamefont {R.}~\bibnamefont {Grobe}},\ }\bibfield
  {title} {\enquote {\bibinfo {title} {Relativistic spin operators in various
  electromagnetic environments},}\ }\href {\doibase 10.1103/physreva.89.052101}
  {\bibfield  {journal} {\bibinfo  {journal} {Phys. Rev. A}\ }\textbf {\bibinfo
  {volume} {89}},\ \bibinfo {pages} {052101} (\bibinfo {year}
  {2014}{\natexlab{a}})}\BibitemShut {NoStop}%
\bibitem [{\citenamefont {Bauke}\ \emph
  {et~al.}(2014{\natexlab{b}})\citenamefont {Bauke}, \citenamefont {Ahrens},
  \citenamefont {Keitel},\ and\ \citenamefont
  {Grobe}}]{Bauke:etal:2014:What_is}%
  \BibitemOpen
  \bibfield  {author} {\bibinfo {author} {\bibfnamefont {H.}~\bibnamefont
  {Bauke}}, \bibinfo {author} {\bibfnamefont {S.}~\bibnamefont {Ahrens}},
  \bibinfo {author} {\bibfnamefont {C.~H.}\ \bibnamefont {Keitel}}, \ and\
  \bibinfo {author} {\bibfnamefont {R.}~\bibnamefont {Grobe}},\ }\bibfield
  {title} {\enquote {\bibinfo {title} {What is the relativistic spin
  operator?}}\ }\href {\doibase 10.1088/1367-2630/16/4/043012} {\bibfield
  {journal} {\bibinfo  {journal} {New J. Phys.}\ }\textbf {\bibinfo {volume}
  {16}},\ \bibinfo {pages} {043012} (\bibinfo {year}
  {2014}{\natexlab{b}})}\BibitemShut {NoStop}%
\bibitem [{\citenamefont {Foldy}\ and\ \citenamefont
  {Wouthuysen}(1950)}]{Foldy:Wouthuysen:1950:Non-Relativistic_Limit}%
  \BibitemOpen
  \bibfield  {author} {\bibinfo {author} {\bibfnamefont {L.~L.}\ \bibnamefont
  {Foldy}}\ and\ \bibinfo {author} {\bibfnamefont {S.~A.}\ \bibnamefont
  {Wouthuysen}},\ }\bibfield  {title} {\enquote {\bibinfo {title} {On the
  {Dirac} theory of spin 1/2 particles and its non-relativistic limit},}\
  }\href {\doibase 10.1103/PhysRev.78.29} {\bibfield  {journal} {\bibinfo
  {journal} {Phys. Rev.}\ }\textbf {\bibinfo {volume} {78}},\ \bibinfo {pages}
  {29} (\bibinfo {year} {1950})}\BibitemShut {NoStop}%
\bibitem [{\citenamefont {Mane}\ \emph {et~al.}(2005)\citenamefont {Mane},
  \citenamefont {Shatunov},\ and\ \citenamefont
  {Yokoya}}]{Mane:Shatunov:etal:2005:Spin-polarized_charged}%
  \BibitemOpen
  \bibfield  {author} {\bibinfo {author} {\bibfnamefont {S.~R.}\ \bibnamefont
  {Mane}}, \bibinfo {author} {\bibfnamefont {Y.~M.}\ \bibnamefont {Shatunov}},
  \ and\ \bibinfo {author} {\bibfnamefont {K.}~\bibnamefont {Yokoya}},\
  }\bibfield  {title} {\enquote {\bibinfo {title} {Spin-polarized charged
  particle beams in high-energy accelerators},}\ }\href {\doibase
  10.1088/0034-4885/68/9/r01} {\bibfield  {journal} {\bibinfo  {journal} {Rep.
  Prog. Phys.}\ }\textbf {\bibinfo {volume} {68}},\ \bibinfo {pages} {1997}
  (\bibinfo {year} {2005})}\BibitemShut {NoStop}%
\bibitem [{\citenamefont
  {Schwinger}(1948)}]{Schwinger:1948:On_Quantum-Electrodynamics}%
  \BibitemOpen
  \bibfield  {author} {\bibinfo {author} {\bibfnamefont {J.}~\bibnamefont
  {Schwinger}},\ }\bibfield  {title} {\enquote {\bibinfo {title} {On
  quantum-electrodynamics and the magnetic moment of the electron},}\ }\href
  {\doibase 10.1103/physrev.73.416} {\bibfield  {journal} {\bibinfo  {journal}
  {Phys. Rev.}\ }\textbf {\bibinfo {volume} {73}},\ \bibinfo {pages} {416}
  (\bibinfo {year} {1948})}\BibitemShut {NoStop}%
\bibitem [{\citenamefont {Lobanov}\ and\ \citenamefont
  {Pavlova}(1999)}]{Lobanov:Pavlova:1999:Solutions_of}%
  \BibitemOpen
  \bibfield  {author} {\bibinfo {author} {\bibfnamefont {A.}~\bibnamefont
  {Lobanov}}\ and\ \bibinfo {author} {\bibfnamefont {O.}~\bibnamefont
  {Pavlova}},\ }\bibfield  {title} {\enquote {\bibinfo {title} {Solutions of
  the classical equation of motion for a spin in electromagnetic fields},}\
  }\href {\doibase 10.1007/BF02557213} {\bibfield  {journal} {\bibinfo
  {journal} {Theor. Math. Phys.}\ }\textbf {\bibinfo {volume} {121}},\ \bibinfo
  {pages} {1691} (\bibinfo {year} {1999})}\BibitemShut {NoStop}%
\bibitem [{\citenamefont {Yang}\ and\ \citenamefont
  {Hirschfelder}(1980)}]{Yang:Hirschfelder:1980:Generalizations_of}%
  \BibitemOpen
  \bibfield  {author} {\bibinfo {author} {\bibfnamefont {K.-H.}\ \bibnamefont
  {Yang}}\ and\ \bibinfo {author} {\bibfnamefont {J.~O.}\ \bibnamefont
  {Hirschfelder}},\ }\bibfield  {title} {\enquote {\bibinfo {title}
  {Generalizations of classical {Poisson} brackets to include spin},}\ }\href
  {\doibase 10.1103/physreva.22.1814} {\bibfield  {journal} {\bibinfo
  {journal} {Phys. Rev. A}\ }\textbf {\bibinfo {volume} {22}},\ \bibinfo
  {pages} {1814} (\bibinfo {year} {1980})}\BibitemShut {NoStop}%
\bibitem [{\citenamefont {Hegstrom}\ and\ \citenamefont
  {Lhuillier}(1977)}]{Hegstrom:Lhuillier:1977:Reinterpretation_of}%
  \BibitemOpen
  \bibfield  {author} {\bibinfo {author} {\bibfnamefont {R.~A.}\ \bibnamefont
  {Hegstrom}}\ and\ \bibinfo {author} {\bibfnamefont {C.}~\bibnamefont
  {Lhuillier}},\ }\bibfield  {title} {\enquote {\bibinfo {title}
  {Reinterpretation of the ``relativistic mass'' correction to the spin
  magnetic moment of a moving particle},}\ }\href {\doibase
  10.1103/physreva.15.1797} {\bibfield  {journal} {\bibinfo  {journal} {Phys.
  Rev. A}\ }\textbf {\bibinfo {volume} {15}},\ \bibinfo {pages} {1797}
  (\bibinfo {year} {1977})}\BibitemShut {NoStop}%
\bibitem [{\citenamefont {Walser}\ and\ \citenamefont
  {Keitel}(2001)}]{Walse:Keitel:2001:Geometric_and}%
  \BibitemOpen
  \bibfield  {author} {\bibinfo {author} {\bibfnamefont {M.~W.}\ \bibnamefont
  {Walser}}\ and\ \bibinfo {author} {\bibfnamefont {C.~H.}\ \bibnamefont
  {Keitel}},\ }\bibfield  {title} {\enquote {\bibinfo {title} {Geometric and
  algebraic approach to classical dynamics of a particle with spin},}\ }\href
  {\doibase 10.1023/a:1010957006903} {\bibfield  {journal} {\bibinfo  {journal}
  {Lett. Math. Phys.}\ }\textbf {\bibinfo {volume} {55}},\ \bibinfo {pages}
  {53} (\bibinfo {year} {2001})}\BibitemShut {NoStop}%
\bibitem [{\citenamefont {Asenjo}\ \emph {et~al.}(2012)\citenamefont {Asenjo},
  \citenamefont {Zamanian}, \citenamefont {Marklund}, \citenamefont {Brodin},\
  and\ \citenamefont
  {Johansson}}]{Asenjo:Zamanian:etal:2012:Semi-relativistic_effects}%
  \BibitemOpen
  \bibfield  {author} {\bibinfo {author} {\bibfnamefont {F.~A.}\ \bibnamefont
  {Asenjo}}, \bibinfo {author} {\bibfnamefont {J.}~\bibnamefont {Zamanian}},
  \bibinfo {author} {\bibfnamefont {M.}~\bibnamefont {Marklund}}, \bibinfo
  {author} {\bibfnamefont {G.}~\bibnamefont {Brodin}}, \ and\ \bibinfo {author}
  {\bibfnamefont {P.}~\bibnamefont {Johansson}},\ }\bibfield  {title} {\enquote
  {\bibinfo {title} {Semi-relativistic effects in spin-1/2 quantum plasmas},}\
  }\href {\doibase 10.1088/1367-2630/14/7/073042} {\bibfield  {journal}
  {\bibinfo  {journal} {New J. Phys.}\ }\textbf {\bibinfo {volume} {14}},\
  \bibinfo {pages} {073042} (\bibinfo {year} {2012})}\BibitemShut {NoStop}%
\bibitem [{\citenamefont {Eriksen}\ and\ \citenamefont
  {Kolsrud}(1960)}]{Eriksen:Kolsrud:1960:Canonical_transformations}%
  \BibitemOpen
  \bibfield  {author} {\bibinfo {author} {\bibfnamefont {E.}~\bibnamefont
  {Eriksen}}\ and\ \bibinfo {author} {\bibfnamefont {M.}~\bibnamefont
  {Kolsrud}},\ }\bibfield  {title} {\enquote {\bibinfo {title} {Canonical
  transformations of {Dirac's} equation to even forms. {Expansion} in terms of
  the external fields},}\ }\href {\doibase 10.1007/bf02782145} {\bibfield
  {journal} {\bibinfo  {journal} {Il Nuovo Cimento}\ }\textbf {\bibinfo
  {volume} {18}},\ \bibinfo {pages} {1} (\bibinfo {year} {1960})}\BibitemShut
  {NoStop}%
\bibitem [{\citenamefont {Dixon}(1965)}]{Dixon:1965:Classical_theory}%
  \BibitemOpen
  \bibfield  {author} {\bibinfo {author} {\bibfnamefont {W.~G.}\ \bibnamefont
  {Dixon}},\ }\bibfield  {title} {\enquote {\bibinfo {title} {Classical theory
  of charged particles with spin and the classical limit of the {Dirac}
  equation},}\ }\href {\doibase 10.1007/bf02750084} {\bibfield  {journal}
  {\bibinfo  {journal} {Il Nuovo Cimento}\ }\textbf {\bibinfo {volume} {38}},\
  \bibinfo {pages} {1616} (\bibinfo {year} {1965})}\BibitemShut {NoStop}%
\bibitem [{\citenamefont {Suttorp}\ and\ \citenamefont
  {De~Groot}(1970)}]{Suttorp:De-Groot:1970:Covariant_equations}%
  \BibitemOpen
  \bibfield  {author} {\bibinfo {author} {\bibfnamefont {L.~G.}\ \bibnamefont
  {Suttorp}}\ and\ \bibinfo {author} {\bibfnamefont {S.~R.}\ \bibnamefont
  {De~Groot}},\ }\bibfield  {title} {\enquote {\bibinfo {title} {Covariant
  equations of motion, for a charged particle with a magnetic dipole moment},}\
  }\href {\doibase 10.1007/bf02752917} {\bibfield  {journal} {\bibinfo
  {journal} {Il Nuovo Cimento A}\ }\textbf {\bibinfo {volume} {65}},\ \bibinfo
  {pages} {245} (\bibinfo {year} {1970})}\BibitemShut {NoStop}%
\bibitem [{Note2()}]{Note2}%
  \BibitemOpen
  \bibinfo {note} {Note that the model considered does not have a commonly used
  standard name. Although it was not derived by Foldy and Wouthuysen, we call
  it the ``classical Foldy-Wouthuysen model'' here because it originates from a
  Dirac Hamiltonian in the Foldy-Wouthuysen representation. Alternatively, one
  might just speak of the semiclassical equations of motion for the Dirac
  theory.}\BibitemShut {Stop}%
\bibitem [{\citenamefont
  {Takabayasi}(1981)}]{Takabayasi:1981:Relativistic_Dynamics}%
  \BibitemOpen
  \bibfield  {author} {\bibinfo {author} {\bibfnamefont {T.}~\bibnamefont
  {Takabayasi}},\ }\bibfield  {title} {\enquote {\bibinfo {title} {Relativistic
  dynamics of classical particle with spin in electromagnetic field},}\ }\href
  {\doibase 10.1143/ptp.66.736} {\bibfield  {journal} {\bibinfo  {journal}
  {Prog. Theor. Phys.}\ }\textbf {\bibinfo {volume} {66}},\ \bibinfo {pages}
  {736} (\bibinfo {year} {1981})}\BibitemShut {NoStop}%
\bibitem [{\citenamefont {Kassandrov}\ \emph {et~al.}(2009)\citenamefont
  {Kassandrov}, \citenamefont {Markova}, \citenamefont {Sch\"afer},\ and\
  \citenamefont {Wipf}}]{Kassandrov:Markova:etal:2009:On_model}%
  \BibitemOpen
  \bibfield  {author} {\bibinfo {author} {\bibfnamefont {V.}~\bibnamefont
  {Kassandrov}}, \bibinfo {author} {\bibfnamefont {N.}~\bibnamefont {Markova}},
  \bibinfo {author} {\bibfnamefont {G.}~\bibnamefont {Sch\"afer}}, \ and\
  \bibinfo {author} {\bibfnamefont {A.}~\bibnamefont {Wipf}},\ }\bibfield
  {title} {\enquote {\bibinfo {title} {On a model of a classical relativistic
  particle of constant and universal mass and spin},}\ }\href {\doibase
  10.1088/1751-8113/42/31/315204} {\bibfield  {journal} {\bibinfo  {journal}
  {J. Phys. A: Math. Theor.}\ }\textbf {\bibinfo {volume} {42}},\ \bibinfo
  {pages} {315204} (\bibinfo {year} {2009})}\BibitemShut {NoStop}%
\bibitem [{\citenamefont {Lebedev}(2016)}]{Lebedev:2016:Spin_radiative}%
  \BibitemOpen
  \bibfield  {author} {\bibinfo {author} {\bibfnamefont {S.~L.}\ \bibnamefont
  {Lebedev}},\ }\bibfield  {title} {\enquote {\bibinfo {title} {Spin radiative
  corrections to the radiation probability and power in classical and quantum
  electrodynamics},}\ }\href {\doibase 10.1134/s1063776116020084} {\bibfield
  {journal} {\bibinfo  {journal} {J. Exp. Theor. Phys.}\ }\textbf {\bibinfo
  {volume} {122}},\ \bibinfo {pages} {650} (\bibinfo {year}
  {2016})}\BibitemShut {NoStop}%
\bibitem [{\citenamefont {Bauke}\ and\ \citenamefont
  {Keitel}(2011)}]{Bauke:Keitel:2011:Accelerating_Fourier}%
  \BibitemOpen
  \bibfield  {author} {\bibinfo {author} {\bibfnamefont {H.}~\bibnamefont
  {Bauke}}\ and\ \bibinfo {author} {\bibfnamefont {C.~H.}\ \bibnamefont
  {Keitel}},\ }\bibfield  {title} {\enquote {\bibinfo {title} {Accelerating the
  {Fourier} split operator method via graphics processing units},}\ }\href
  {\doibase 10.1016/j.cpc.2011.07.003} {\bibfield  {journal} {\bibinfo
  {journal} {Comput. Phys. Commun.}\ }\textbf {\bibinfo {volume} {182}},\
  \bibinfo {pages} {2454} (\bibinfo {year} {2011})}\BibitemShut {NoStop}%
\bibitem [{\citenamefont {Bagrov}\ \emph {et~al.}(1975)\citenamefont {Bagrov},
  \citenamefont {Bukhbinder},\ and\ \citenamefont
  {Gitman}}]{Bagrov:Bukhbinder:etal:1975:Coherent_states}%
  \BibitemOpen
  \bibfield  {author} {\bibinfo {author} {\bibfnamefont {V.~G.}\ \bibnamefont
  {Bagrov}}, \bibinfo {author} {\bibfnamefont {I.~L.}\ \bibnamefont
  {Bukhbinder}}, \ and\ \bibinfo {author} {\bibfnamefont {D.~M.}\ \bibnamefont
  {Gitman}},\ }\bibfield  {title} {\enquote {\bibinfo {title} {Coherent states
  of relativistic particles},}\ }\href {\doibase 10.1007/bf01110051} {\bibfield
   {journal} {\bibinfo  {journal} {Sov. Phys. J.}\ }\textbf {\bibinfo {volume}
  {18}},\ \bibinfo {pages} {1180} (\bibinfo {year} {1975})}\BibitemShut
  {NoStop}%
\bibitem [{\citenamefont {Bagrov}\ and\ \citenamefont
  {Gitman}(2014)}]{Bagrov:Gitman:2014:Dirac_Equation}%
  \BibitemOpen
  \bibfield  {author} {\bibinfo {author} {\bibfnamefont {V.~G.}\ \bibnamefont
  {Bagrov}}\ and\ \bibinfo {author} {\bibfnamefont {D.}~\bibnamefont
  {Gitman}},\ }\href@noop {} {\emph {\bibinfo {title} {{The Dirac Equation and
  Its Solutions}}}},\ \bibinfo {series} {De Gruyter Studies in Mathematical
  Physics}, Vol.~\bibinfo {volume} {4}\ (\bibinfo  {publisher} {De Gruyter},\
  \bibinfo {address} {Berlin},\ \bibinfo {year} {2014})\BibitemShut {NoStop}%
\bibitem [{\citenamefont {Rakhimov}\ \emph {et~al.}(2011)\citenamefont
  {Rakhimov}, \citenamefont {Chaves}, \citenamefont {Farias},\ and\
  \citenamefont {Peeters}}]{Rakhimov:JOPCM:2011}%
  \BibitemOpen
  \bibfield  {author} {\bibinfo {author} {\bibfnamefont {K.~Y.}\ \bibnamefont
  {Rakhimov}}, \bibinfo {author} {\bibfnamefont {A.}~\bibnamefont {Chaves}},
  \bibinfo {author} {\bibfnamefont {G.~A.}\ \bibnamefont {Farias}}, \ and\
  \bibinfo {author} {\bibfnamefont {F.~M.}\ \bibnamefont {Peeters}},\
  }\bibfield  {title} {\enquote {\bibinfo {title} {Wavepacket scattering of
  {Dirac} and {Schr\"odinger} particles on potential and magnetic barriers},}\
  }\href {\doibase 10.1088/0953-8984/23/27/275801} {\bibfield  {journal}
  {\bibinfo  {journal} {J. Phys.: Condens. Matter}\ }\textbf {\bibinfo {volume}
  {23}},\ \bibinfo {pages} {275801} (\bibinfo {year} {2011})}\BibitemShut
  {NoStop}%
\bibitem [{\citenamefont {Ruiz}\ \emph {et~al.}(2015)\citenamefont {Ruiz},
  \citenamefont {Ellison},\ and\ \citenamefont
  {Dodin}}]{Ruiz:Ellison:etal:2015:Relativistic_ponderomotive}%
  \BibitemOpen
  \bibfield  {author} {\bibinfo {author} {\bibfnamefont {D.~E.}\ \bibnamefont
  {Ruiz}}, \bibinfo {author} {\bibfnamefont {C.~L.}\ \bibnamefont {Ellison}}, \
  and\ \bibinfo {author} {\bibfnamefont {I.~Y.}\ \bibnamefont {Dodin}},\
  }\bibfield  {title} {\enquote {\bibinfo {title} {Relativistic ponderomotive
  {Hamiltonian} of a {Dirac} particle in a vacuum laser field},}\ }\href
  {\doibase 10.1103/physreva.92.062124} {\bibfield  {journal} {\bibinfo
  {journal} {Phys. Rev. A}\ }\textbf {\bibinfo {volume} {92}},\ \bibinfo
  {pages} {062124} (\bibinfo {year} {2015})}\BibitemShut {NoStop}%
\bibitem [{Note3()}]{Note3}%
  \BibitemOpen
  \bibinfo {note} {The longer pulse length in combination with the wave-packet
  spreading leads to a $\lambda ^3$ or $\lambda ^4$ scaling of the
  computational demand to solve the Dirac equation in two or three dimensions,
  respectively.}\BibitemShut {Stop}%
\bibitem [{\citenamefont {Salamin}\ and\ \citenamefont
  {Keitel}(2002)}]{Salamin:PhysRevLett.88.095005:2002}%
  \BibitemOpen
  \bibfield  {author} {\bibinfo {author} {\bibfnamefont {Y.~I.}\ \bibnamefont
  {Salamin}}\ and\ \bibinfo {author} {\bibfnamefont {C.~H.}\ \bibnamefont
  {Keitel}},\ }\bibfield  {title} {\enquote {\bibinfo {title} {Electron
  acceleration by a tightly focused laser beam},}\ }\href {\doibase
  10.1103/physrevlett.88.095005} {\bibfield  {journal} {\bibinfo  {journal}
  {Phys. Rev. Lett.}\ }\textbf {\bibinfo {volume} {88}},\ \bibinfo {pages}
  {095005} (\bibinfo {year} {2002})}\BibitemShut {NoStop}%
\bibitem [{\citenamefont
  {Batelaan}(2000)}]{Batelaan:Contemporary_Physics:2000}%
  \BibitemOpen
  \bibfield  {author} {\bibinfo {author} {\bibfnamefont {H.}~\bibnamefont
  {Batelaan}},\ }\bibfield  {title} {\enquote {\bibinfo {title} {The
  {Kapitza-Dirac} effect},}\ }\href {\doibase 10.1080/00107510010001220}
  {\bibfield  {journal} {\bibinfo  {journal} {Contemp. Phys.}\ }\textbf
  {\bibinfo {volume} {41}},\ \bibinfo {pages} {369} (\bibinfo {year}
  {2000})}\BibitemShut {NoStop}%
\bibitem [{\citenamefont {Hartemann}(2001)}]{hartemann2001high}%
  \BibitemOpen
  \bibfield  {author} {\bibinfo {author} {\bibfnamefont {F.~V.}\ \bibnamefont
  {Hartemann}},\ }\href@noop {} {\emph {\bibinfo {title} {High-Field
  Electrodynamics}}},\ Pure and Applied Physics\ (\bibinfo  {publisher} {CRC
  Press},\ \bibinfo {address} {Boca Raton},\ \bibinfo {year}
  {2001})\BibitemShut {NoStop}%
\bibitem [{\citenamefont {Plateau}\ \emph {et~al.}(2012)\citenamefont
  {Plateau}, \citenamefont {Geddes}, \citenamefont {Thorn}, \citenamefont
  {Chen}, \citenamefont {Benedetti}, \citenamefont {Esarey}, \citenamefont
  {Gonsalves}, \citenamefont {Matlis}, \citenamefont {Nakamura}, \citenamefont
  {Schroeder}, , \citenamefont {Shiraishi}, \citenamefont {Sokollik},
  \citenamefont {van Tilborg}, \citenamefont {Toth}, \citenamefont {Trotsenko},
  \citenamefont {Kim}, \citenamefont {Battaglia}, \citenamefont {St\"ohlker},\
  and\ \citenamefont {Leemans}}]{PhysRevLett.109.064802}%
  \BibitemOpen
  \bibfield  {author} {\bibinfo {author} {\bibfnamefont {G.~R.}\ \bibnamefont
  {Plateau}}, \bibinfo {author} {\bibfnamefont {C.~G.~R.}\ \bibnamefont
  {Geddes}}, \bibinfo {author} {\bibfnamefont {D.~B.}\ \bibnamefont {Thorn}},
  \bibinfo {author} {\bibfnamefont {M.}~\bibnamefont {Chen}}, \bibinfo {author}
  {\bibfnamefont {C.}~\bibnamefont {Benedetti}}, \bibinfo {author}
  {\bibfnamefont {E.}~\bibnamefont {Esarey}}, \bibinfo {author} {\bibfnamefont
  {A.~J.}\ \bibnamefont {Gonsalves}}, \bibinfo {author} {\bibfnamefont {N.~H.}\
  \bibnamefont {Matlis}}, \bibinfo {author} {\bibfnamefont {K.}~\bibnamefont
  {Nakamura}}, \bibinfo {author} {\bibfnamefont {C.~B.}\ \bibnamefont
  {Schroeder}}, , \bibinfo {author} {\bibfnamefont {S.}~\bibnamefont
  {Shiraishi}}, \bibinfo {author} {\bibfnamefont {T.}~\bibnamefont {Sokollik}},
  \bibinfo {author} {\bibfnamefont {J.}~\bibnamefont {van Tilborg}}, \bibinfo
  {author} {\bibfnamefont {C.}~\bibnamefont {Toth}}, \bibinfo {author}
  {\bibfnamefont {S.}~\bibnamefont {Trotsenko}}, \bibinfo {author}
  {\bibfnamefont {T.~S.}\ \bibnamefont {Kim}}, \bibinfo {author} {\bibfnamefont
  {M.}~\bibnamefont {Battaglia}}, \bibinfo {author} {\bibfnamefont
  {T.}~\bibnamefont {St\"ohlker}}, \ and\ \bibinfo {author} {\bibfnamefont
  {W.~P.}\ \bibnamefont {Leemans}},\ }\bibfield  {title} {\enquote {\bibinfo
  {title} {Low-emittance electron bunches from a laser-plasma accelerator
  measured using single-shot x-ray spectroscopy},}\ }\href {\doibase
  10.1103/physrevlett.109.064802} {\bibfield  {journal} {\bibinfo  {journal}
  {Phys. Rev. Lett.}\ }\textbf {\bibinfo {volume} {109}},\ \bibinfo {pages}
  {064802} (\bibinfo {year} {2012})}\BibitemShut {NoStop}%
\bibitem [{\citenamefont {Burke}\ \emph {et~al.}(1997)\citenamefont {Burke},
  \citenamefont {Field}, \citenamefont {Horton-Smith}, \citenamefont {Spencer},
  \citenamefont {Walz}, \citenamefont {Berridge}, \citenamefont {Bugg},
  \citenamefont {Shmakov}, \citenamefont {Weidemann}, \citenamefont {Bula},\
  and\ \citenamefont {et~al.}}]{Burke:Field:etal:1997:Positron_Production}%
  \BibitemOpen
  \bibfield  {author} {\bibinfo {author} {\bibfnamefont {D.~L.}\ \bibnamefont
  {Burke}}, \bibinfo {author} {\bibfnamefont {R.~C.}\ \bibnamefont {Field}},
  \bibinfo {author} {\bibfnamefont {G.}~\bibnamefont {Horton-Smith}}, \bibinfo
  {author} {\bibfnamefont {J.~E.}\ \bibnamefont {Spencer}}, \bibinfo {author}
  {\bibfnamefont {D.}~\bibnamefont {Walz}}, \bibinfo {author} {\bibfnamefont
  {S.~C.}\ \bibnamefont {Berridge}}, \bibinfo {author} {\bibfnamefont {W.~M.}\
  \bibnamefont {Bugg}}, \bibinfo {author} {\bibfnamefont {K.}~\bibnamefont
  {Shmakov}}, \bibinfo {author} {\bibfnamefont {A.~W.}\ \bibnamefont
  {Weidemann}}, \bibinfo {author} {\bibfnamefont {C.}~\bibnamefont {Bula}}, \
  and\ \bibinfo {author} {\bibnamefont {et~al.}},\ }\bibfield  {title}
  {\enquote {\bibinfo {title} {Positron production in multiphoton
  light-by-light scattering},}\ }\href {\doibase 10.1103/physrevlett.79.1626}
  {\bibfield  {journal} {\bibinfo  {journal} {Phys. Rev. Lett.}\ }\textbf
  {\bibinfo {volume} {79}},\ \bibinfo {pages} {1626} (\bibinfo {year}
  {1997})}\BibitemShut {NoStop}%
\bibitem [{\citenamefont {Ta~Phuoc}\ \emph {et~al.}(2012)\citenamefont
  {Ta~Phuoc}, \citenamefont {Corde}, \citenamefont {Thaury}, \citenamefont
  {Malka}, \citenamefont {Tafzi}, \citenamefont {Goddet}, \citenamefont {Shah},
  \citenamefont {Sebban},\ and\ \citenamefont
  {Rousse}}]{Ta-Phuoc:Corde:etal:2012:All-optical_Compton}%
  \BibitemOpen
  \bibfield  {author} {\bibinfo {author} {\bibfnamefont {K.}~\bibnamefont
  {Ta~Phuoc}}, \bibinfo {author} {\bibfnamefont {S.}~\bibnamefont {Corde}},
  \bibinfo {author} {\bibfnamefont {C.}~\bibnamefont {Thaury}}, \bibinfo
  {author} {\bibfnamefont {V.}~\bibnamefont {Malka}}, \bibinfo {author}
  {\bibfnamefont {A.}~\bibnamefont {Tafzi}}, \bibinfo {author} {\bibfnamefont
  {J.~P.}\ \bibnamefont {Goddet}}, \bibinfo {author} {\bibfnamefont {R.~C.}\
  \bibnamefont {Shah}}, \bibinfo {author} {\bibfnamefont {S.}~\bibnamefont
  {Sebban}}, \ and\ \bibinfo {author} {\bibfnamefont {A.}~\bibnamefont
  {Rousse}},\ }\bibfield  {title} {\enquote {\bibinfo {title} {All-optical
  compton gamma-ray source},}\ }\href {\doibase 10.1038/nphoton.2012.82}
  {\bibfield  {journal} {\bibinfo  {journal} {Nat. Photon.}\ }\textbf {\bibinfo
  {volume} {6}},\ \bibinfo {pages} {308} (\bibinfo {year} {2012})}\BibitemShut
  {NoStop}%
\end{thebibliography}%

\end{document}